\documentclass{vldb}
\usepackage{enumitem}
\usepackage{graphicx}
\usepackage{balance}  
\usepackage{algorithm}
\usepackage{algpseudocode} 
\usepackage{tikz}
\usepackage{pgfplots}
\usetikzlibrary{patterns,shadows}
\usepgfplotslibrary{external}
\colorlet{darkgreen}{green!50!black}
\colorlet{darkred}{red!90!black}
\colorlet{darkyellow}{yellow!90!black}

\tikzset{
  gnode/.style={
    fill=white,
    draw=black,
    circle,
    very thick, 
    inner sep=3.5,
    drop shadow={shadow xshift=0.3ex,shadow yshift=-0.5ex, path
      fading={circle with fuzzy edge 20 percent}}
  }
}

\tikzset{
  rnode/.style={
    fill=black,
    draw=black,
    circle,
    very thick, 
    inner sep=3.5,
    drop shadow={shadow xshift=0.3ex,shadow yshift=-0.5ex, path
      fading={circle with fuzzy edge 20 percent}}
  }
}

\tikzset{
  ynode/.style={
    fill=black!50!white,
    draw=black,
    circle,
    very thick, 
    inner sep=3.5,
    drop shadow={shadow xshift=0.3ex,shadow yshift=-0.5ex, path
      fading={circle with fuzzy edge 20 percent}}
  }
}

\usepackage{caption}
\DeclareCaptionType{copyrightbox}
\usepackage{subcaption}
\usepackage{hyperref}
\usepackage{color}
\usepackage{amsmath}

\usepackage[sort&compress,numbers]{natbib}

\input{Definitions}

\begin{document}

  \def\gridwidth{2}
  \def\nodenum{4}
  \def\asnwidth{0.5}
  \def\asnmargin{0.05}

\newcommand{\asnbox}[3]{
  \fill [#3!25!lightgray]
  (#1 * \asnwidth + \asnmargin, 
  \gridwidth * 3 - #2 * \gridwidth + \asnmargin)
  rectangle
  (#1 * \asnwidth + \asnwidth - \asnmargin, 
  \gridwidth * 3 - #2 * \gridwidth + \gridwidth - \asnmargin);
  \draw
  (#1 * \asnwidth + \asnmargin, 
  \gridwidth * 3 - #2 * \gridwidth + \asnmargin)
  rectangle
  (#1 * \asnwidth + \asnwidth - \asnmargin, 
  \gridwidth * 3 - #2 * \gridwidth + \gridwidth - \asnmargin);
}

  \newcommand{\dasnbox}[2]{
    \draw[dashed]
    (#1 * \asnwidth + \asnmargin, 
    \gridwidth * 3 - #2 * \gridwidth + \asnmargin)
    rectangle
    (#1 * \asnwidth + \asnwidth - \asnmargin, 
    \gridwidth * 3 - #2 * \gridwidth + \gridwidth - \asnmargin);
  }

  \newcommand{\dtpt}[2]{
    \draw (#1 * \asnwidth + 0.5 * \asnwidth, 
    #2 * \asnwidth + 0.5 * \asnwidth) 
    node {\tiny{x}};
  }

  \newcommand{\mvarr}[4]{
    \draw [->]
    (#1 * \asnwidth + 0.5 * \asnwidth, 
    3 * \gridwidth - #2 * \gridwidth + 0.5 * \gridwidth) 
    --
    (#3 * \asnwidth + 0.5 * \asnwidth, 
    3 * \gridwidth - #4 * \gridwidth + 0.5 * \gridwidth);
  }
  
  \newcommand{\boxlayout}{
    \draw (0,0) rectangle (\gridwidth * \nodenum, \gridwidth * \nodenum);

    \fill [red!10] (0, 3 * \gridwidth) 
    rectangle (\gridwidth * \nodenum, 4 * \gridwidth);

    \fill [green!10] (0, 2 * \gridwidth) 
    rectangle (\gridwidth * \nodenum, 3 * \gridwidth);

    \fill [blue!10] (0, 1 * \gridwidth) 
    rectangle (\gridwidth * \nodenum, 2 * \gridwidth);

    \fill [brown!10] (0,0) 
    rectangle (\gridwidth * \nodenum, 1 * \gridwidth);

    \foreach \y in {1,2,3}
    {
      \draw[densely dashed] 
      (-\gridwidth * 0.1, \gridwidth * \y)
      --
      (\gridwidth * \nodenum + \gridwidth * 0.1,  \gridwidth * \y);
    }

    \dtpt{5}{1}  \dtpt{0}{15}  \dtpt{13}{12}  \dtpt{13}{6}  \dtpt{14}{8}  \dtpt{0}{8}  \dtpt{7}{10}  \dtpt{15}{2}  \dtpt{10}{5}  \dtpt{8}{13}  \dtpt{5}{3}  \dtpt{12}{5}  \dtpt{10}{7}  \dtpt{11}{6}  \dtpt{15}{3}  \dtpt{6}{9}  \dtpt{13}{5}  \dtpt{1}{3}  \dtpt{8}{15}  \dtpt{1}{6}  \dtpt{1}{7}  \dtpt{8}{9}  \dtpt{12}{4}  \dtpt{12}{1}  \dtpt{13}{10}  \dtpt{9}{0}  \dtpt{8}{4}  \dtpt{15}{12}  \dtpt{3}{6}  \dtpt{10}{13} 
  }

  \newcommand{\ptrna}{crosshatch}
  \newcommand{\ptrnb}{north west lines}
  \newcommand{\ptrnc}{north east lines}
  \newcommand{\ptrnd}{horizontal lines}

  \newcommand{\ptcla}{red}
  \newcommand{\ptclb}{green}
  \newcommand{\ptclc}{blue}
  \newcommand{\ptcld}{brown}

  \newcommand{\drawrowpart}{
    \draw[pattern=\ptrnd, pattern color=\ptcld] 
    (-\gridwidth - \gridwidth * 0.25, 0 * \gridwidth) 
    rectangle 
    (- \gridwidth * 0.25, 
    0 * \gridwidth + 1 * \gridwidth);
    \draw[pattern=\ptrnc, pattern color=\ptclc] 
    (-\gridwidth - \gridwidth * 0.25, 
    1 * \gridwidth) 
    rectangle 
    (- \gridwidth * 0.25, 
    1 * \gridwidth + 1 * \gridwidth);
    
    \draw[pattern=\ptrnb, pattern color=\ptclb] 
    (-\gridwidth - \gridwidth * 0.25, 
    2 * \gridwidth) 
    rectangle 
    (- \gridwidth * 0.25, 
    2 * \gridwidth + 1 * \gridwidth);
    
    \draw[pattern=\ptrna, pattern color=\ptcla] 
    (-\gridwidth - \gridwidth * 0.25, 
    3 * \gridwidth) 
    rectangle 
    (- \gridwidth * 0.25, 
    3 * \gridwidth + 1 * \gridwidth);
  }
  
  \newcommand{\drawcolpart}[4]{
    \draw[pattern=#3, pattern color=#4] 
    (#1 * \asnwidth, 
    4 * \gridwidth + 0.25 * \gridwidth) 
    rectangle 
    (#2 * \asnwidth + \asnwidth, 
    4 * \gridwidth + 0.25 * \gridwidth + \gridwidth);
  }

\title{NOMAD: Non-locking, stOchastic Multi-machine algorithm
  for Asynchronous and Decentralized matrix completion}

\numberofauthors{5} 
\author{
  \alignauthor
  Hyokyun Yun \\ 
  \affaddr{Purdue University}\\
  \email{yun3@purdue.edu}
  \alignauthor
  Hsiang-Fu Yu\\ 
  \affaddr{University of Texas, Austin}\\
  \email{rofuyu@cs.utexas.edu}
  \alignauthor
  Cho-Jui Hsieh \\ 
  \affaddr{University of Texas, Austin}\\
  \email{cjhsieh@cs.utexas.edu}
  \and 
  \alignauthor
  S V N Vishwanathan\\ 
  \affaddr{Purdue University}\\
  \email{vishy@stat.purdue.edu}
  \alignauthor
  Inderjit Dhillon \\ 
  \affaddr{University of Texas, Austin}\\
  \email{inderjit@cs.utexas.edu}
}
\date{\today}

\maketitle

\begin{abstract}
  We develop an efficient parallel distributed algorithm for matrix
  completion, named NOMAD (Non-locking, stOchastic Multi-machine
  algorithm for Asynchronous and Decentralized matrix completion).
  NOMAD is a decentralized algorithm with non-blocking communication
  between processors. One of the key features of NOMAD is that the
  ownership of a variable is asynchronously transferred between
  processors in a decentralized fashion. As a consequence it is a
  lock-free parallel algorithm. In spite of being an asynchronous
  algorithm, the variable updates of NOMAD are serializable, that is,
  there is an equivalent update ordering in a serial implementation.
  NOMAD outperforms synchronous algorithms which require explicit bulk
  synchronization after every iteration: our extensive empirical
  evaluation shows that not only does our algorithm perform well in
  distributed setting on commodity hardware, but also outperforms
  state-of-the-art algorithms on a HPC cluster both in multi-core and
  distributed memory settings.
\end{abstract}

\section{Introduction}
\label{sec:Introduction}


The aim of this paper is to develop an efficient parallel distributed
algorithm for matrix completion.  We are specifically interested in
solving large industrial scale matrix completion problems on commodity
hardware with limited computing power, memory, and interconnect speed,
such as the ones found in data centers. The widespread availability of
cloud computing platforms such as Amazon Web Services (AWS) make the
deployment of such systems feasible.

However, existing algorithms for matrix completion are designed for
conventional high performance computing (HPC) platforms. In order to
deploy them on commodity hardware we need to employ a large number of
machines, which increases inter-machine communication. Since the network
bandwidth in data centers is significantly lower and less-reliable than
the high-speed interconnects typically found in HPC hardware, this can
often have disastrous consequences in terms of convergence speed or the
quality of the solution. 

In this paper, we present NOMAD (Non-locking, stOchastic Multi-machine
algorithm for Asynchronous and Decentralized matrix completion), a new
parallel algorithm for matrix completion with the following properties: 
\begin{itemize}
\item Non-blocking communication: Processors exchange messages in an
  asynchronous fashion \citep{BerTsi97a}, and there is no bulk
  synchronization.
\item Decentralized: Processors are symmetric to each other, and each
  processor does the same amount of computation and communication.
\item Lock free: Using an \emph{owner computes} paradigm, we completely
  eliminate the need for locking variables. 
\item Fully asynchronous computation: Because of the lock free nature
  of our algorithm, the variable updates in individual processors are
  fully asynchronous.
\item Serializability: There is an equivalent update ordering in a
  serial implementation. In our algorithm \emph{stale} parameters are
  never used and this empirically leads to faster convergence
  \citep{LowGonKyrBicetal12}.
\end{itemize}
Our extensive empirical evaluation shows that not only does our
algorithm perform well in distributed setting on commodity hardware, but
also outperforms state-of-the-art algorithms on a HPC cluster both in
multi-core and distributed memory settings. We show that our algorithm
is significantly better than existing multi-core and multi-machine
algorithms for the matrix completion problem.

This paper is organized as follows: Section~\ref{sec:prior_work}
establishes some notation and introduces the matrix completion problem
formally. Section~\ref{sec:NOMAD} is devoted to describing NOMAD. We
contrast NOMAD with existing work in Section~\ref{sec:RelatedWork}.  In
Section~\ref{sec:Experiments} we present extensive empirical comparison
of NOMAD with various existing
algorithms. Section~\ref{sec:ConclusionFutureWork} concludes the paper
with a discussion.

\section{Background}
\label{sec:prior_work}

Let $A \in \RR^{m \times n}$ be a rating matrix, where $m$ denotes the
number of users and $n$ the number of items. Typically $m \gg n$,
although the algorithms we consider in this paper do not depend on such
an assumption. Furthermore, let $\Omega \subseteq \cbr{1\ldots m} \times
\cbr{1,\ldots,n}$ denote the observed entries of $A$, that is, $(i,j)
\in \Omega$ implies that user $i$ gave item $j$ a rating of $A_{ij}$.
The goal here is to predict accurately the unobserved ratings.  For
convenience, we define $\Omega_{i}$ to be the set of items rated by the
$i$-th user, i.e., $ \Omega_{i} := \{ j : (i,j) \in \Omega \}$.
Analogously $\Omegabar_j := \{ i : (i,j) \in \Omega \}$ is the set of
users who have rated item $j$.  Also, let $\ab_{i}^{\top}$ denote the
$i$-th row of $A$.

One popular model for matrix completion finds matrices $W \in \RR^{m
  \times k}$ and $H \in \RR^{n \times k}$, with $k \ll \min(m, n)$, such
that $A \approx W H^{\top}$.  One way to understand this model is to
realize that each row $\wb_{i}^{\top} \in \RR^{k}$ of $W$ can be thought
of as a $k$-dimensional embedding of the user. Analogously, each row
$\hb_{j}^{\top} \in \RR^{k}$ of $H$ is an embedding of the item in the
same $k$-dimensional space. In order to predict the $(i,j)$-th entry of
$A$ we simply use $\inner{\wb_{i}}{\hb_{j}}$, where
$\inner{\cdot}{\cdot}$ denotes the Euclidean inner product of two
vectors. The goodness of fit of the model is measured by a loss
function. While our optimization algorithm can work with an arbitrary
separable loss, for ease of exposition we will only discuss the square
loss: $\frac{1}{2} \rbr{A_{ij} -
  \inner{\wb_{i}}{\hb_{j}}}^{2}$. Furthermore, we need to enforce
regularization to prevent over-fitting, and to predict well on the
unknown entries of $A$. Again, a variety of regularizers can be handled
by our algorithm, but we will only focus on the following weighted
square norm-regularization in this paper: $ \frac \lambda 2
\sum_{i=1}^{m} \abr{\Omega_{i}} \cdot \nbr{\wb_{i}}^{2} + \frac \lambda
2 \sum_{j=1}^{n} \abr{\Omegabar_{j}} \cdot \nbr{\hb_{i}}^{2}$, where
$\lambda > 0$ is a regularization parameter.  Here, $| \cdot |$ denotes
the cardinality of a set, and $\nbr{\cdot}^{2}$ is the $L_2$ norm of a
vector. Putting everything together yields the following objective
function:
\begin{align}
  \label{eq:obj_fcn}
  \min_{
    \tiny{
      \begin{array}{cc}
        W \in \RR^{m \times k} \\
        H \in \RR^{n \times k}
      \end{array}
    }
  }
  & J(W,H) := \frac{1}{2} \sum_{(i,j) \in \Omega}
  \rbr{A_{ij} - \inner{\wb_{i}}{\hb_{j}}}^{2} \nonumber \\
  & + \frac{\lambda}{2} \rbr{ \sum_{i=1}^{m} \abr{\Omega_{i}} \cdot
    \nbr{\wb_{i}}^{2} +
    \sum_{j=1}^{n} \abr{\Omegabar_{j}} \cdot \nbr{\hb_{i}}^{2} }. 
\end{align}
This can be further simplified and written as 
\begin{align*}
  J(W,H) = \frac{1}{2} \sum_{(i,j) \in \Omega} \cbr{ \rbr{A_{ij} -
      \inner{\wb_{i}}{\hb_{j}}}^2 + \lambda \rbr{\nbr{\wb_{i}}^{2} +
      \nbr{\hb_{j}}^{2}}}.
\end{align*}
In the above equations, $\lambda > 0$ is a scalar which trades off the
loss function with the regularizer.

In the sequel we will let $w_{il}$ and $h_{il}$ for $1 \leq l \leq k$
denote the $l$-th coordinate of the column vectors $\wb_i$ and $\hb_j$,
respectively.  Furthermore, $H_{\Omega_i}$ (resp.\ $W_{\Omegabar_{j}}$)
will be used to denote the sub-matrix of $H$ (resp.\ $W$) formed by
collecting rows corresponding to $\Omega_i$ (resp.\ $\Omegabar_{j}$).

Note the following property of the above objective function
\eqref{eq:obj_fcn}: If we fix $H$ then the problem decomposes to $m$
independent convex optimization problems, each of which has the
following form:
\begin{align}
  \label{eq:fix-h}
  \min_{\wb_i \in \RR^{k}} J_{i}(\wb_i) = \frac 1 2 \sum_{j \in
    \Omega_{i}} \rbr{A_{ij} - \inner{\wb_{i}}{\hb_{j}}}^{2} +
  \lambda \nbr{\wb_i}^{2}.
\end{align}
Analogously, if we fix $W$ then \eqref{eq:obj_fcn} decomposes into
$n$ independent convex optimization problems, each of which has the
following form:
\begin{align*}
  \min_{\hb_j \in \RR^{k}} \Jbar_{j}(\hb_j) = 
  \frac 1 2 \sum_{i \in \bar{\Omega}_{j}}
  \rbr{A_{ij} - \inner{\wb_{i}}{\hb_{j}}}^{2} + \lambda
  \nbr{\hb_j}^{2}.
\end{align*}
The gradient and Hessian of $J_{i}(\wb)$ can be easily computed:
\begin{align*}
  \nabla J_{i}(\wb_i) = M \wb_i - \bb, \text{ and } \nabla^{2}
  J_{i}(\wb_i) = M,
\end{align*}
where we have defined
$ M := H_{\Omega_{i}}^{\top} H_{\Omega_{i}} + \lambda I \text{ and } \bb
:= H^{\top} \ab_{i}$.

We will now present three well known optimization strategies for solving
\eqref{eq:obj_fcn}, which essentially differ in only two characteristics
namely, the sequence in which updates to the variables in $W$ and $H$
are carried out, and the level of approximation in the update.

\subsection{Alternating Least Squares}
\label{sec:AlternLeastSquar}

A simple version of the Alternating Least Squares (ALS) algorithm
updates variables as follows: $\wb_{1}$, $\wb_{2}$, $\ldots$, $\wb_{m}$,
$\hb_{1}$, $\hb_{2}$, $\ldots$, $\hb_{n}$, $\wb_{1}$, $\ldots$ and so
on. Updates to $\wb_{i}$ are computed by solving \eqref{eq:fix-h} which
is in fact a least squares problem, and thus the following Newton update
gives us: 
\begin{align}
  \label{eq:newton-fix-h}
  \wb_{i} & \leftarrow \wb_{i} - \sbr{\nabla^{2}
    J_{i}(\wb_{i})}^{-1} \nabla J_{i}(\wb_{i}),
\end{align}
which can be rewritten using $M$ and $\bb$ as $ \wb_{i} \leftarrow
M^{-1} \bb$.  Updates to $\hb_j$'s are analogous.

\subsection{Coordinate Descent}
\label{sec:CoordinateDescent}

The ALS update involves formation of the Hessian and its inversion.  In
order to reduce the computational complexity, one can replace the
Hessian by its diagonal approximation:
\begin{align}
  \label{eq:ccd-newton}
  \wb_{i} & \leftarrow 
  \wb_{i} -
  \sbr{\diag\rbr{\nabla^{2} J_{i}\rbr{\wb_{i}}}}^{-1} \nabla
  J_{i}\rbr{\wb_{i}},
\end{align}
which can be rewritten using $M$ and $\bb$ as 
\begin{align}
  \label{eq:ccd-newton-rewrite}
  \wb_{i} & \leftarrow \wb_{i} - \diag(M)^{-1} \sbr{M \wb_{i} - \bb}.
\end{align}
If we update one component of $\wb_i$ at a time, the update
\eqref{eq:ccd-newton-rewrite} can be written as:
\begin{align}
  \label{eq:ccd-coord-update}
  w_{il} \leftarrow w_{il} - \frac{\inner{\mb_{l}}{\wb_{i}} -
    b_{l}}{m_{ll}},
\end{align}
where $\mb_{l}$ is $l$-th row of matrix $M$, $b_{l}$ is $l$-th component
of $\bb$ and $m_{ll}$ is the $l$-th coordinate of $\mb_{l}$.

If we choose the update sequence $w_{11}$, $\ldots$, $w_{1k}$, $w_{21}$,
$\ldots$, $w_{2k}$, $\ldots$, $w_{m1}$, $\ldots$, $w_{mk}$, $h_{11}$,
$\ldots$, $h_{1k}$, $h_{21}$, $\ldots$, $h_{2k}$, \ldots, $h_{n1}$,
$\ldots$, $h_{nk}$, $w_{11}$, $\ldots$, $w_{1k}$, and so on, then this
recovers Cyclic Coordinate Descent (CCD)~\cite{HsiDhi11}. On the other
hand, the update sequence $w_{11}$, $\ldots$, $w_{m1}$, $h_{11}$,
$\ldots$, $h_{n1}$, $w_{12}$, $\ldots$, $w_{m2}$, $h_{12}$, $\ldots$,
$h_{n2}$ and so on, recovers the CCD++ algorithm
of~\citet{YuHsiSiDhi12}. The CCD++ updates can be performed more
efficiently than the CCD updates by maintaining a residual
matrix~\citep{YuHsiSiDhi12}.

\subsection{Stochastic Gradient Descent}
\label{sec:StochGradDesc}

The stochastic gradient descent (SGD) algorithm for matrix completion
can be motivated from its more classical version, gradient
descent. Given an objective function $f(\theta) = \frac{1}{m}
\sum_{i=1}^{m} f_{i}(\theta)$, the gradient descent update is
\begin{align}
  \label{eq:grd_desc}
  \theta \leftarrow \theta - s_{t} \cdot \nabla_{\theta} f(\theta),
\end{align}
where $t$ denotes the iteration number and $\cbr{s_{t}}$ is a sequence
of step sizes. The stochastic gradient descent update replaces
$\nabla_{\theta} f(\theta)$ by its unbiased estimate $\nabla_{\theta}
f_{i}(\theta)$, which yields
\begin{align}
  \label{eq:stoch_grd_desc}
  \theta \leftarrow \theta - s_{t} \cdot \nabla_{\theta} f_{i}(\theta). 
\end{align}
It is significantly cheaper to evaluate $\nabla_{\theta} f_{i}(\theta)$
as compared to $\nabla_{\theta} f(\theta)$. One can show that for
sufficiently large $t$, the above updates will converge to a fixed point
of $f$ \citep{KusCla78,RobMon51}. The above update also enjoys desirable
properties in terms of sample complexity, and hence is widely used in
machine learning \citep{BotBou11,ShaSre08}. 

For the matrix completion problem, note that for a fixed $(i,j)$ pair,
the gradient of \eqref{eq:obj_fcn} can be written as 
\begin{align*}
  \nabla_{\wb_{i}} J(W, H) & = (A_{ij} - \inner{\wb_{i}}{\hb_{j}}) \;
  \hb_{j} + \lambda \wb_{i}
  \text { and } \\
  \nabla_{\hb_{j}} J(W, H) & = (A_{ij} - \inner{\wb_{i}}{\hb_{j}}) \;
  \wb_{i} + \lambda \hb_{j},
\end{align*}
Therefore the SGD updates require sampling a random index $\rbr{i_{t},
  j_{t}}$ uniformly from the set of nonzero indicies $\Omega$, and
performing the update
\begin{align}
  \label{eq:w_update}
  \wb_{i_t} &\leftarrow
  \wb_{i_t} - 
  s_t \cdot
  \sbr{(A_{{i_t}{j_t}} - \wb_{i_t} \hb_{j_t}) \hb_{j_t}
    + \lambda \wb_{i_t}} \text { and }
  \\
  \label{eq:h_update}
  \hb_{j_t} &\leftarrow
  \hb_{j_t} - 
  s_t \cdot
  \sbr{(A_{{i_t}{j_t}} - \wb_{i_t} \hb_{j_t}) \wb_{j_t}
    + \lambda \hb_{j_t}}.
\end{align}

\section{NOMAD}
\label{sec:NOMAD}

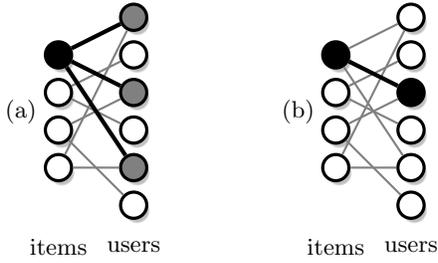
\begin{figure}[tbp]
  \centering
  \begin{tikzpicture}[scale=0.5]
    \draw[help lines,white] (0,0) grid (6,3);
    \node at (2.5,-0.6) {users};
    \node at (0.5,-0.6) {items};
    \node[ynode] (u1) at (2.5,5.5) {};
    \node[gnode] (u2) at (2.5,4.5) {};
    \node[ynode] (u3) at (2.5,3.5) {};
    \node[gnode] (u4) at (2.5,2.5) {};
    \node[ynode] (u5) at (2.5,1.5) {};
    \node[gnode] (u6) at (2.5,0.5) {};

    \node[rnode] (m1) at (0.5,4.5) {};
    \node[gnode] (m2) at (0.5,3.5) {};
    \node[gnode] (m3) at (0.5,2.5) {};
    \node[gnode] (m4) at (0.5,1.5) {};

    \draw[-,ultra thick,black] (u1) -- (m1);
    \draw[-,thick,gray] (u1) -- (m4);

    \draw[-,thick,gray] (u2) -- (m2);

    \draw[-,ultra thick,black] (u3) -- (m1);
    \draw[-,thick,gray] (u3) -- (m3);

    \draw[-,thick,gray] (u4) -- (m2);

    \draw[-,ultra thick,black] (u5) -- (m1);
    \draw[-,thick,gray] (u5) -- (m4);

    \draw[-,thick,gray] (u6) -- (m3);

    \node at (-0.5,3.0) {(a)};
  \end{tikzpicture}
  \begin{tikzpicture}[scale=0.5]
    \draw[help lines,white] (0,0) grid (6,3);
    \node at (2.5, -0.6) {users};
    \node at (0.5, -0.6) {items};
    \node[gnode] (u1) at (2.5,5.5) {};
    \node[gnode] (u2) at (2.5,4.5) {};
    \node[rnode] (u3) at (2.5,3.5) {};
    \node[gnode] (u4) at (2.5,2.5) {};
    \node[gnode] (u5) at (2.5,1.5) {};
    \node[gnode] (u6) at (2.5,0.5) {};

    \node[rnode] (m1) at (0.5,4.5) {};
    \node[gnode] (m2) at (0.5,3.5) {};
    \node[gnode] (m3) at (0.5,2.5) {};
    \node[gnode] (m4) at (0.5,1.5) {};

    \draw[-,thick,gray] (u1) -- (m1);
    \draw[-,thick,gray] (u1) -- (m4);

    \draw[-,thick,gray] (u2) -- (m2);

    \draw[-,ultra thick,black] (u3) -- (m1);
    \draw[-,thick,gray] (u3) -- (m3);

    \draw[-,thick,gray] (u4) -- (m2);

    \draw[-,thick,gray] (u5) -- (m1);
    \draw[-,thick,gray] (u5) -- (m4);

    \draw[-,thick,gray] (u6) -- (m3);

    \node at (-0.5,3.0) {(b)};
  \end{tikzpicture}
  \caption{Illustration of updates used in matrix completion. Three
    algorithms are shown here: (a) alternating least squares and
    coordinate descent, (b) stochastic gradient descent. Black indicates
    that the value of the node is being updated, gray
    indicates that the value of the node is being read.  White nodes
    are neither being read nor updated. }
  \label{fig:mf}
\end{figure}

In NOMAD, we use an optimization scheme based on SGD. In order to
justify this choice, we find it instructive to first understand the
updates performed by ALS, coordinate descent, and SGD on a bipartite
graph which is constructed as follows: the $i$-th user node
corresponds to $\wb_{i}$, the $j$-th item node corresponds to
$\hb_{j}$, and an edge $(i,j)$ indicates that user $i$ has rated item
$j$ (see Figure~\ref{fig:mf}). Both the ALS update
\eqref{eq:newton-fix-h} and coordinate descent update
\eqref{eq:ccd-coord-update} for $\wb_{i}$ require us to access the
values of $\hb_{j}$ for all $j \in \Omega_{i}$. This is shown in
Figure~\ref{fig:mf} (a), where the black node corresponds to
$\wb_{i}$, while the gray nodes correspond to $\hb_{j}$ for $j \in
\Omega_{i}$.  On the other hand, the SGD update to $\wb_{i}$
\eqref{eq:w_update} only requires us to retrieve the value of
$\hb_{j}$ for a single random $j \in \Omega_{i}$ (Figure~\ref{fig:mf}
(b)). What this means is that in contrast to ALS or CCD, multiple SGD
updates can be carried out simultaneously in parallel, without
interfering with each other.  Put another way, SGD has higher
potential for finer-grained parallelism than other approaches, and
therefore we use it as our optimization scheme in NOMAD.

\subsection{Description}

For now, we will denote each parallel computing unit as a \emph{worker};
in a shared memory setting a worker is a thread and in a distributed
memory architecture a worker is a machine. This abstraction allows us to
present NOMAD in a unified manner. Of course, NOMAD can be used in a
hybrid setting where there are multiple threads spread across multiple
machines, and this will be discussed in Section~\ref{sec:multi_multi}.

In NOMAD, the users $\cbr{1,2,\ldots,m}$ are split into $p$ disjoint
sets $I_1,I_2,\ldots,I_p$ which are of approximately equal
size\footnote{An alternative strategy is to split the users such that
  each set has approximately the same number of ratings.}. This
induces a partition of the rows of the ratings matrix $A$. The $q$-th
worker stores $n$ sets of indices $\Omegabar_{j}^{(q)}$, for $j \in
\cbr{1, \ldots, n}$, which are defined as
\begin{align*}
  \Omegabar_{j}^{(q)} := \cbr{(i,j) \in \Omegabar_j; i \in I_q },
\end{align*}
as well as the corresponding values of $A$. Note that once the data is
partitioned and distributed to the workers, it is never moved during
the execution of the algorithm. 

Recall that there are two types of parameters in matrix completion:
user parameters $\wb_i$'s, and item parameters $\hb_j$'s.  In NOMAD,
$\wb_i$'s are partitioned according to $I_1, I_2, \ldots, I_p$, that
is, the $q$-th worker stores and updates $\wb_{i}$ for $i \in I_{q}$.
The variables in $W$ are partitioned at the beginning, and never move
across workers during the execution of the algorithm.  On the other
hand, the $\hb_{j}$'s are split randomly into $p$ partitions at the
beginning, and their ownership changes as the algorithm progresses. At
each point of time an $\hb_j$ variable resides in one and only worker,
and it moves to another worker after it is processed, independent of
other item variables.  Hence these are \emph{nomadic}
variables\footnote{Due to symmetry in the formulation of the matrix
  completion problem, one can also make the $\wb_{i}$'s nomadic and
  partition the $\hb_{j}$'s. Since usually the number of users is much
  larger than the number of items, this leads to more communication
  and therefore we make the $\hb_{j}$ variables nomadic.}.

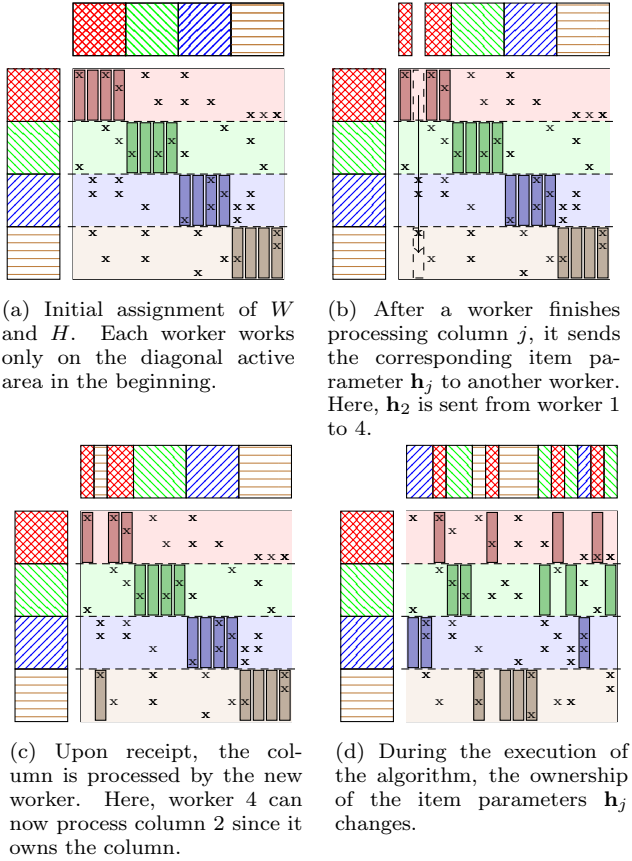
\begin{figure}[htbp]

      






  

  \begin{subfigure}[t]{0.22\textwidth}
    \centering
    \begin{tikzpicture}[scale=0.35]
      
      \draw (0,0) rectangle (\gridwidth * \nodenum, \gridwidth * \nodenum);
  
    \fill [red!10] (0, 3 * \gridwidth) 
    rectangle (\gridwidth * \nodenum, 4 * \gridwidth);

    \fill [green!10] (0, 2 * \gridwidth) 
    rectangle (\gridwidth * \nodenum, 3 * \gridwidth);

    \fill [blue!10] (0, 1 * \gridwidth) 
    rectangle (\gridwidth * \nodenum, 2 * \gridwidth);

    \fill [brown!10] (0,0) 
    rectangle (\gridwidth * \nodenum, 1 * \gridwidth);

    \foreach \y in {1,2,3}
    {
      \draw[densely dashed] 
      (-\gridwidth * 0.1, \gridwidth * \y)
      --
      (\gridwidth * \nodenum + \gridwidth * 0.1,  \gridwidth * \y);
    }
    
    \drawrowpart

    \drawcolpart{0}{3}{\ptrna}{\ptcla}
    \drawcolpart{4}{7}{\ptrnb}{\ptclb}
    \drawcolpart{8}{11}{\ptrnc}{\ptclc}
    \drawcolpart{12}{15}{\ptrnd}{\ptcld}

    \dtpt{5}{1} \dtpt{0}{15} \dtpt{13}{12} \dtpt{13}{6} \dtpt{14}{8}
    \dtpt{0}{8} \dtpt{7}{10} \dtpt{15}{2} \dtpt{10}{5} \dtpt{8}{13}
    \dtpt{5}{3} \dtpt{12}{5} \dtpt{10}{7} \dtpt{11}{6} \dtpt{15}{3}
    \dtpt{6}{9} \dtpt{13}{5} \dtpt{1}{3} \dtpt{8}{15} \dtpt{1}{6}
    \dtpt{1}{7} \dtpt{8}{9} \dtpt{12}{4} \dtpt{12}{1} \dtpt{13}{10}
    \dtpt{9}{0} \dtpt{8}{4} \dtpt{15}{12} \dtpt{3}{6} \dtpt{10}{13}
    \dtpt{6}{13} \dtpt{2}{11} \dtpt{5}{5} \dtpt{2}{1} \dtpt{3}{7}
    \dtpt{9}{3} \dtpt{11}{1} \dtpt{3}{14} \dtpt{11}{11} \dtpt{5}{15}

    \drawcolpart{0}{3}{\ptrna}{\ptcla}
    \drawcolpart{4}{7}{\ptrnb}{\ptclb}
    \drawcolpart{8}{11}{\ptrnc}{\ptclc}
    \drawcolpart{12}{15}{\ptrnd}{\ptcld}

    \asnbox{0}{0}{red} \asnbox{1}{0}{red} \asnbox{2}{0}{red} \asnbox{3}{0}{red}
    \asnbox{4}{1}{green} \asnbox{5}{1}{green} \asnbox{6}{1}{green} \asnbox{7}{1}{green}
    \asnbox{8}{2}{blue} \asnbox{9}{2}{blue} \asnbox{10}{2}{blue} \asnbox{11}{2}{blue}
    \asnbox{12}{3}{brown} \asnbox{13}{3}{brown} \asnbox{14}{3}{brown} \asnbox{15}{3}{brown}
      
    \dtpt{5}{1} \dtpt{0}{15} \dtpt{13}{12} \dtpt{13}{6} \dtpt{14}{8}
    \dtpt{0}{8} \dtpt{7}{10} \dtpt{15}{2} \dtpt{10}{5} \dtpt{8}{13}
    \dtpt{5}{3} \dtpt{12}{5} \dtpt{10}{7} \dtpt{11}{6} \dtpt{15}{3}
    \dtpt{6}{9} \dtpt{13}{5} \dtpt{1}{3} \dtpt{8}{15} \dtpt{1}{6}
    \dtpt{1}{7} \dtpt{8}{9} \dtpt{12}{4} \dtpt{12}{1} \dtpt{13}{10}
    \dtpt{9}{0} \dtpt{8}{4} \dtpt{15}{12} \dtpt{3}{6} \dtpt{10}{13}
    \dtpt{6}{13} \dtpt{2}{11} \dtpt{5}{5} \dtpt{2}{1} \dtpt{3}{7}
    \dtpt{9}{3} \dtpt{11}{1} \dtpt{3}{14} \dtpt{11}{11} \dtpt{5}{15}

    \dtpt{2}{15} \dtpt{3}{10} \dtpt{4}{9} \dtpt{15}{12} \dtpt{14}{12}

    \end{tikzpicture}
    \caption{Initial assignment of $W$ and $H$. Each worker works
      only on the diagonal active area in the beginning.}
  \end{subfigure}
  \quad
  \begin{subfigure}[t]{0.22\textwidth}
    \centering
    \begin{tikzpicture}[scale=0.35]

      \boxlayout
      
      \drawrowpart

      \drawcolpart{0}{0}{\ptrna}{\ptcla}
      \drawcolpart{2}{3}{\ptrna}{\ptcla}
      \drawcolpart{4}{7}{\ptrnb}{\ptclb}
      \drawcolpart{8}{11}{\ptrnc}{\ptclc}
      \drawcolpart{12}{15}{\ptrnd}{\ptcld}

      \asnbox{0}{0}{red} \asnbox{2}{0}{red} \asnbox{3}{0}{red}
      \asnbox{4}{1}{green} \asnbox{5}{1}{green} \asnbox{6}{1}{green} \asnbox{7}{1}{green}
      \asnbox{8}{2}{blue} \asnbox{9}{2}{blue} \asnbox{10}{2}{blue} \asnbox{11}{2}{blue}
      \asnbox{12}{3}{brown} \asnbox{13}{3}{brown} \asnbox{14}{3}{brown} \asnbox{15}{3}{brown}
      
      \dasnbox{1}{0} \dasnbox{1}{3}
      \mvarr{1}{0}{1}{3}

      \dtpt{5}{1} \dtpt{0}{15} \dtpt{13}{12} \dtpt{13}{6} \dtpt{14}{8}
      \dtpt{0}{8} \dtpt{7}{10} \dtpt{15}{2} \dtpt{10}{5} \dtpt{8}{13}
      \dtpt{5}{3} \dtpt{12}{5} \dtpt{10}{7} \dtpt{11}{6} \dtpt{15}{3}
      \dtpt{6}{9} \dtpt{13}{5} \dtpt{1}{3} \dtpt{8}{15} \dtpt{1}{6}
      \dtpt{1}{7} \dtpt{8}{9} \dtpt{12}{4} \dtpt{12}{1} \dtpt{13}{10}
      \dtpt{9}{0} \dtpt{8}{4} \dtpt{15}{12} \dtpt{3}{6} \dtpt{10}{13}
      \dtpt{6}{13} \dtpt{2}{11} \dtpt{5}{5} \dtpt{2}{1} \dtpt{3}{7}
      \dtpt{9}{3} \dtpt{11}{1} \dtpt{3}{14} \dtpt{11}{11} \dtpt{5}{15}
      \dtpt{2}{15} \dtpt{3}{10} \dtpt{4}{9} \dtpt{15}{12} \dtpt{14}{12}

    \end{tikzpicture}
    \caption{After a worker finishes processing column $j$, it sends
      the corresponding item parameter $\hb_j$ to another worker.
      Here, $\hb_2$ is sent from worker $1$ to $4$.  }
  \end{subfigure}

  \centering
  \begin{subfigure}[t]{0.22\textwidth}
    \centering
    \begin{tikzpicture}[scale=0.35]

      \boxlayout
    
      \drawrowpart

      \drawcolpart{0}{0}{\ptrna}{\ptcla}
      \drawcolpart{1}{1}{\ptrnd}{\ptcld}
      \drawcolpart{2}{3}{\ptrna}{\ptcla}
      \drawcolpart{4}{7}{\ptrnb}{\ptclb}
      \drawcolpart{8}{11}{\ptrnc}{\ptclc}
      \drawcolpart{12}{15}{\ptrnd}{\ptcld}

      \asnbox{0}{0}{red} \asnbox{2}{0}{red} \asnbox{3}{0}{red}
      \asnbox{4}{1}{green} \asnbox{5}{1}{green} \asnbox{6}{1}{green} \asnbox{7}{1}{green}
      \asnbox{8}{2}{blue} \asnbox{9}{2}{blue} \asnbox{10}{2}{blue} \asnbox{11}{2}{blue}
      \asnbox{12}{3}{brown} \asnbox{13}{3}{brown} \asnbox{14}{3}{brown} \asnbox{15}{3}{brown}
      \asnbox{1}{3}{brown}

      \dtpt{5}{1} \dtpt{0}{15} \dtpt{13}{12} \dtpt{13}{6} \dtpt{14}{8}
      \dtpt{0}{8} \dtpt{7}{10} \dtpt{15}{2} \dtpt{10}{5} \dtpt{8}{13}
      \dtpt{5}{3} \dtpt{12}{5} \dtpt{10}{7} \dtpt{11}{6} \dtpt{15}{3}
      \dtpt{6}{9} \dtpt{13}{5} \dtpt{1}{3} \dtpt{8}{15} \dtpt{1}{6}
      \dtpt{1}{7} \dtpt{8}{9} \dtpt{12}{4} \dtpt{12}{1} \dtpt{13}{10}
      \dtpt{9}{0} \dtpt{8}{4} \dtpt{15}{12} \dtpt{3}{6} \dtpt{10}{13}
      \dtpt{6}{13} \dtpt{2}{11} \dtpt{5}{5} \dtpt{2}{1} \dtpt{3}{7}
      \dtpt{9}{3} \dtpt{11}{1} \dtpt{3}{14} \dtpt{11}{11} \dtpt{5}{15}
      \dtpt{2}{15} \dtpt{3}{10} \dtpt{4}{9} \dtpt{15}{12}
      \dtpt{14}{12}

    \end{tikzpicture}
    \caption{Upon receipt, the column is processed by the new worker.
      Here, worker $4$ can now process column $2$ since it owns the
      column.}
  \end{subfigure}
  \quad
  \begin{subfigure}[t]{0.22\textwidth}
    \centering
    \begin{tikzpicture}[scale=0.35]

      \boxlayout

      \drawrowpart

      \asnbox{0}{2}{blue} \asnbox{1}{2}{blue} \asnbox{2}{0}{red} \asnbox{3}{1}{green}
      \asnbox{4}{1}{green} \asnbox{5}{3}{brown} \asnbox{6}{0}{red} \asnbox{7}{3}{brown}
      \asnbox{8}{3}{brown} \asnbox{9}{3}{brown} \asnbox{10}{1}{green} \asnbox{11}{0}{red}
      \asnbox{12}{1}{green} \asnbox{13}{2}{blue} \asnbox{14}{0}{red} \asnbox{15}{1}{green}

      \drawcolpart{0}{1}{\ptrnc}{\ptclc}
      \drawcolpart{2}{2}{\ptrna}{\ptcla}
      \drawcolpart{3}{4}{\ptrnb}{\ptclb}
      \drawcolpart{5}{5}{\ptrnd}{\ptcld}
      \drawcolpart{6}{6}{\ptrna}{\ptcla}
      \drawcolpart{7}{9}{\ptrnd}{\ptcld}
      \drawcolpart{10}{10}{\ptrnb}{\ptclb}
      \drawcolpart{11}{11}{\ptrna}{\ptcla}
      \drawcolpart{12}{12}{\ptrnb}{\ptclb}
      \drawcolpart{13}{13}{\ptrnc}{\ptclc}
      \drawcolpart{14}{14}{\ptrna}{\ptcla}
      \drawcolpart{15}{15}{\ptrnb}{\ptclb}
      
      \dtpt{5}{1} \dtpt{0}{15} \dtpt{13}{12} \dtpt{13}{6} \dtpt{14}{8}
      \dtpt{0}{8} \dtpt{7}{10} \dtpt{15}{2} \dtpt{10}{5} \dtpt{8}{13}
      \dtpt{5}{3} \dtpt{12}{5} \dtpt{10}{7} \dtpt{11}{6} \dtpt{15}{3}
      \dtpt{6}{9} \dtpt{13}{5} \dtpt{1}{3} \dtpt{8}{15} \dtpt{1}{6}
      \dtpt{1}{7} \dtpt{8}{9} \dtpt{12}{4} \dtpt{12}{1} \dtpt{13}{10}
      \dtpt{9}{0} \dtpt{8}{4} \dtpt{15}{12} \dtpt{3}{6} \dtpt{10}{13}
      \dtpt{6}{13} \dtpt{2}{11} \dtpt{5}{5} \dtpt{2}{1} \dtpt{3}{7}
      \dtpt{9}{3} \dtpt{11}{1} \dtpt{3}{14} \dtpt{11}{11} \dtpt{5}{15}

      \dtpt{2}{15} \dtpt{3}{10} \dtpt{4}{9} \dtpt{15}{12} \dtpt{14}{12}

    \end{tikzpicture}
    \caption{During the execution of the algorithm, the ownership of the
      item parameters $\hb_j$ changes.}
  \end{subfigure}
  \caption{Illustration of the NOMAD algorithm}
  \label{fig:nomad_scheme}
\end{figure}

Processing an item variable $\hb_j$ at the $q$-th worker entails
executing SGD updates \eqref{eq:w_update} and \eqref{eq:h_update} on the
ratings in the set $\Omegabar_{j}^{(q)}$.  Note that these updates only
require access to $\hb_{j}$ and $\wb_i$ for $i \in I_q$; since $I_q$'s
are disjoint, each $\wb_i$ variable in the set is accessed by only one
worker.  This is why the communication of $\wb_i$ variables is not
necessary.  On the other hand, $\hb_j$ is updated only by the worker
that currently owns it, so there is no need for a lock; this is the
popular \emph{owner-computes} rule in parallel computing.  See
Figure~\ref{fig:nomad_scheme}.




We now formally define the NOMAD algorithm (see
Algorithm~\ref{alg:nomad} for detailed pseudo-code).  Each worker $q$
maintains its own concurrent queue, \texttt{queue}$[q]$, which
contains a list of items it has to process.  Each element of the list
consists of the index of the item $j$ ($1 \leq j \leq n$), and a
corresponding $k$-dimensional parameter vector $\hb_j$; this pair is
denoted as $(j, \hb_j)$.  Each worker $q$ pops a $(j, \hb_j)$ pair
from its own queue, \texttt{queue}$[q]$, and runs stochastic gradient
descent update on $\Omegabar_j^{(q)}$, which is the set of ratings on
item $j$ locally stored in worker $q$ (line~\ref{alg:sgd_for_start} to
\ref{alg:sgd_for_end}).  This changes values of $\wb_i$ for $i \in
I_q$ and $\hb_j$.  After all the updates on item $j$ are done, a
uniformly random worker $q'$ is sampled (line \ref{alg:sample_worker})
and the updated $(j, \hb_j)$ pair is pushed into the queue of that
worker, $q'$ (line~\ref{alg:push_to_queue}).  Note that this is the
only time where a worker communicates with another worker. Also note
that the nature of this communication is asynchronous and
non-blocking. Furthermore, as long as there are items in the queue,
the computations are completely asynchronous and
decentralized. Moreover, all workers are symmetric, that is, there is
no designated master or slave.

\algblockdefx[parallel]{Parstart}{Parend}
[1][]{\textbf{Parallel Foreach} #1}
{\textbf{Parallel End}}

\begin{algorithm}[htbp]
  \begin{algorithmic}[1]
    \State $\lambda$: regularization parameter
    \State $\cbr{s_t}$: step size sequence
    \State \texttt{// initialize parameters} 
    \State $w_{il} \sim \text{UniformReal}\rbr{0,\frac{1}{\sqrt{k}}}$ for 
      $1 \leq i \leq m, 1 \leq l \leq k$
    \State $h_{jl} \sim \text{UniformReal}\rbr{0,\frac{1}{\sqrt{k}}}$ for 
      $1 \leq j \leq n, 1 \leq l \leq k$
    \State \texttt{// initialize queues} 
    \For {$j \in \cbr{1,2,\ldots,n}$} 
      \State $q \sim \text{UniformDiscrete}\cbr{1,2,\ldots,p}$ 
      \State $\texttt{queue}[q] \texttt{.push}((j,\hb_j))$
    \EndFor
    \State \texttt{// start} $p$ \texttt{workers} 
    \Parstart{$q \in \cbr{1,2,\ldots,p}$}
      \While{stop signal is not yet received}
        \If {$\texttt{queue}[q]$ not empty}
          \State $(j, \hb_j) \gets \texttt{queue}[q]\texttt{.pop()}$ \label{alg:pop}
          \For{$(i,j) \in \Omegabar_j^{(q)}$} \label{alg:sgd_for_start}
            \State \texttt{// SGD update} 
            \State $t \leftarrow$ number of updates on $(i,j)$
            \State
            $\wb_{i} \leftarrow \wb_{i} - s_t \cdot
            \sbr{(A_{{i}{j}} - \wb_{i} \hb_{j}) \hb_{j}
              + \lambda \wb_{i}}$
            \State
            $\hb_{j} \leftarrow
            \hb_{j} - 
            s_t \cdot
            \sbr{(A_{{i}{j}} - \wb_{i} \hb_{j}) \wb_{j}
              + \lambda \hb_{j}}.$
          \EndFor  \label{alg:sgd_for_end}
          \State $q' \sim \text{UniformDiscrete}\cbr{1,2,\ldots,p}$
          \label{alg:sample_worker}
          \State $\texttt{queue}[q'] \texttt{.push}((j,\hb_j))$
          \label{alg:push_to_queue}
        \EndIf
      \EndWhile
    \Parend
  \end{algorithmic}
  \caption{the basic NOMAD algorithm}
  \label{alg:nomad}
\end{algorithm}

\subsection{Complexity Analysis}
\label{sec:complexity_analysis}


First, we consider the case when the problem is distributed across $p$
workers, and study how the space and time complexity behaves as a
function of $p$. Each worker has to store $1/p$ fraction of the $m$ user
parameters, and approximately $1/p$ fraction of the $n$ item
parameters. Furthermore, each worker also stores approximately $1/p$
fraction of the $\abr{\Omega}$ ratings. Since storing a row of $W$ or
$H$ requires $O(k)$ space the space complexity per worker is
$O((mk+nk+\abr{\Omega})/p)$. As for time complexity, we find it useful
to use the following assumptions: performing the SGD updates in
line~\ref{alg:sgd_for_start} to \ref{alg:sgd_for_end} takes $a \cdot k$
time and communicating a $(j, \hb_j)$ to another worker takes $c \cdot
k$ time, where $a$ and $c$ are hardware dependent constants.  On the
average, each $(j, \hb_j)$ pair contains $O\rbr{\abr{\Omega}/np}$
non-zero entries.  Therefore when a $(j, \hb_{j})$ pair is popped from
\texttt{queue}$[q]$ in line~\ref{alg:pop} of Algorithm~\ref{alg:nomad},
on the average it takes $a \cdot \rbr{\abr{\Omega} k/np}$ time to
process the pair.  Since computation and communication can be done in
parallel, as long as $a \cdot \rbr{\abr{\Omega} k/np}$ is higher than $c
\cdot k$ a worker thread is always busy and NOMAD scales linearly.


Suppose that $\abr{\Omega}$ is fixed but the number of workers $p$
increases; that is, we take a fixed size dataset and distribute it
across $p$ workers. As expected, for a large enough value of $p$
(which is determined by hardware dependent constants $a$ and $b$) the
cost of communication will overwhelm the cost of processing an item,
thus leading to slowdown.

On the other hand, suppose the work per worker is fixed, that is,
$\abr{\Omega}$ increases and the number of workers $p$ increases
proportionally. The average time $a \cdot \rbr{\abr{\Omega} k /np}$ to
process an item remains constant, and NOMAD scales linearly.

Finally, we discuss the communication complexity of NOMAD. For this
discussion we focus on a single item parameter $\hb_j$ which consists
of $O(k)$ numbers. In order to be processed by all the $p$ workers
once, it needs to be communicated $p$ times. This requires $O(kp)$
communication per item. There are $n$ items, and if we make a
simplifying assumption that during the execution of NOMAD each item is
processed a constant $c$ number of times by each processor, then the
total communication complexity is $O(nkp)$.

\subsection{Dynamic Load Balancing}
\label{sec:dynamic_load_balancing}

As different workers have different number of ratings per item, the
speed at which a worker processes a set of ratings $\Omegabar_j^{(q)}$
for an item $j$ also varies among workers.  Furthermore, in the
distributed memory setting different workers might process updates at
different rates dues to differences in hardware and system load.
NOMAD can handle this by dynamically balancing the workload of
workers: in line \ref{alg:sample_worker} of Algorithm~\ref{alg:nomad},
instead of sampling the recipient of a message uniformly at random we
can preferentially select a worker which has fewer items in its queue
to process. To do this, a payload carrying information about the size
of the \texttt{queue}$[q]$ is added to the messages that the workers
send each other. The overhead of passing the payload information is
just a single integer per message. This scheme allows us to
dynamically load balance, and ensures that a slower worker will
receive smaller amount of work compared to others.

\subsection{Hybrid Architecture}
\label{sec:multi_multi}

In a hybrid architecture we have multiple threads on a single machine as
well as multiple machines distributed across the network. In this case,
we make two improvements to the basic algorithm. First, in order to
amortize the communication costs we reserve two additional threads per
machine for sending and receiving $(j, \hb_{j})$ pairs over the
network. Intra-machine communication is much cheaper than
machine-to-machine communication, since the former does not involve a
network hop.  Therefore, whenever a machine receives a $(j, \hb_{j})$
pair, it circulates the pair among all of its threads before sending the
pair over the network. This is done by uniformly sampling a random
permutation whose size equals to the number of worker threads, and
sending the item variable to each thread according to this
permutation. Circulating a variable more than once was found to not
improve convergence, and hence is not used in our algorithm.

\subsection{Implementation Details}
\label{sec:ImplDeta}

Multi-threaded MPI was used for inter-machine communication. Instead
of communicating single $(j, \hb_{j})$ pairs, we follow the strategy
of \citep{SmoNar10}, and accumulate a fixed number of pairs (e.g.,
100) before transmitting them over the network.

NOMAD can be implemented with lock-free data structures since the only
interaction between threads is via operations on the queue.  We used
the concurrent queue provided by Intel Thread Building Blocks (TBB)
\citep{TBB}. Although technically not lock-free, the TBB concurrent
queue nevertheless scales almost linearly with the number of threads.

Since there is very minimal sharing of memory across threads in NOMAD,
by making memory assignments in each thread carefully aligned with
cache lines we can exploit memory locality and avoid cache
ping-pong. This results in near linear scaling for the multi-threaded
setting.


\section{Related Work}
\label{sec:RelatedWork}

\subsection{Map-Reduce and Friends}
\label{sec:MapReduceFriends}

Since many machine learning algorithms are iterative in nature, a
popular strategy to distribute them across multiple machines is to use
bulk synchronization after every iteration. Typically, one partitions
the data into chunks that are distributed to the workers at the
beginning. A master communicates the current parameters which are used
to perform computations on the slaves. The slaves return the solutions
during the bulk synchronization step, which are used by the master to
update the parameters. The popularity of this strategy is partly thanks
to the widespread availability of Hadoop
\citep{Hadoop}, an open source implementation of the MapReduce framework
\cite{DeaGhe08}.

\begin{figure}
  \centering
  \begin{tikzpicture}[scale=0.35]
    
    \boxlayout
    
    \drawrowpart
    
    \drawcolpart{0}{3}{\ptrna}{\ptcla}
    \drawcolpart{4}{7}{\ptrnb}{\ptclb}
    \drawcolpart{8}{11}{\ptrnc}{\ptclc}
    \drawcolpart{12}{15}{\ptrnd}{\ptcld}

    \fill [red!25!lightgray]
    (0 * \gridwidth + \asnmargin, 
    \gridwidth * 4 + \asnmargin)
    rectangle
    (1 * \gridwidth - \asnmargin, 
    \gridwidth * 3 - \asnmargin);    
    \draw
    (0 * \gridwidth + \asnmargin, 
    \gridwidth * 4 + \asnmargin)
    rectangle
    (1 * \gridwidth - \asnmargin, 
    \gridwidth * 3 - \asnmargin);

    \fill [green!25!lightgray]
    (1 * \gridwidth + \asnmargin, 
    \gridwidth * 3 + \asnmargin)
    rectangle
    (2 * \gridwidth - \asnmargin, 
    \gridwidth * 2 - \asnmargin);
    \draw
    (1 * \gridwidth + \asnmargin, 
    \gridwidth * 3 + \asnmargin)
    rectangle
    (2 * \gridwidth - \asnmargin, 
    \gridwidth * 2 - \asnmargin);
    
    \fill [blue!25!lightgray]
    (2 * \gridwidth + \asnmargin, 
    \gridwidth * 2 + \asnmargin)
    rectangle
    (3 * \gridwidth - \asnmargin, 
    \gridwidth * 1 - \asnmargin);
    \draw
    (2 * \gridwidth + \asnmargin, 
    \gridwidth * 2 + \asnmargin)
    rectangle
    (3 * \gridwidth - \asnmargin, 
    \gridwidth * 1 - \asnmargin);
    
    \fill [brown!25!lightgray]
    (3 * \gridwidth + \asnmargin, 
    \gridwidth * 1 + \asnmargin)
    rectangle
    (4 * \gridwidth - \asnmargin, 
    \gridwidth * 0 - \asnmargin);
    \draw
    (3 * \gridwidth + \asnmargin, 
    \gridwidth * 1 + \asnmargin)
    rectangle
    (4 * \gridwidth - \asnmargin, 
    \gridwidth * 0 - \asnmargin);
    
    \dtpt{5}{1} \dtpt{0}{15} \dtpt{13}{12} \dtpt{13}{6} \dtpt{14}{8}
    \dtpt{0}{8} \dtpt{7}{10} \dtpt{15}{2} \dtpt{10}{5} \dtpt{8}{13}
    \dtpt{5}{3} \dtpt{12}{5} \dtpt{10}{7} \dtpt{11}{6} \dtpt{15}{3}
    \dtpt{6}{9} \dtpt{13}{5} \dtpt{1}{3} \dtpt{8}{15} \dtpt{1}{6}
    \dtpt{1}{7} \dtpt{8}{9} \dtpt{12}{4} \dtpt{12}{1} \dtpt{13}{10}
    \dtpt{9}{0} \dtpt{8}{4} \dtpt{15}{12} \dtpt{3}{6} \dtpt{10}{13}
    \dtpt{6}{13} \dtpt{2}{11} \dtpt{5}{5} \dtpt{2}{1} \dtpt{3}{7}
    \dtpt{9}{3} \dtpt{11}{1} \dtpt{3}{14} \dtpt{11}{11} \dtpt{5}{15}

  \end{tikzpicture}
  \quad \quad 
  \begin{tikzpicture}[scale=0.35]
      
    \boxlayout
    
    \drawrowpart
    
    \drawcolpart{8}{11}{\ptrna}{\ptcla}
    \drawcolpart{0}{3}{\ptrnb}{\ptclb}
    \drawcolpart{12}{15}{\ptrnc}{\ptclc}
    \drawcolpart{4}{7}{\ptrnd}{\ptcld}
    
    \fill [red!25!lightgray]
    (2 * \gridwidth + \asnmargin, 
    \gridwidth * 4 + \asnmargin)
    rectangle
    (3 * \gridwidth - \asnmargin, 
    \gridwidth * 3 - \asnmargin);
    \draw
    (2 * \gridwidth + \asnmargin, 
    \gridwidth * 4 + \asnmargin)
    rectangle
    (3 * \gridwidth - \asnmargin, 
    \gridwidth * 3 - \asnmargin);
    
    \fill [green!25!lightgray]
    (0 * \gridwidth + \asnmargin, 
    \gridwidth * 3 + \asnmargin)
    rectangle
    (1 * \gridwidth - \asnmargin, 
    \gridwidth * 2 - \asnmargin);
    \draw
    (0 * \gridwidth + \asnmargin, 
    \gridwidth * 3 + \asnmargin)
    rectangle
    (1 * \gridwidth - \asnmargin, 
    \gridwidth * 2 - \asnmargin);
    
    \fill [blue!25!lightgray]
    (3 * \gridwidth + \asnmargin, 
    \gridwidth * 2 + \asnmargin)
    rectangle
    (4 * \gridwidth - \asnmargin, 
    \gridwidth * 1 - \asnmargin);
    \draw
    (3 * \gridwidth + \asnmargin, 
    \gridwidth * 2 + \asnmargin)
    rectangle
    (4 * \gridwidth - \asnmargin, 
    \gridwidth * 1 - \asnmargin);
    
    \fill [brown!25!lightgray]
    (1 * \gridwidth + \asnmargin, 
    \gridwidth * 1 + \asnmargin)
    rectangle
    (2 * \gridwidth - \asnmargin, 
    \gridwidth * 0 - \asnmargin);
    \draw
    (1 * \gridwidth + \asnmargin, 
    \gridwidth * 1 + \asnmargin)
    rectangle
    (2 * \gridwidth - \asnmargin, 
    \gridwidth * 0 - \asnmargin);
    
    \dtpt{5}{1} \dtpt{0}{15} \dtpt{13}{12} \dtpt{13}{6} \dtpt{14}{8}
    \dtpt{0}{8} \dtpt{7}{10} \dtpt{15}{2} \dtpt{10}{5} \dtpt{8}{13}
    \dtpt{5}{3} \dtpt{12}{5} \dtpt{10}{7} \dtpt{11}{6} \dtpt{15}{3}
    \dtpt{6}{9} \dtpt{13}{5} \dtpt{1}{3} \dtpt{8}{15} \dtpt{1}{6}
    \dtpt{1}{7} \dtpt{8}{9} \dtpt{12}{4} \dtpt{12}{1} \dtpt{13}{10}
    \dtpt{9}{0} \dtpt{8}{4} \dtpt{15}{12} \dtpt{3}{6} \dtpt{10}{13}
    \dtpt{6}{13} \dtpt{2}{11} \dtpt{5}{5} \dtpt{2}{1} \dtpt{3}{7}
    \dtpt{9}{3} \dtpt{11}{1} \dtpt{3}{14} \dtpt{11}{11} \dtpt{5}{15}

  \end{tikzpicture}
  \caption{Illustration of DSGD algorithm with 4 workers. Initially $W$
    and $H$ are partitioned as shown on the left. Each worker runs SGD
    on its active area as indicated. After each worker completes
    processing data points in its own active area, the columns of item
    parameters $H^{\top}$ are exchanged randomly, and the active area
    changes.  This process is repeated for each iteration.}
  \label{fig:dsgd_scheme}
\end{figure}
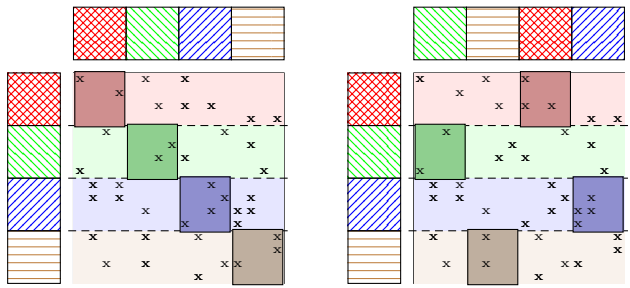

All three optimization schemes for matrix completion namely ALS, CCD++,
and SGD, can be parallelized using a bulk synchronization strategy. This
is relatively simple for ALS \citep{ZhoWilSchPan08} and CCD++
\citep{YuHsiSiDhi12}, but a bit more involved for SGD
\citep{GemNijHaaSis11,RecRe13}.  Suppose $p$ machines are available.
Then, the Distributed Stochastic Gradient Descent (DSGD) algorithm of
\citet{GemNijHaaSis11} partitions the indices of users
$\cbr{1,2,\ldots,m}$ into mutually exclusive sets $I_1, I_2, \ldots,
I_p$ and the indices of items into $J_1, J_2, \ldots, J_p$.  Now, define
\begin{align*}
  \Omega^{(q)} := \cbr{ (i,j) \in \Omega; i \in I_q, j \in J_q }, 
  \;\; 1 \leq q \leq p,
\end{align*}
and suppose that each machine runs SGD updates \eqref{eq:w_update} and
\eqref{eq:h_update} independently, but machine $q$ samples $(i,j)$ pairs
only from $\Omega^{(q)}$. By construction, $\Omega^{(q)}$'s are disjoint
and hence these updates can be run in parallel. A similar observation
was also made by \citet{RecRe13}. A bulk synchronization step
redistributes the sets $J_1, J_2, \ldots,J_p$ and corresponding rows of
$H$, which in turn changes the $\Omega^{(q)}$ processed by each machine,
and the iteration proceeds (see Figure~\ref{fig:dsgd_scheme})

Unfortunately, bulk synchronization based algorithms have two major
drawbacks: First, the communication and computation steps are done in
sequence. What this means is that when the CPU is busy, the network is
idle and vice versa. The second issue is that they suffer from what is
widely known as the \emph{the curse of last reducer}
\citep{SurVas11,AgaChaDudLan11}. In other words, all machines have to
wait for the slowest machine to finish before proceeding to the next
iteration. \citet{ZhuChiJuaLin13} report that DSGD suffers from this
problem even in the shared memory setting.

DSGD++ is an algorithm proposed by \citet{TefMakGem12} to address the
first issue discussed above. Instead of using $p$ partitions, DSGD++
uses $2p$ partitions. While the $p$ workers are processing $p$
partitions, the other $p$ partitions are sent over the network. This
keeps both the network and CPU busy simultaneously. However, DSGD++
also suffers from the curse of the last reducer. 

Another attempt to alleviate the problems of bulk synchronization in the
shared memory setting is the FPSGD** algorithm of
\citet{ZhuChiJuaLin13}; given $p$ threads, FPSGD** partitions the
parameters into more than $p$ sets, and uses a task manager thread to
distribute the partitions. When a thread finishes updating one
partition, it requests for another partition from the task manager. It
is unclear how to extend this idea to the distributed memory setting.

\begin{figure}
  \centering
  \begin{subfigure}[t]{0.22\textwidth}
    \centering
    \begin{tikzpicture}[scale=1.0]

      \fill [black!20!white] (0,2) rectangle (1,3);
      \fill [black!20!white] (2,1) rectangle (3,2);
      \fill [black!20!white] (1,0) rectangle (2,1);

      \begin{scope}[scale=0.375]
        \dtpt{5}{1}  \dtpt{0}{15}  \dtpt{13}{12}  \dtpt{13}{6}  \dtpt{14}{8}  \dtpt{0}{8}  \dtpt{7}{10}  \dtpt{15}{2}  \dtpt{10}{5}  \dtpt{8}{13}  \dtpt{5}{3}  \dtpt{12}{5}  \dtpt{10}{7}  \dtpt{11}{6}  \dtpt{15}{3}  \dtpt{6}{9}  \dtpt{13}{5}  \dtpt{1}{3}  \dtpt{8}{15}  \dtpt{1}{6}  \dtpt{1}{7}  \dtpt{8}{9}  \dtpt{12}{4}  \dtpt{12}{1}  \dtpt{13}{10}  \dtpt{9}{0}  \dtpt{8}{4}  \dtpt{15}{12}  \dtpt{3}{6}  \dtpt{10}{13} 
      \end{scope}

      \draw [thick, step=1.0] (0,0) grid (3,3);
    \end{tikzpicture}
    \caption{DSGD}
  \end{subfigure}
  \quad
  \begin{subfigure}[t]{0.22\textwidth}
    \centering
    \begin{tikzpicture}[scale=1.0]

      \fill [black!20!white] (0.5,2) rectangle (1,3);
      \fill [black!20!white] (2,1) rectangle (2.5,2);
      \fill [black!20!white] (1,0) rectangle (1.5,1);

      \begin{scope}[scale=0.375]
        \dtpt{5}{1}  \dtpt{0}{15}  \dtpt{13}{12}  \dtpt{13}{6}  \dtpt{14}{8}  \dtpt{0}{8}  \dtpt{7}{10}  \dtpt{15}{2}  \dtpt{10}{5}  \dtpt{8}{13}  \dtpt{5}{3}  \dtpt{12}{5}  \dtpt{10}{7}  \dtpt{11}{6}  \dtpt{15}{3}  \dtpt{6}{9}  \dtpt{13}{5}  \dtpt{1}{3}  \dtpt{8}{15}  \dtpt{1}{6}  \dtpt{1}{7}  \dtpt{8}{9}  \dtpt{12}{4}  \dtpt{12}{1}  \dtpt{13}{10}  \dtpt{9}{0}  \dtpt{8}{4}  \dtpt{15}{12}  \dtpt{3}{6}  \dtpt{10}{13} 
      \end{scope}

      \draw [dashed, xstep=0.5, ystep=1.0] (0,0) grid (3,3);
      \draw [thick, step=1.0] (0,0) grid (3,3);
    \end{tikzpicture}
    \caption{DSGD++}
  \end{subfigure}

  \vspace{0.1in}

  \centering
  \begin{subfigure}[t]{0.22\textwidth}
    \centering
    \begin{tikzpicture}[scale=1.0]

      \fill [black!20!white] (0,2.25) rectangle (0.75,3);
      \fill [black!20!white] (1.5,1.5) rectangle (2.25,2.25);
      \fill [black!20!white] (2.25,0) rectangle (3,0.75);

      \begin{scope}[scale=0.375]
        \dtpt{5}{1}  \dtpt{0}{15}  \dtpt{13}{12}  \dtpt{13}{6}  \dtpt{14}{8}  \dtpt{0}{8}  \dtpt{7}{10}  \dtpt{15}{2}  \dtpt{10}{5}  \dtpt{8}{13}  \dtpt{5}{3}  \dtpt{12}{5}  \dtpt{10}{7}  \dtpt{11}{6}  \dtpt{15}{3}  \dtpt{6}{9}  \dtpt{13}{5}  \dtpt{1}{3}  \dtpt{8}{15}  \dtpt{1}{6}  \dtpt{1}{7}  \dtpt{8}{9}  \dtpt{12}{4}  \dtpt{12}{1}  \dtpt{13}{10}  \dtpt{9}{0}  \dtpt{8}{4}  \dtpt{15}{12}  \dtpt{3}{6}  \dtpt{10}{13} 
      \end{scope}
      
      \draw [thick, step=0.75] (0,0) grid (3,3);
    \end{tikzpicture}
    \caption{FPSGD**}
  \end{subfigure}
  \quad
  \begin{subfigure}[t]{0.22\textwidth}
    \centering
    \begin{tikzpicture}[scale=1.0]

      \fill [black!20!white] (2.4,2) rectangle (2.6,3);
      \fill [black!20!white] (0.2,1) rectangle (0.4,2);
      \fill [black!20!white] (1.6,0) rectangle (1.8,1);

      \begin{scope}[scale=0.375]
        \dtpt{5}{1}  \dtpt{0}{15}  \dtpt{13}{12}  \dtpt{13}{6}  \dtpt{14}{8}  \dtpt{0}{8}  \dtpt{7}{10}  \dtpt{15}{2}  \dtpt{10}{5}  \dtpt{8}{13}  \dtpt{5}{3}  \dtpt{12}{5}  \dtpt{10}{7}  \dtpt{11}{6}  \dtpt{15}{3}  \dtpt{6}{9}  \dtpt{13}{5}  \dtpt{1}{3}  \dtpt{8}{15}  \dtpt{1}{6}  \dtpt{1}{7}  \dtpt{8}{9}  \dtpt{12}{4}  \dtpt{12}{1}  \dtpt{13}{10}  \dtpt{9}{0}  \dtpt{8}{4}  \dtpt{15}{12}  \dtpt{3}{6}  \dtpt{10}{13} 
      \end{scope}

      \draw [xstep=0.2, ystep=1.0] (0,0) grid (3,3);
      \draw [thick] (0,0) rectangle (3,3);
    \end{tikzpicture}
    \caption{NOMAD}
  \end{subfigure}

  \caption{Comparison of data partitioning schemes between algorithms.
    Example active area of stochastic gradient sampling is marked as
    gray.}
  \label{fig:split_scheme}
\end{figure}
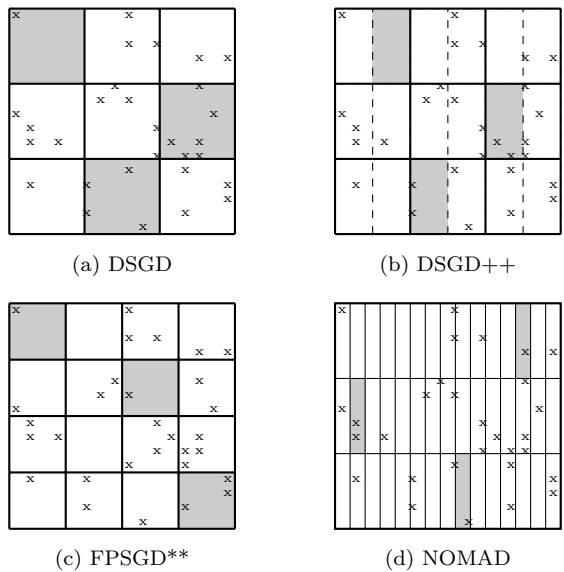

In NOMAD we sidestep all the drawbacks of bulk synchronization. Like
DSGD++ we also simultaneously keep the network and CPU busy. On the
other hand, like FPSGD** we effectively load balance between the
threads. To understand why NOMAD enjoys both these benefits, it is
instructive to contrast the data partitioning schemes underlying DSGD,
DSGD++, FPSGD**, and NOMAD (see Figure~\ref{fig:split_scheme}).  Given
$p$ number of workers, DSGD divides the rating matrix $A$ into $p
\times p$ number of blocks; DSGD++ improves upon DSGD by further
dividing each block to $1 \times 2$ sub-blocks
(Figure~\ref{fig:split_scheme} (a) and (b)).  On the other hand,
FPSGD** splits $A$ into $p' \times p'$ blocks with $p' > p$
(Figure~\ref{fig:split_scheme} (c)), while NOMAD uses $p \times n$
blocks (Figure~\ref{fig:split_scheme} (d)). In terms of communication
there is no difference between various partitioning schemes; all of
them require $O(nkp)$ communication for each item to be processed a
constant $c$ number of times. However, having smaller blocks means
that NOMAD has much more flexibility in assigning blocks to
processors, and hence better ability to exploit parallelism. Because
NOMAD operates at the level of individual item parameters, $\hb_{j}$,
it can dynamically load balance by assigning fewer columns to a slower
worker. A pleasant side effect of such a fine grained partitioning
coupled with the lock free nature of updates is that one does not
require sophisticated scheduling algorithms to achieve good
performance. Consequently, NOMAD outperforms DSGD, DSGD++, and
FPSGD**.

\subsection{Asynchronous Algorithms}
\label{sec:AsynchrAlgor}

There is growing interest in designing machine learning algorithms that
do not perform bulk synchronization. See, for instance, the randomized
(block) coordinate descent methods of \citet{RicTak13} and the Hogwild!\
algorithm of \citet{RecReWriNiu11}. A relatively new approach to
asynchronous parallelism is to use a so-called parameter server. A
parameter server is either a single machine or a distributed set of
machines which caches the current values of the parameters. Workers
store local copies of the parameters and perform updates on them, and
periodically synchronize their local copies with the parameter
server. The parameter server receives updates from all workers,
aggregates them, and communicates them back to the workers. The earliest
work on a parameter server, that we are aware of, is due to
\citet{SmoNar10}, who propose using a parameter server for collapsed
Gibbs sampling in Latent Dirichlet Allocation. PowerGraph
\citep{GonLowGuBicetal12}, upon which the latest version of the GraphLab
toolkit is based, is also essentially based on the idea of a parameter
server. However, the difference in case of PowerGraph is that the
responsibility of parameters is distributed across multiple machines,
but at the added expense of synchronizing the copies.


Very roughly speaking, the asynchronously parallel version of the ALS
algorithm in GraphLab works as follows: $\wb_i$ and $\hb_j$ variables
are distributed across multiple machines, and whenever $\wb_i$ is being
updated with equation \eqref{eq:newton-fix-h}, the values of $\hb_j$'s
for $j \in \Omega_i$ are retrieved across the network and read-locked
until the update is finished.  GraphLab provides functionality such as
network communication and a distributed locking mechanism to implement
this. However, frequently acquiring read-locks over the network can be
expensive.  In particular, a popular user who has rated many items will
require read locks on a large number of items, and this will lead to
vast amount of communication and delays in updates on those items.
GraphLab provides a complex job scheduler which attempts to minimize
this cost, but then the efficiency of parallelization depends on the
difficulty of the scheduling problem and the effectiveness of the
scheduler.

In our empirical evaluation NOMAD performs significantly better than
GraphLab. The reasons are not hard to see. First, because of the lock
free nature of NOMAD, we completely avoid acquiring expensive network
locks. Second, we use SGD which allows us to exploit finer grained
parallelism as compared to ALS, and also leads to faster convergence. In
fact, the GraphLab framework is not well suited for SGD (personal
communication with the developers of GraphLab). Finally, because of the
finer grained data partitioning scheme used in NOMAD, unlike GraphLab
whose performance heavily depends on the underlying scheduling
algorithms, we do not require a complicated scheduling mechanism.

\subsection{Numerical Linear Algebra}
\label{sec:NumerLineAlgebra}

The concepts of asynchronous and non-blocking updates have also been
studied in numerical linear algebra. To avoid the load balancing problem
and to reduce processor idle time, asynchronous numerical methods were
first proposed over four decades ago by \citet{ChaMir69}.  Given an
operator $\mathcal{H}: {\mathbb{R}}^m \rightarrow {\mathbb{R}}^m$, to
find the fixed point solution $x^*$ such that $\mathcal{H}(x^*)=x^*$, a
standard Gauss-Seidel-type procedure performs the update
$x_i=\left(\mathcal{H}(x)\right)_i$ sequentially (or randomly).  Using
the asynchronous procedure, each computational node asynchronously
conducts updates on each variable (or a subset) $x_i^{\text{new}} =
\left(\mathcal{H}(x)\right)_i$ and then overwrites $x_i$ in common
memory by $x_i^{\text{new}}$.  Theory and applications of this
asynchronous method have been widely studied (see the literature review
of \citet{FroSzy00} and the seminal textbook by \citet{BerTsi97a}).  The
concept of this asynchronous fixed-point update is very closely related
to the Hogwild algorithm of \citet{RecReWriNiu11} or the so-called
Asynchronous SGD (ASGD) method proposed by
\citet{TefMakGem12}. Unfortunately, such algorithms are
\emph{non-serializable}, that is, there may not exist an equivalent
update ordering in a serial implementation.  In contrast, our NOMAD
algorithm is not only asynchronous but also serializable, and therefore
achieves faster convergence in practice.

On the other hand, non-blocking communication has also been proposed to
accelerate iterative solvers in a distributed setting.  For example,
\citet{HoeGotRehLum06} presented a distributed conjugate gradient
implementation with non-blocking collective MPI operations for solving
linear systems. However, this algorithm still requires synchronization
at each CG iteration, so it is very different from our NOMAD algorithm.


\subsection{Discussion}

We remark that among algorithms we have discussed so far, NOMAD is the
only distributed-memory algorithm which is both asynchronous and
lock-free.  Other parallelizations of SGD such as DSGD and DSGD++ are
lock-free, but not fully asynchronous; therefore, the cost of
synchronization will increase as the number of machines grows
\citep{ZhuChiJuaLin13}.  On the other hand, GraphLab implementation of
ALS \citep{LowGonKyrBicetal12} is asynchronous but not lock-free,
therefore depends on a complex job scheduler to reduce the side-effect
of using locks.

\section{Experiments}
\label{sec:Experiments}

In this section, we evaluate the empirical performance of NOMAD with
extensive experiments.  For the distributed memory experiments we
compare NOMAD with DSGD~\citep{GemNijHaaSis11},
DSGD++~\citep{TefMakGem12} and CCD++~\citep{YuHsiSiDhi12}.  We also
compare against GraphLab, but the quality of results produced by
GraphLab are significantly worse than the other methods, and therefore
the plots for this experiment are delegated to
Appendix~\ref{sec:comp-with-graphl}. For the shared memory experiments
we pitch NOMAD against FPSGD**~\citep{ZhuChiJuaLin13} (which is shown to
outperform DSGD in single machine experiments) as well as CCD++. Our
experiments are designed to answer the following:
\begin{itemize}
\item How does NOMAD scale with the number of cores on a single machine?
  (Section \ref{sec:ScalingNumberCores})
\item How does NOMAD scale as a fixed size dataset is distributed across
  multiple machines? (Section \ref{sec:ScalingasFixed})
\item How does NOMAD perform on a commodity hardware cluster?
  (Section~\ref{sec:ScalDiffHardw})
\item How does NOMAD scale when both the size of the data as well
  as the number of machines grow? (Section
  \ref{sec:Scalingasboth})
\end{itemize}

Since the objective function \eqref{eq:obj_fcn} is non-convex, different
optimizers will converge to different solutions.  Factors which affect
the quality of the final solution include 1) initialization strategy, 2)
the sequence in which the ratings are accessed, and 3) the step size
decay schedule.  It is clearly not feasible to consider the
combinatorial effect of all these factors on each algorithm.  However,
we believe that the overall trend of our results is not affected by
these factors. 


\subsection{Experimental Setup}
\label{sec:ExperimentalSetup}

Publicly available code for
FPSGD**\footnote{\url{http://www.csie.ntu.edu.tw/~cjlin/libmf/}} and
CCD++\footnote{\url{http://www.cs.utexas.edu/~rofuyu/libpmf/}} was used
in our experiments.  For DSGD and DSGD++, which we had to implement
ourselves because the code is not publicly available, we closely
followed the recommendations of \citet{GemNijHaaSis11} and
\citet{TefMakGem12}, and in some cases made improvements based on our
experience.  For a fair comparison all competing algorithms were tuned
for optimal performance on our hardware.  The code and scripts required
for reproducing the experiments are readily available for download from
\url{https://sites.google.com/site/hyokunyun/software}. Parameters used
in our experiments are summarized in Table~\ref{tab:parameters}.

\begin{table}[htbp]
  \centering
  \caption{Dimensionality parameter $k$, regularization parameter
    $\lambda$ \eqref{eq:obj_fcn} and step-size schedule
    parameters $\alpha, \beta$ \eqref{eq:stepsize_method}}
  \begin{tabular}[htbp]{|l|c|c|c|c|}
    \hline 
    Name & $k$ & $\lambda$ & $\alpha$ & $\beta$\\
    \hline
    Netflix & 100 & 0.05 & 0.012 & 0.05 \\
    \hline
    Yahoo!\ Music & 100 & 1.00 & 0.00075 & 0.01 \\
    \hline
    Hugewiki & 100 & 0.01 & 0.001 & 0 \\
    \hline
  \end{tabular}
  \label{tab:parameters}
\end{table}

\begin{table}[htbp]
  \centering
  \caption{Dataset Details}
  \begin{tabular}[htbp]{|l|c|c|c|}
    \hline
    Name & Rows & Columns & Non-zeros \\
    \hline
    Netflix~\citep{BelKor07} & 2,649,429 & 17,770 & 99,072,112  \\
    \hline
    Yahoo!\ Music~\citep{DroKoeKorWei11} & 1,999,990 & 624,961 & 252,800,275  \\
    \hline
    Hugewiki~\citep{GraphLabDatasets} & 50,082,603 & 39,780 & 2,736,496,604 \\
    \hline
  \end{tabular}
  \label{tab:data}
\end{table}

For all experiments, except the ones in Section~\ref{sec:Scalingasboth},
we will work with three benchmark datasets namely Netflix, Yahoo!\
Music, and Hugewiki (see Table~\ref{tab:data} for more details).  The
same training and test dataset partition is used consistently for all
algorithms in every experiment. Since our goal is to compare
optimization algorithms, we do very minimal parameter tuning. For
instance, we used the same regularization parameter $\lambda$ for each
dataset as reported by \citet{YuHsiSiDhi12}, and shown in
Table~\ref{tab:parameters}; we study the effect of the regularization
parameter on the convergence of NOMAD in
Appendix~\ref{sec:EffectRegulParam}.  By default we use $k = 100$ for
the dimension of the latent space; we study how the dimension of the
latent space affects convergence of NOMAD in
Appendix~\ref{sec:EffectLatentDimens}.  All algorithms were initialized
with the same initial parameters; we set each entry of $W$ and $H$ by
independently sampling a uniformly random variable in the range
$(0,\frac 1 {\sqrt{k}})$ \citep{YuHsiSiDhi12, ZhuChiJuaLin13}.

We compare solvers in terms of Root Mean Square Error (RMSE) on the test
set, which is defined as:
\begin{align*}
  \sqrt{\frac{\sum_{(i,j) \in \Omega^{\text{test}}} \rbr{A_{ij} -
        \inner{\wb_i}{\hb_j}}^2}
    {\abr{\Omega^{\text{test}}}}},
\end{align*}
where $\Omega^{\text{test}}$ denotes the ratings in the test set.

All experiments, except the ones reported in
Section~\ref{sec:ScalDiffHardw}, are run using the Stampede Cluster at
University of Texas, a Linux cluster where each node is outfitted with
2 Intel Xeon E5 (Sandy Bridge) processors and an Intel Xeon Phi
Coprocessor (MIC Architecture).  For single-machine experiments
(Section~\ref{sec:ScalingNumberCores}), we used nodes in the
\texttt{largemem} queue which are equipped with 1TB of RAM and 32
cores. For all other experiments, we used the nodes in the
\texttt{normal} queue which are equipped with 32 GB of RAM and 16
cores (only 4 out of the 16 cores were used for computation).
Inter-machine communication on this system is handled by MVAPICH2.

For the commodity hardware experiments in
Section~\ref{sec:ScalDiffHardw} we used \texttt{m1.xlarge} instances of
Amazon Web Services, which are equipped with 15GB of RAM and four cores.
We utilized all four cores in each machine; NOMAD and DSGD++ uses two
cores for computation and two cores for network communication, while
DSGD and CCD++ use all four cores for both computation and
communication.  Inter-machine communication on this system is handled by
MPICH2.

Since FPSGD** uses single precision arithmetic, the experiments in
Section~\ref{sec:ScalingNumberCores} are performed using single
precision arithmetic, while all other experiments use double precision
arithmetic.  All algorithms are compiled with Intel C++ compiler, with
the exception of experiments in Section~\ref{sec:ScalDiffHardw} where we
used \texttt{gcc} which is the only compiler toolchain available on the
commodity hardware cluster.  For ready reference, exceptions to the
experimental settings specific to each section are summarized in
Table~\ref{tab:exceptions}.

\begin{table}[htbp]
  \centering
  \caption{Exceptions to each experiment}
  {\small
    \begin{tabular}[htbp]{|c|p{0.35\textwidth}|}
      \hline 
    Section & Exception \\
    \hline
    Section~\ref{sec:ScalingNumberCores} & 
    \mbox{}\par\vspace{-\baselineskip}
    \begin{itemize}[noitemsep,leftmargin=*,topsep=0pt,partopsep=0pt] 
    \item run on \texttt{largemem} queue (32 cores, 1TB RAM)
    \item single precision floating point used
    \end{itemize}
    \mbox{}\par\vspace{-2\baselineskip}
    \\
    \hline
    Section~\ref{sec:ScalDiffHardw} & 
    \mbox{}\par\vspace{-\baselineskip}
    \begin{itemize}[noitemsep,leftmargin=*,topsep=0pt,partopsep=0pt]
    \item run on \texttt{m1.xlarge} (4 cores, 15GB RAM)
    \item compiled with \texttt{gcc}
    \item \texttt{MPICH2} for MPI implementation
    \end{itemize}
    \mbox{}\par\vspace{-2\baselineskip}\\
    \hline
    Section~\ref{sec:Scalingasboth} & 
    \mbox{}\par\vspace{-\baselineskip}
    \begin{itemize}[noitemsep,leftmargin=*,topsep=0pt,partopsep=0pt]
    \item Synthetic datasets
    \end{itemize}
    \mbox{}\par\vspace{-2\baselineskip}\\
    \hline
  \end{tabular}
  }
  \label{tab:exceptions}
\end{table}

The convergence speed of stochastic gradient descent methods depends on
the choice of the step size schedule.  The schedule we used for NOMAD
is
\begin{align}
  \label{eq:stepsize_method}
  s_t = \frac{\alpha}{1 + \beta \cdot t^{1.5}},
\end{align}
where $t$ is the number of SGD updates that were performed on a
particular user-item pair $(i,j)$.  DSGD and DSGD++, on the other
hand, use an alternative strategy called bold-driver
\citep{GemNijHaaSis11}; here, the step size is adapted by monitoring
the change of the objective function.

\subsection{Scaling in Number of Cores}
\label{sec:ScalingNumberCores}

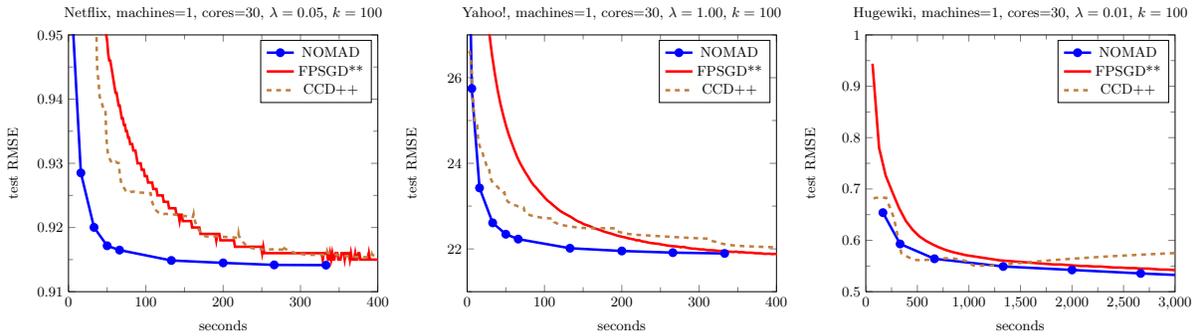
\begin{figure*}[htbp]
  \centering
  \begin{tikzpicture}[scale=0.6]
    \begin{axis}[minor tick num=1,
      title={Netflix, machines=1, cores=30, $\lambda=0.05$, $k=100$},
      xlabel={seconds}, ylabel={test RMSE},
      ymin={0.91}, ymax={0.95}, xmin={0}, xmax={400}]
      
      \addplot[ultra thick, color=blue, mark=*] table [x index=2, y index=1, header=false]
      {../../Results/stam_cpus_netflix_30.txt};
      
      \addplot[ultra thick, color=red, mark=none] table [x index=0, y index=1, header=false]
      {../../Plots/stampede_single_fpsgd_netflix_1_30_100_0.050000_0.002000.txt};
      
      \addplot[ultra thick, color=brown, mark=none, dashed] table [x index=0, y index=1, header=false]
      {../../Plots/stampede_single_ccdpp_netflix_1_30_100_0.050000.txt};

      \legend{NOMAD, FPSGD**, CCD++}
    \end{axis}
  \end{tikzpicture}
  \begin{tikzpicture}[scale=0.6]
    \begin{axis}[minor tick num=1,
      title={Yahoo!, machines=1, cores=30, $\lambda=1.00$, $k=100$},
      xlabel={seconds}, ylabel={test RMSE},
      ymin={21}, ymax={27}, xmin={0}, xmax={400}]
      
      \addplot[ultra thick, color=blue, mark=*] table [x index=2, y index=1, header=false]
      {../../Results/stam_cpus_yahoo_30.txt};
      
      \addplot[ultra thick, color=red, mark=none] table [x index=0, y index=1, header=false]
      {../../Plots/stampede_single_fpsgd_yahoo_1_30_100_1.000000_0.001000.txt};
      
      \addplot[ultra thick, color=brown, mark=none, dashed] table [x index=0, y index=1, header=false]
      {../../Plots/stampede_single_ccdpp_yahoo_1_30_100_1.000000.txt};

      \legend{NOMAD, FPSGD**, CCD++}
    \end{axis}
  \end{tikzpicture}
  \begin{tikzpicture}[scale=0.6]
    \begin{axis}[minor tick num=1,
      title={Hugewiki, machines=1, cores=30, $\lambda=0.01$, $k=100$},
      xlabel={seconds}, ylabel={test RMSE},
      ymin={0.5}, ymax={1}, xmin={0}, xmax={3000}]
      

      \addplot[ultra thick, color=blue, mark=*] table 
      [x index=3, y index=5, header=false, col sep=comma]
      {../../Plots/stampede_single_nomad_const_hugewiki_1_30_100_0.010000_0.000667_0.000000.txt};

      \addplot[ultra thick, color=red, mark=none] table [x index=0, y index=1, header=false]
      {../../Plots/stampede_single_fpsgd_hugewiki_1_30_100_0.010000_0.001000.txt};
      
      \addplot[ultra thick, color=brown, mark=none, dashed] table [x index=0, y index=1, header=false]
      {../../Plots/stampede_single_ccdpp_hugewiki_1_30_100_0.010000.txt};

      \legend{NOMAD, FPSGD**, CCD++}
    \end{axis}
  \end{tikzpicture}
  \caption{Comparison of NOMAD, FPSGD**, and CCD++ on a single-machine
    with 30 computation cores.}
  \label{fig:single_compare}
\end{figure*}

For the first experiment we fixed the number of cores to 30, and
compared the performance of NOMAD vs FPSGD**\footnote{Since the current
  implementation of FPSGD** in LibMF only reports CPU execution time, we
  divide this by the number of threads and use this as a proxy for wall
  clock time.} and CCD++ (Figure~\ref{fig:single_compare}).  On Netflix
(left) NOMAD not only converges to a slightly better quality solution
(RMSE 0.914 vs 0.916 of others), but is also able to reduce the RMSE
rapidly right from the beginning.  On Yahoo!\ Music (middle), NOMAD
converges to a slightly worse solution than FPSGD** (RMSE 21.894 vs
21.853) but as in the case of Netflix, the initial convergence is more
rapid.  On Hugewiki, the difference is smaller but NOMAD still
outperforms.  The initial speed of CCD++ on Hugewiki is comparable to
NOMAD, but the quality of the solution starts to deteriorate in the
middle.  Note that the performance of CCD++ here is better than what was
reported in \citet{ZhuChiJuaLin13} since they used double-precision
floating point arithmetic for CCD++.  In other experiments (not reported
here) we varied the number of cores and found that the relative
difference in performance between NOMAD, FPSGD** and CCD++ are very
similar to that observed in Figure~\ref{fig:single_compare}.

\begin{figure*}[htbp]
  \centering
  \begin{tikzpicture}[scale=0.6]

    \begin{axis}[minor tick num=1,
      title={Netflix, machines=1, $\lambda=0.05$, $k=100$},
      xlabel={seconds $\times$ cores}, ylabel={test RMSE}, xmax={6000}]
      
      \addplot
      table [x expr=\thisrowno{3}*4, y
      index=5, header=false, col sep=comma]
      {../../Plots/stampede_single_nomad_netflix_1_4_100_0.050000_0.008000_0.010000.txt};

      \addplot
      table [x expr=\thisrowno{3}*8, y
      index=5, header=false, col sep=comma]
      {../../Plots/stampede_single_nomad_netflix_1_8_100_0.050000_0.008000_0.010000.txt};

      \addplot
      table [x expr=\thisrowno{3} * 16, y
      index=5, header=false, col sep=comma]
      {../../Plots/stampede_single_nomad_netflix_1_16_100_0.050000_0.008000_0.010000.txt};

      \addplot
      table [x expr=\thisrowno{3} * 30, y
      index=5, header=false, col sep=comma]
      {../../Plots/stampede_single_nomad_netflix_1_30_100_0.050000_0.008000_0.010000.txt};
      
      \legend{\# cores=4, \# cores=8,\# cores=16,\# cores=30}
    \end{axis}
  \end{tikzpicture}
  \begin{tikzpicture}[scale=0.6]
    \begin{axis}[minor tick num=1,
      title={Yahoo!, machines=1, $\lambda=1.00$, $k=100$},
      xlabel={seconds $\times$ cores}, ylabel={test RMSE}, xmax={8000}]
      
      \addplot
      table [x expr=\thisrowno{3} * 4, y
      index=5, header=false, col sep=comma]
      {../../Plots/stampede_single_nomad_yahoo_1_4_100_1.000000_0.000500_0.050000.txt};

      \addplot 
      table [x expr=\thisrowno{3} * 8, y
      index=5, header=false, col sep=comma]
      {../../Plots/stampede_single_nomad_yahoo_1_8_100_1.000000_0.000500_0.050000.txt};

      \addplot 
      table [x expr=\thisrowno{3} * 16, y
      index=5, header=false, col sep=comma]
      {../../Plots/stampede_single_nomad_yahoo_1_16_100_1.000000_0.000500_0.050000.txt};

      \addplot 
      table [x expr=\thisrowno{3} * 30, y
      index=5, header=false, col sep=comma]
      {../../Plots/stampede_single_nomad_yahoo_1_30_100_1.000000_0.000500_0.050000.txt};
      
      \legend{\# cores=4, \# cores=8,\# cores=16,\# cores=30}
    \end{axis}
  \end{tikzpicture}
  \begin{tikzpicture}[scale=0.6]
    \begin{axis}[minor tick num=1,
      title={Hugewiki, machines=1, $\lambda=0.01$, $k=100$},
      xlabel={seconds $\times$ cores}, ylabel={test RMSE}, xmax={200000}]
      
      \addplot
      table [x expr=\thisrowno{3} * 4, y
      index=5, header=false, col sep=comma]
      {../../Plots/stampede_single_nomad_hugewiki_1_4_100_0.010000_0.000667_0.000000.txt};

      \addplot
      table [x expr=\thisrowno{3} * 8, y
      index=5, header=false, col sep=comma]
      {../../Plots/stampede_single_nomad_hugewiki_1_8_100_0.010000_0.000667_0.000000.txt};

      \addplot
      table [x expr=\thisrowno{3} * 16, y
      index=5, header=false, col sep=comma]
      {../../Plots/stampede_single_nomad_hugewiki_1_16_100_0.010000_0.000667_0.000000.txt};

      \addplot
      table [x expr=\thisrowno{3} * 30, y
      index=5, header=false, col sep=comma]
      {../../Plots/stampede_single_nomad_hugewiki_1_30_100_0.010000_0.000667_0.000000.txt};

      \legend{\# cores=4,\# cores=8,\# cores=16,\# cores=30}
    \end{axis}
  \end{tikzpicture}
  \caption{Test RMSE of NOMAD as a function of computation time (time in
    seconds $\times$ the number of cores), when the number of cores is
    varied.}
  \label{fig:single_converge_time}
\end{figure*}
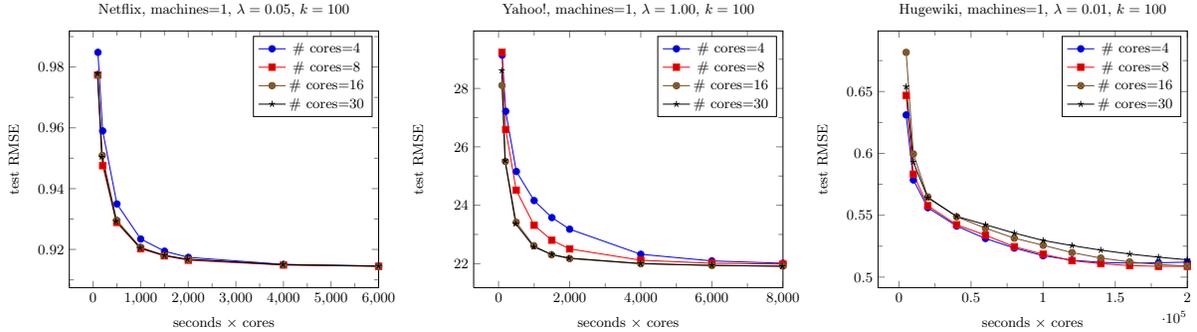

For the second experiment we varied the number of cores from 4 to 30,
and plot the scaling behavior of NOMAD
(Figures~\ref{fig:single_collection} and
\ref{fig:single_converge_time}).  Figure~\ref{fig:single_collection}
(left) shows how test RMSE changes as a function of the number of
updates on Yahoo!~Music.  Interestingly, as we increased the number of
cores, the test RMSE decreased faster.  We believe this is because when
we increase the number of cores, the rating matrix $A$ is partitioned
into smaller blocks; recall that we split $A$ into $p \times n$ blocks,
where $p$ is the number of parallel workers.  Therefore, the
communication between workers becomes more frequent, and each SGD update
is based on fresher information (see also
Section~\ref{sec:complexity_analysis} for mathematical analysis).  This
effect was more strongly observed on Yahoo!~Music than others, since
Yahoo!~Music has much larger number of items (624,961 vs. 17,770 of
Netflix and 39,780 of Hugewiki) and therefore more amount of
communication is needed to circulate the new information to all workers.
Results for other datasets are provided in
Figure~\ref{fig:single_converge_update} in Appendix~\ref{sec:test-rmse}.

On the other hand, to assess the efficiency of computation we define
\emph{average throughput} as the average number of ratings processed
per core per second, and plot it for each dataset in
Figure~\ref{fig:single_collection} (right), while varying the number
of cores.  If NOMAD exhibits linear scaling in terms of the speed it
processes ratings, the average throughput should remain
constant\footnote{Note that since we use single-precision floating
  point arithmetic in this section to match the implementation of
  FPSGD**, the throughput of NOMAD is about 50\% higher than that in
  other experiments.}.  On Netflix, the average throughput
indeed remains almost constant as the number of cores changes.  On
Yahoo!~Music and Hugewiki, the throughput decreases to about
50\% as the number of cores is increased to 30.  We believe this is
mainly due to cache locality effects. 

Now we study how much speed-up NOMAD can achieve by increasing the
number of cores.  In Figure~\ref{fig:single_converge_time}, we set
$y$-axis to be test RMSE and $x$-axis to be the total CPU time expended
which is given by the number of seconds elapsed multiplied by the number
of cores.  We plot the convergence curves by setting the \# cores=4, 8,
16, and 30. If the curves overlap, then this shows that we achieve
linear speed up as we increase the number of cores. This is indeed the
case for Netflix and Hugewiki. In the case of Yahoo!~Music we observe
that the speed of convergence increases as the number of cores
increases. This, we believe, is again due to the decrease in the block
size which leads to faster convergence.

\begin{figure}
  \begin{tikzpicture}[scale=0.48]
    \begin{axis}[minor tick num=1,
      title={Yahoo!, machines=1, $\lambda=1.00$, $k=100$},
      xlabel={number of updates}, ylabel={test RMSE}, xmax={30000000000}]
      
      \addplot
      table [x index=4, y
      index=5, header=false, col sep=comma]
      {../../Plots/stampede_single_nomad_yahoo_1_4_100_1.000000_0.000500_0.050000.txt};

      \addplot 
      table [x index=4, y
      index=5, header=false, col sep=comma]
      {../../Plots/stampede_single_nomad_yahoo_1_8_100_1.000000_0.000500_0.050000.txt};

      \addplot 
      table [x index=4, y
      index=5, header=false, col sep=comma]
      {../../Plots/stampede_single_nomad_yahoo_1_16_100_1.000000_0.000500_0.050000.txt};

      \addplot 
      table [x index=4, y
      index=5, header=false, col sep=comma]
      {../../Plots/stampede_single_nomad_yahoo_1_30_100_1.000000_0.000500_0.050000.txt};
      
      \legend{\# cores=4, \# cores=8,\# cores=16,\# cores=30}
    \end{axis}
  \end{tikzpicture}
  \begin{tikzpicture}[scale=0.48]
    \begin{axis}[minor tick num=1, only marks, 
      title={machines=1, $\lambda=0.05$, $k=100$},
      xlabel={number of cores}, ylabel={updates per core per sec},
      legend style={legend pos=south west}, ymin={0} ]
      
      \addplot [color=blue, mark=o] table [x index=2, y
      expr={\thisrowno{4}/\thisrowno{3}/\thisrowno{2}/\thisrowno{1}}, header=false,
      col sep=comma]
      {../../Plots/stampede_single_nomad_netflix_1_all_100_0.050000_0.008000_0.010000.txt};

      \addplot [color=red, mark=*] table [x index=2, y
      expr={\thisrowno{4}/\thisrowno{3}/\thisrowno{2}/\thisrowno{1}}, header=false,
      col sep=comma]
      {../../Plots/stampede_single_nomad_yahoo_1_all_100_1.000000_0.000500_0.050000.txt};

      \addplot [color=brown, mark=+] table [x index=2, y
      expr={\thisrowno{4}/\thisrowno{3}/\thisrowno{2}/\thisrowno{1}}, header=false,
      col sep=comma]
      {../../Plots/stampede_single_nomad_hugewiki_1_all_100_0.010000_0.000667_0.000000.txt};

      \legend{Netflix, Yahoo!, Hugewiki}
    \end{axis}
  \end{tikzpicture}

  \caption{Left: Test RMSE of NOMAD as a function of the number of
    updates on Yahoo!~Music, when the number of cores is varied.
    Right: Number of updates of NOMAD per core per second as a
    function of the number of cores.
  }
  \label{fig:single_collection}

\end{figure}
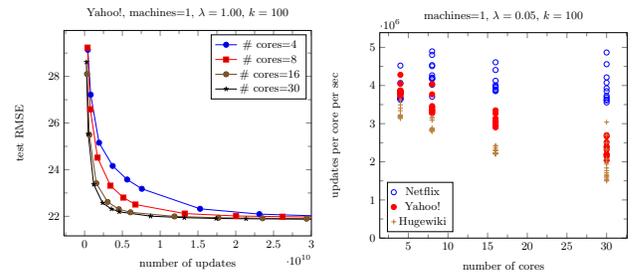

\subsection{Scaling as a Fixed Dataset is Distributed Across Workers}
\label{sec:ScalingasFixed}

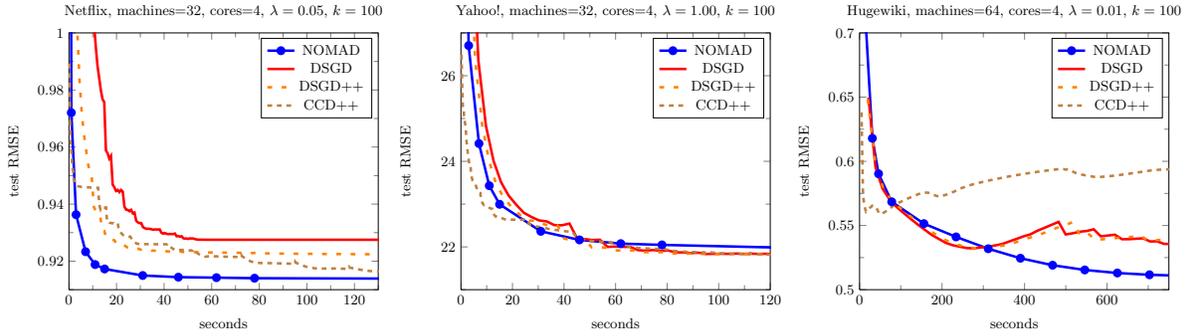
\begin{figure*}[htbp]
  \centering
  \begin{tikzpicture}[scale=0.6]
    \begin{axis}[minor tick num=1,
      title={Netflix, machines=32, cores=4, $\lambda=0.05$, $k=100$},
      xlabel={seconds}, ylabel={test RMSE},
      ymin={0.91}, ymax={1.00}, xmin={0}, xmax={130}
      ]
      
      \addplot[ultra thick, color=blue, mark=*] table [x index=3, y
      index=5, header=false, col sep=comma]
      {../../Plots/stampede_multi_nomad_netflix_32_4_100_0.050000_0.008000_0.010000.txt};
      
      \addplot[ultra thick, color=red, mark=none] table [x index=0, y index=1, header=false]
      {../../Plots/stampede_multi_dsgd_bold_netflix_32_4_100_0.050000_0.008000_0.010000.txt};

      \addplot[ultra thick, loosely dashed, color=orange, mark=none] table [x index=0, y index=1, header=false]
      {../../Plots/stampede_multi_dsgdpp_netflix_32_4_100_0.050000_0.008000_0.500000.txt};

      \addplot[ultra thick, color=brown, mark=none, dashed] table [x index=0, y index=1, header=false]
      {../../Plots/stampede_multi_ccdpp_netflix_32_4_100_0.050000.txt};

      \legend{NOMAD, DSGD, DSGD++, CCD++}
    \end{axis}
  \end{tikzpicture}
  \begin{tikzpicture}[scale=0.6]
    \begin{axis}[minor tick num=1,
      title={Yahoo!, machines=32, cores=4, $\lambda=1.00$, $k=100$},
      xlabel={seconds}, ylabel={test RMSE},
      ymin={21}, ymax={27}, xmin={0}, xmax={120}]
      
      \addplot[ultra thick, color=blue, mark=*] table [x index=3, y
      index=5, header=false, col sep=comma]
      {../../Plots/stampede_multi_nomad_yahoo_32_4_100_1.000000_0.000500_0.050000.txt};
      
      \addplot[ultra thick, color=red, mark=none] table [x index=0, y index=1, header=false]
      {../../Plots/stampede_multi_dsgd_bold_wor_yahoo_32_4_100_1.000000_0.000500_0.500000.txt};

      \addplot[ultra thick, loosely dashed, color=orange, mark=none] table [x index=0, y index=1, header=false]
      {../../Plots/stampede_multi_dsgdpp_yahoo_32_4_100_1.000000_0.000500_0.500000.txt};

      \addplot[ultra thick, color=brown, mark=none, dashed] table [x index=0, y index=1, header=false]
      {../../Plots/stampede_multi_ccdpp_yahoo_32_4_100_1.000000.txt};

      \legend{NOMAD, DSGD, DSGD++, CCD++}
    \end{axis}
  \end{tikzpicture}
  \begin{tikzpicture}[scale=0.6]
    \begin{axis}[minor tick num=1,
      title={Hugewiki, machines=64, cores=4, $\lambda=0.01$, $k=100$},
      xlabel={seconds}, ylabel={test RMSE},
      ymin={0.5}, ymax={0.7}, xmin={0}, xmax={750}]
      
      \addplot[ultra thick, color=blue, mark=*] table [x index=3, y
      index=5, header=false, col sep=comma]
      {../../Plots/stampede_multi_nomad_hugewiki_64_4_100_0.010000_0.000667_0.000000.txt};
      
      \addplot[ultra thick, color=red, mark=none] table [x index=0, y index=1, header=false]      
      {../../Plots/stampede_multi_dsgd_bold_wor_hugewiki_64_4_100_0.010000_0.008000_0.500000.txt};

      \addplot[ultra thick, loosely dashed, color=orange, mark=none] table [x index=0, y index=1, header=false]
      {../../Plots/stampede_multi_dsgdpp_hugewiki_64_4_100_0.010000_0.008000_0.500000.txt};

      \addplot[ultra thick, color=brown, mark=none, dashed] table [x index=0, y index=1, header=false]
      {../../Plots/stampede_multi_ccdpp_hugewiki_64_4_100_0.010000.txt};

      \legend{NOMAD, DSGD, DSGD++, CCD++}
    \end{axis}
  \end{tikzpicture}

  \caption{Comparison of NOMAD, DSGD, DSGD++, and CCD++ on a HPC cluster.}
  \label{fig:multi_compare}
\end{figure*}

In this subsection, we use 4 computation threads per machine.  For the
first experiment we fix the number of machines to 32 (64 for hugewiki),
and compare the performance of NOMAD with DSGD, DSGD++ and CCD++
(Figure~\ref{fig:multi_compare}).  On Netflix and Hugewiki, NOMAD
converges much faster than its competitors; not only initial convergence
is faster, it also discovers a better quality solution.  On Yahoo!
Music, four methods perform almost the same to each other. This is
because the cost of network communication relative to the size of the
data is much higher for Yahoo!\ Music; while Netflix and Hugewiki have
5,575 and 68,635 non-zero ratings per each item respectively, Yahoo!\
Music has only 404 ratings per item.  Therefore, when Yahoo!~Music
is divided equally across 32 machines, each item has only 10
ratings on average per each machine.  Hence the cost of sending and
receiving item parameter vector $\hb_j$ for one item $j$ across the
network is higher than that of executing SGD updates on the ratings of
the item locally stored within the machine, $\Omegabar^{(q)}_j$.  As a
consequence, the cost of network communication dominates the overall
execution time of all algorithms, and little difference in convergence
speed is found between them.

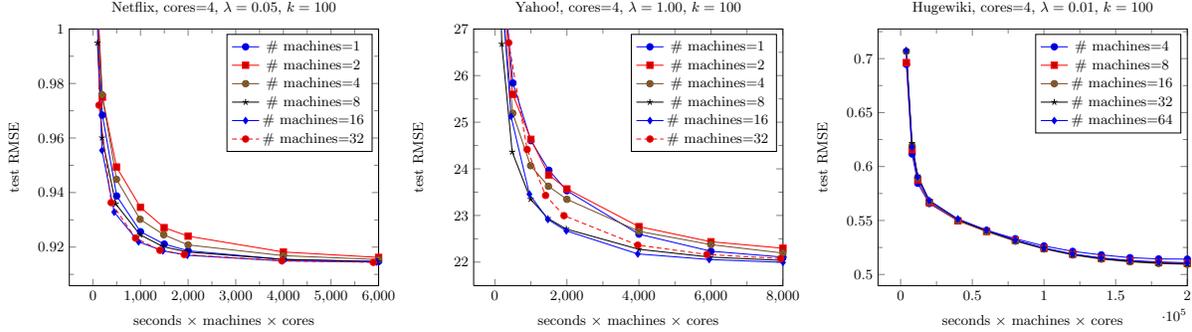
\begin{figure*}[htbp]
  \centering
  \begin{tikzpicture}[scale=0.6]

    \begin{axis}[minor tick num=1,
      title={Netflix, cores=4, $\lambda=0.05$, $k=100$},
      xlabel={seconds $\times$ machines $\times$ cores}, ylabel={test RMSE},
      xmax={6000}, ymax={1}]
      
      \addplot
      table [x expr=\thisrowno{3} * 4, y
      index=5, header=false, col sep=comma]
      {../../Plots/stampede_multi_nomad_netflix_1_4_100_0.050000_0.008000_0.010000.txt};

      \addplot
      table [x expr=\thisrowno{3}*8, y
      index=5, header=false, col sep=comma]
      {../../Plots/stampede_multi_nomad_netflix_2_4_100_0.050000_0.008000_0.010000.txt};

      \addplot
      table [x expr=\thisrowno{3}*16, y
      index=5, header=false, col sep=comma]
      {../../Plots/stampede_multi_nomad_netflix_4_4_100_0.050000_0.008000_0.010000.txt};

      \addplot
      table [x expr=\thisrowno{3}*32, y
      index=5, header=false, col sep=comma]
      {../../Plots/stampede_multi_nomad_netflix_8_4_100_0.050000_0.008000_0.010000.txt};

      \addplot
      table [x expr=\thisrowno{3}*64, y
      index=5, header=false, col sep=comma]
      {../../Plots/stampede_multi_nomad_netflix_16_4_100_0.050000_0.008000_0.010000.txt};

      \addplot
      table [x expr=\thisrowno{3}*128, y
      index=5, header=false, col sep=comma]
      {../../Plots/stampede_multi_nomad_netflix_32_4_100_0.050000_0.008000_0.010000.txt};
      
      \legend{\# machines=1, \# machines=2, \# machines=4,\#
        machines=8,\# machines=16, \# machines=32}
    \end{axis}
  \end{tikzpicture}
  \begin{tikzpicture}[scale=0.6]
    \begin{axis}[minor tick num=1,
      title={Yahoo!, cores=4, $\lambda=1.00$, $k=100$},
      xlabel={seconds $\times$ machines $\times$ cores}, ylabel={test RMSE},
      xmax={8000}, ymax={27}]
      
      \addplot
      table [x expr=\thisrowno{3}*4, y
      index=5, header=false, col sep=comma]
      {../../Plots/stampede_multi_nomad_yahoo_1_4_100_1.000000_0.000500_0.050000.txt};

      \addplot 
      table [x expr=\thisrowno{3}*8, y
      index=5, header=false, col sep=comma]
      {../../Plots/stampede_multi_nomad_yahoo_2_4_100_1.000000_0.000500_0.050000.txt};

      \addplot 
      table [x expr=\thisrowno{3}*16, y
      index=5, header=false, col sep=comma]
      {../../Plots/stampede_multi_nomad_yahoo_4_4_100_1.000000_0.000500_0.050000.txt};

      \addplot 
      table [x expr=\thisrowno{3}*32, y
      index=5, header=false, col sep=comma]
      {../../Plots/stampede_multi_nomad_yahoo_8_4_100_1.000000_0.000500_0.050000.txt};

      \addplot 
      table [x expr=\thisrowno{3}*64, y
      index=5, header=false, col sep=comma]
      {../../Plots/stampede_multi_nomad_yahoo_16_4_100_1.000000_0.000500_0.050000.txt};

      \addplot 
      table [x expr=\thisrowno{3}*128, y
      index=5, header=false, col sep=comma]
      {../../Plots/stampede_multi_nomad_yahoo_32_4_100_1.000000_0.000500_0.050000.txt};
      
      \legend{\# machines=1, \# machines=2,\# machines=4,\#
        machines=8, \# machines=16, \# machines=32}
    \end{axis}
  \end{tikzpicture}
  \begin{tikzpicture}[scale=0.6]
    \begin{axis}[minor tick num=1,
      title={Hugewiki, cores=4, $\lambda=0.01$, $k=100$},
      xlabel={seconds $\times$ machines $\times$ cores}, ylabel={test RMSE}, xmax={200000}]
      
      \addplot
      table [x expr=\thisrowno{3}*16, y
      index=5, header=false, col sep=comma]
      {../../Plots/stampede_multi_nomad_hugewiki_4_4_100_0.010000_0.000667_0.000000.txt};

      \addplot
      table [x expr=\thisrowno{3}*32, y
      index=5, header=false, col sep=comma]
      {../../Plots/stampede_multi_nomad_hugewiki_8_4_100_0.010000_0.000667_0.000000.txt};

      \addplot
      table [x expr=\thisrowno{3}*64, y
      index=5, header=false, col sep=comma]
      {../../Plots/stampede_multi_nomad_hugewiki_16_4_100_0.010000_0.000667_0.000000.txt};

      \addplot
      table [x expr=\thisrowno{3}*128, y
      index=5, header=false, col sep=comma]
      {../../Plots/stampede_multi_nomad_hugewiki_32_4_100_0.010000_0.000667_0.000000.txt};

      \addplot
      table [x expr=\thisrowno{3}*256, y
      index=5, header=false, col sep=comma]
      {../../Plots/stampede_multi_nomad_hugewiki_64_4_100_0.010000_0.000667_0.000000.txt};

      \legend{\# machines=4,\# machines=8,
      \# machines=16, \# machines=32, \# machines=64}
    \end{axis}
  \end{tikzpicture}
  \caption{Test RMSE of NOMAD as a function of computation time (time in
    seconds $\times$ the number of machines $\times$ the number of cores
    per each machine) on a HPC cluster, when the number of machines is
    varied.}
  \label{fig:multi_converge_time}
\end{figure*}

For the second experiment we varied the number of machines from 1 to 32,
and plot the scaling behavior of NOMAD
(Figures~\ref{fig:multi_collection} and \ref{fig:multi_converge_time}).
Figure~\ref{fig:multi_collection} (left) shows how test RMSE decreases
as a function of the number of updates on Yahoo!~Music.  Again, if NOMAD
scales linearly the average throughput has to remain constant; here we
observe improvement in convergence speed when 8 or more machines are
used.  This is again the effect of smaller block sizes which was
discussed in Section~\ref{sec:ScalingNumberCores}.  On Netflix, a
similar effect was present but was less significant; on Hugewiki we did
not see any notable difference between configurations (see
Figure~\ref{fig:multi_converge_update} in
Appendix~\ref{sec:test-rmse}). 

In Figure~\ref{fig:multi_collection} (right) we plot the average
throughput (the number of updates per machine per core per second) as a
function of the number of machines.  On Yahoo!~Music the average
throughput goes down as we increase the number of machines, because as
mentioned above, each item has a small number of ratings.  On Hugewiki
we observe almost linear scaling, and on Netflix the average throughput
even improves as we increase the number of machines; we believe this is
because of cache locality effects.  As we partition users into smaller
and smaller blocks, the probability of cache miss on user parameters
$\wb_i$'s within the block decrease, and on Netflix this makes a
meaningful difference: indeed, there are only 480,189 users in Netflix
who have at least one rating.  When this is equally divided into 32
machines, each machine contains only 11,722 active users on average.
Therefore the $\wb_i$ variables only take 11MB of memory, which is
smaller than the size of L3 cache (20MB) of the machine we used and
therefore leads to increase in the number of updates per machine per
core per second.

Now we study how much speed-up NOMAD can achieve by increasing the
number of machines.  In Figure~\ref{fig:multi_converge_time}, we set
$y$-axis to be test RMSE and $x$-axis to be the number of seconds
elapsed multiplied by the total number of cores used in the
configuration.  Again, all lines will coincide with each other if NOMAD
shows linear scaling.  On Netflix, with 2 and 4 machines we observe mild
slowdown, but with more than 4 machines NOMAD exhibits super-linear
scaling.  On Yahoo!~Music we observe super-linear scaling with respect
to the speed of a single machine on all configurations, but the highest
speedup is seen with 16 machines.  On Hugewiki, linear scaling is
observed in every configuration.

\subsection{Scaling on Commodity Hardware}
\label{sec:ScalDiffHardw}

In this subsection, we want to analyze the scaling behavior of NOMAD
on commodity hardware.  Using Amazon Web Services (AWS), we set up a
computing cluster that consists of 32 machines; each machine is of
type \texttt{m1.xlarge} and equipped with quad-core Intel Xeon E5430
CPU and 15GB of RAM.  Network bandwidth among these machines is
reported to be approximately
1Gb/s\footnote{\url{http://epamcloud.blogspot.com/2013/03/testing-amazon-ec2-network-speed.html}}.

\begin{figure*}[htbp]
  \centering
  \begin{tikzpicture}[scale=0.6]
    \begin{axis}[minor tick num=1,
      title={Netflix, machines=32, cores=4, $\lambda=0.05$, $k=100$},
      xlabel={seconds}, ylabel={test RMSE},
      ymin={0.91}, ymax={1.0}, xmin={0}, xmax={500}]
      
      \addplot[ultra thick, color=blue, mark=*] table [x index=3, y
      index=5, header=false, col sep=comma]
      {../../Plots/aws_nomad_netflix_32_2_100_0.050000_0.008000_0.010000.txt};
      
      \addplot[ultra thick, color=red, mark=none] table [x index=0, y index=1, header=false]
      {../../Plots/aws_dsgd_bold_wor_netflix_32_4_100_0.050000_0.008000_0.500000.txt};

      \addplot[ultra thick, loosely dashed, color=orange, mark=none] table [x index=0, y index=1, header=false]
      {../../Plots/aws_dsgdpp_netflix_32_4_100_0.050000_0.008000_0.500000.txt};

      \addplot[ultra thick, color=brown, mark=none, dashed] table [x index=0, y index=1, header=false]
      {../../Plots/aws_ccdpp_netflix_32_4_100_0.050000.txt};
      
      \legend{NOMAD, DSGD, DSGD++, CCD++}
    \end{axis}
  \end{tikzpicture}
  \begin{tikzpicture}[scale=0.6]
    \begin{axis}[minor tick num=1,
      title={Yahoo!, machines=32, cores=4, $\lambda=1.00$, $k=100$},
      xlabel={seconds}, ylabel={test RMSE},
      ymin={21}, ymax={27}, xmin={0}, xmax={600}]
      
      \addplot[ultra thick, color=blue, mark=*] table [x index=3, y
      index=5, header=false, col sep=comma]
      {../../Plots/aws_nomad_yahoo_32_2_100_1.000000_0.000500_0.050000.txt};
      
      \addplot[ultra thick, color=red, mark=none] table [x index=0, y index=1, header=false]
      {../../Plots/aws_dsgd_bold_wor_yahoo_32_4_100_1.000000_0.000500_0.500000.txt};

      \addplot[ultra thick, loosely dashed, color=orange, mark=none] table [x index=0, y index=1, header=false]
      {../../Plots/aws_dsgdpp_yahoo_32_4_100_1.000000_0.000500_0.500000.txt};

      \addplot[ultra thick, color=brown, mark=none, dashed] table [x index=0, y index=1, header=false]
      {../../Plots/aws_ccdpp_yahoo_32_4_100_1.000000.txt};
      
      \legend{NOMAD, DSGD, DSGD++, CCD++}
    \end{axis}
  \end{tikzpicture}
  \begin{tikzpicture}[scale=0.6]
    \begin{axis}[minor tick num=1,
      title={Hugewiki, machines=32, cores=4, $\lambda=1.00$, $k=100$},
      xlabel={seconds}, ylabel={test RMSE},
      ymin={0.5}, ymax={0.65}, xmin={0}, xmax={4500}]
      
      \addplot[ultra thick, color=blue, mark=*] table [x index=3, y
      index=5, header=false, col sep=comma]
      {../../Plots/aws_nomad_hugewiki_32_2_100_0.010000_0.000667_0.000000.txt};

      \addplot[ultra thick, color=red, mark=none] table [x index=0, y index=1, header=false]
      {../../Plots/aws_dsgd_bold_wor_hugewiki_32_4_100_0.010000_0.008000_0.500000.txt};

      \addplot[ultra thick, loosely dashed, color=orange, mark=none] table [x index=0, y index=1, header=false]
      {../../Plots/aws_dsgdpp_hugewiki_32_4_100_0.010000_0.008000_0.500000.txt};

      \addplot[ultra thick, color=brown, mark=none, dashed] table [x index=0, y index=1, header=false]
      {../../Plots/aws_ccdpp_hugewiki_32_4_100_0.010000.txt}; 

      \legend{NOMAD, DSGD, DSGD++, CCD++}
    \end{axis}
  \end{tikzpicture}

  \caption{Comparison of NOMAD, DSGD, DSGD++, and CCD++ on a commodity hardware
    cluster.}

  \label{fig:aws_compare}
\end{figure*}
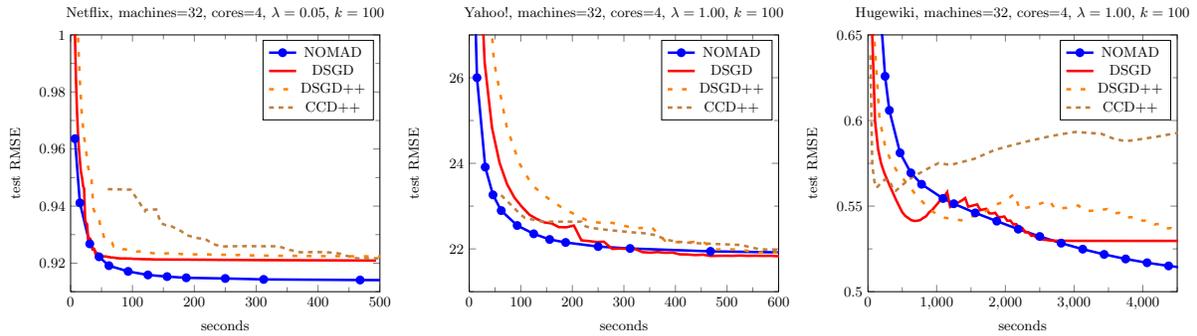

Since NOMAD and DSGD++ dedicates two threads for network
communication, on each machine only two cores are available for
computation\footnote{Since network communication is not
  computation-intensive, for DSGD++ we used four computation threads
  instead of two and got better results; thus we report results with
  four computation threads for DSGD++. }. In contrast, bulk
synchronization algorithms such as DSGD and CCD++ which separate
computation and communication can utilize all four cores for
computation.
In spite of this disadvantage, Figure~\ref{fig:aws_compare} shows that
NOMAD outperforms all other algorithms in this setting as well.  In this
plot, we fixed the number of machines to 32; on Netflix and Hugewiki,
NOMAD converges more rapidly to a better solution.  Recall that on
Yahoo!~Music, all four algorithms performed very similarly on a HPC
cluster in Section~\ref{sec:ScalingasFixed}.  However, on commodity
hardware NOMAD outperforms the other algorithms.  This shows that the
efficiency of network communication plays a very important role in
commodity hardware clusters where the communication is relatively slow.
On Hugewiki, however, the number of columns is very small
compared to the number of ratings and thus network communication plays
smaller role in this dataset compared to others.  Therefore, initial
convergence of DSGD is a bit faster than NOMAD as it uses all four cores
on computation while NOMAD uses only two.  Still, the overall
convergence speed is similar and NOMAD finds a better quality solution.

As in Section~\ref{sec:ScalingasFixed}, we increased the number of
machines from 1 to 32, and studied the scaling behavior of NOMAD.  The
overall trend was identical to what we observed in
Figure~\ref{fig:multi_collection} and \ref{fig:multi_converge_time};
due to page constraints, the plots for this experiment can be found in
the Appendix~\ref{app:ScalDiffHardw}.

\begin{figure}[tbp]
  \centering
  \begin{tikzpicture}[scale=0.48]
    \begin{axis}[minor tick num=1,
      title={Yahoo!, cores=4, $\lambda=1.00$, $k=100$},
      xlabel={number of updates}, ylabel={test RMSE},
      xmax={30000000000}, ymax={27}]
      
      \addplot
      table [x index=4, y
      index=5, header=false, col sep=comma]
      {../../Plots/stampede_multi_nomad_yahoo_1_4_100_1.000000_0.000500_0.050000.txt};

      \addplot 
      table [x index=4, y
      index=5, header=false, col sep=comma]
      {../../Plots/stampede_multi_nomad_yahoo_2_4_100_1.000000_0.000500_0.050000.txt};

      \addplot 
      table [x index=4, y
      index=5, header=false, col sep=comma]
      {../../Plots/stampede_multi_nomad_yahoo_4_4_100_1.000000_0.000500_0.050000.txt};

      \addplot 
      table [x index=4, y
      index=5, header=false, col sep=comma]
      {../../Plots/stampede_multi_nomad_yahoo_8_4_100_1.000000_0.000500_0.050000.txt};

      \addplot 
      table [x index=4, y
      index=5, header=false, col sep=comma]
      {../../Plots/stampede_multi_nomad_yahoo_16_4_100_1.000000_0.000500_0.050000.txt};

      \addplot 
      table [x index=4, y
      index=5, header=false, col sep=comma]
      {../../Plots/stampede_multi_nomad_yahoo_32_4_100_1.000000_0.000500_0.050000.txt};
      
      \legend{\# machines=1, \# machines=2,\# machines=4,\#
        machines=8, \# machines=16, \# machines=32}
    \end{axis}
  \end{tikzpicture}
  \begin{tikzpicture}[scale=0.48]
    \begin{axis}[minor tick num=1, only marks, 
      title={Netflix, cores=4, $\lambda=0.05$, $k=100$},
      xlabel={number of machines}, ylabel={updates per machine per core per sec},
      legend style={legend pos=south west}, ymin={0}, xmax={34} ]

      \addplot [color=blue, mark=o] table [x index=1, y
      expr={\thisrowno{4}/\thisrowno{3}/\thisrowno{2}/\thisrowno{1}}, header=false,
      col sep=comma]
      {../../Plots/stampede_multi_nomad_netflix_all_4_100_0.050000_0.008000_0.010000.txt};

      \addplot [color=red, mark=*] table [x index=1, y
      expr={\thisrowno{4}/\thisrowno{3}/\thisrowno{2}/\thisrowno{1}}, header=false,
      col sep=comma]
      {../../Plots/stampede_multi_nomad_yahoo_all_4_100_1.000000_0.000500_0.050000.txt};

      \addplot [color=brown, mark=+] table [x index=1, y
      expr={\thisrowno{4}/\thisrowno{3}/\thisrowno{2}/\thisrowno{1}}, header=false,
      col sep=comma]
      {../../Plots/stampede_multi_nomad_hugewiki_all_4_100_0.010000_0.000667_0.000000.txt};

      \legend{Netflix, Yahoo!, Hugewiki}

    \end{axis}
  \end{tikzpicture}
  \caption{Results on HPC cluster when the number of machines is
    varied. Left: Test RMSE of NOMAD as a function of the number of
    updates on Netflix and Yahoo!~Music. Right: Number of updates of
    NOMAD per machine per core per second as a function of the number of
    machines.}
  \label{fig:multi_collection}
\end{figure}
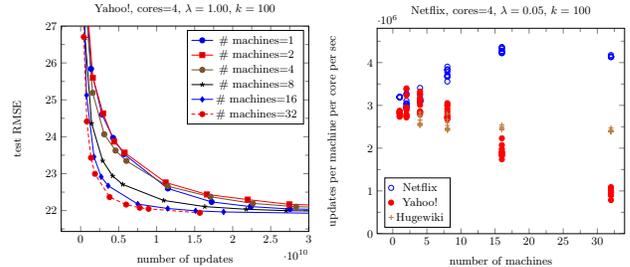

\subsection{Scaling as both Dataset Size and Number of Machines Grows}
\label{sec:Scalingasboth}

In previous sections (Section~\ref{sec:ScalingasFixed} and
Section~\ref{sec:ScalDiffHardw}), we studied the scalability of
algorithms by partitioning a fixed amount of data into increasing
number of machines.  In real-world applications of collaborative
filtering, however, the size of the data should grow over time as new
users are added to the system.  Therefore, to match the increased
amount of data with equivalent amount of physical memory and
computational power, the number of machines should increase as well.
The aim of this section is to compare the scaling behavior of NOMAD
and that of other algorithms in this realistic scenario.

To simulate such a situation, we generated synthetic datasets which
resemble characteristics of real data; the number of ratings for each
user and each item is sampled from the corresponding empirical
distribution of the Netflix data.  As we increase the number of machines
from 4 to 32, we fixed the number of items to be the same to that of
Netflix (17,770), and increased the number of users to be proportional
to the number of machines (480,189 $\times$ the number of
machines\footnote{480,189 is the number of users in Netflix who have at
  least one rating. }).  Therefore, the expected number of ratings in
each dataset is proportional to the number of machines (99,072,112
$\times$ the number of machines) as well.

Conditioned on the number of ratings for each user and item, the
nonzero locations are sampled uniformly at random.  Ground-truth user
parameters $\wb_i$'s and item parameters $\hb_j$'s are generated from
100-dimensional standard isometric Gaussian distribution, and for each
rating $A_{ij}$, Gaussian noise with mean zero and standard deviation
0.1 is added to the ``true'' rating $\inner{\wb_i}{\hb_j}$.

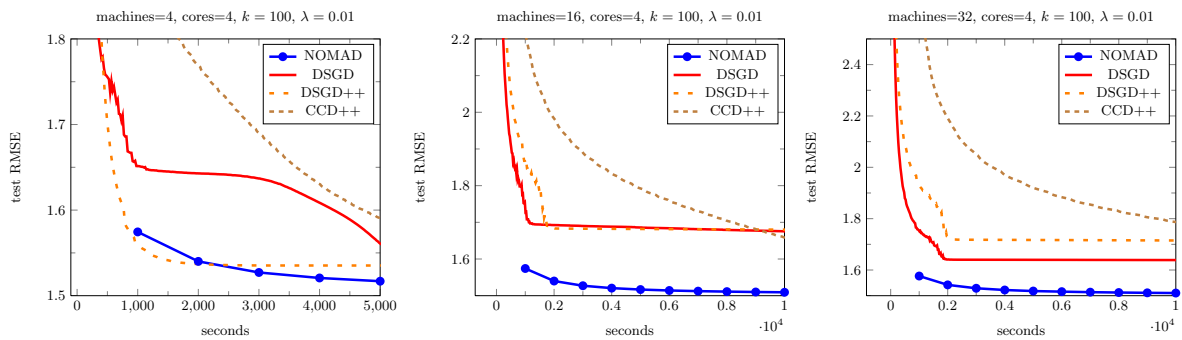
\begin{figure*}[htbp]
  \centering
  \begin{tikzpicture}[scale=0.6]
    \begin{axis}[minor tick num=1,
      title={machines=4, cores=4, $k=100$, $\lambda=0.01$},
      xlabel={seconds}, ylabel={test RMSE}, ymin={1.5}, ymax={1.8}, xmax={5000}
      ]
      
      \addplot[ultra thick, color=blue, mark=*] table [x index=3, y
      index=5, header=false, col sep=comma]
      {../../Plots/stampede_newsynth_nomad_synth_4_4_100_0.010000_0.000800_0.000100.txt};

      
      \addplot[ultra thick, color=red, mark=none] table [x index=0, y
      index=1, header=false]
      {../../Plots/stampede_newsynth_dsgd_bold_wor_4_4_100_0.010000_0.002000_0.500000.txt};

      \addplot[ultra thick, loosely dashed, color=orange, mark=none] table [x index=0, y
      index=1, header=false]
      {../../Plots/stampede_newsynth_dsgdpp_4_4_100_0.010000_0.002000_0.500000.txt};

      \addplot[ultra thick, color=brown, mark=none, dashed] table [x index=0, y index=1, header=false]
      {../../Plots/stampede_newsynth_ccdpp_4_4_100_0.010000.txt};

      \legend{NOMAD, DSGD, DSGD++, CCD++}
    \end{axis}
  \end{tikzpicture}
  \begin{tikzpicture}[scale=0.6]
    \begin{axis}[minor tick num=1,
      title={machines=16, cores=4, $k=100$, $\lambda=0.01$},
      xlabel={seconds}, ylabel={test RMSE}, ymin={1.5}, ymax={2.2}, xmax={10000}
      ]
      
      \addplot[ultra thick, color=blue, mark=*] table [x index=3, y
      index=5, header=false, col sep=comma]
      {../../Plots/stampede_newsynth_nomad_synth_16_4_100_0.010000_0.000800_0.000100.txt};
      
      \addplot[ultra thick, color=red, mark=none] table [x index=0, y
      index=1, header=false]
      {../../Plots/stampede_newsynth_dsgd_bold_wor_16_4_100_0.010000_0.002000_0.500000.txt};

      \addplot[ultra thick, loosely dashed, color=orange, mark=none] table [x index=0, y
      index=1, header=false]
      {../../Plots/stampede_newsynth_dsgdpp_16_4_100_0.010000_0.002000_0.500000.txt};

      \addplot[ultra thick, color=brown, mark=none, dashed] table [x index=0, y index=1, header=false]
      {../../Plots/stampede_newsynth_ccdpp_16_4_100_0.010000.txt};

      \legend{NOMAD, DSGD, DSGD++, CCD++}
    \end{axis}
  \end{tikzpicture}
  \begin{tikzpicture}[scale=0.6]
    \begin{axis}[minor tick num=1,
      title={machines=32, cores=4, $k=100$, $\lambda=0.01$},
      xlabel={seconds}, ylabel={test RMSE}, ymin={1.5}, ymax={2.5}, xmax={10000}
      ]
      
      \addplot[ultra thick, color=blue, mark=*] table [x index=3, y
      index=5, header=false, col sep=comma]
      {../../Plots/stampede_newsynth_nomad_synth_32_4_100_0.010000_0.000800_0.000100.txt};

      
      \addplot[ultra thick, color=red, mark=none] table [x index=0, y
      index=1, header=false]
      {../../Plots/stampede_newsynth_dsgd_bold_wor_32_4_100_0.010000_0.002000_0.500000.txt};

      \addplot[ultra thick, loosely dashed, color=orange, mark=none] table [x index=0, y
      index=1, header=false]
      {../../Plots/stampede_newsynth_dsgdpp_32_4_100_0.010000_0.002000_0.500000.txt};

      \addplot[ultra thick, color=brown, mark=none, dashed] table [x index=0, y index=1, header=false]
      {../../Plots/stampede_newsynth_ccdpp_32_4_100_0.010000.txt};

      \legend{NOMAD, DSGD, DSGD++, CCD++} 
    \end{axis}
  \end{tikzpicture}

  \caption{Comparison of algorithms when both dataset size and the
    number of machines grows.  Left: 4 machines, middle: 16 machines,
    right: 32 machines}
  \label{fig:scaleboth}
\end{figure*}

Figure~\ref{fig:scaleboth} shows that the comparative advantage of
NOMAD against DSGD, DSGD++ and CCD++ increases as we grow the scale of
the problem.  NOMAD clearly outperforms other methods on all
configurations; DSGD++ is very competitive on the small scale, but as
the size of the problem grows NOMAD shows better scaling behavior.

\section{Conclusion and Future Work}
\label{sec:ConclusionFutureWork}

From our experimental study we conclude that
\begin{itemize}
\item On a single machine, NOMAD shows near-linear scaling up to 30
  threads.
\item When a fixed size dataset is distributed across multiple
  machines, NOMAD shows near-linear scaling up to 32 machines.
\item Both in shared-memory and distributed-memory setting, NOMAD
  exhibits superior performance against state-of-the-art competitors;
  in commodity hardware cluster, the comparative advantage is more
  conspicuous.
\item When both the size of the data as well as the number of machines
  grow, the scaling behavior of NOMAD is much nicer than its
  competitors.
\end{itemize}

Although we only discussed the matrix completion problem in this paper,
it is worth noting that the idea of NOMAD is more widely
applicable. Specifically, ideas discussed in this paper can be easily
adapted as long as the objective function can be written as
\begin{align*}
  f(W,H) = \sum_{i,j \in \Omega} f_{ij}(\wb_i, \hb_j).
\end{align*}
As part of our ongoing work we are investigating ways to rewrite
Support Vector Machines (SVMs), binary logistic regression as a saddle
point problem which have the above structure. 

Inference in Latent Dirichlet Allocation (LDA) using a collapsed Gibbs
sampler has a similar structure as the stochastic gradient descent
updates for matrix factorization. An additional complication in LDA is
that the variables need to be normalized. We are investigating how the
NOMAD framework can be used for LDA. 




\balance

\section{Acknowledgements}
\label{sec:Acknowledgements}

We thank the anonymous reviewers for their constructive comments.  We
thank the Texas Advanced Computing Center at University of Texas and
the Research Computing group at Purdue University for providing
infrastructure and timely support for our experiments. Computing
experiments on commodity hardware were made possible by an AWS in
Education Machine Learning Research Grant Award. This material is
partially based upon work supported by the National Science Foundation
under grant no IIS-1219015 and CCF-1117055.


\bibliographystyle{abbrvnat}
\bibliography{nomad}  

\newpage
\appendix

\section{Effect of the Regularization Parameter}
\label{sec:EffectRegulParam}

\begin{figure*}[htbp]
  \centering
  \begin{tikzpicture}[scale=0.68]
    \begin{axis}[minor tick num=1,
      title={Netflix, machines=8, cores=4, $k=100$},
      xlabel={seconds}, ylabel={test RMSE},
      xmin={0}, xmax={150}]
      \addplot[ultra thick, color=blue, mark=*] table [x index=3, y
      index=5, header=false, col sep=comma] {../../Results/reg_netflix_r0.000500.txt};

      \addplot[ultra thick, color=green, mark=x] table [x index=3, y
      index=5, header=false, col sep=comma] {../../Results/reg_netflix_r0.005000.txt};

      \addplot[ultra thick, color=red, mark=diamond*] table [x index=3, y
      index=5, header=false, col sep=comma] {../../Results/reg_netflix_r0.050000.txt};

      \addplot[ultra thick, color=black, mark=square*] table [x index=3, y
      index=5, header=false, col sep=comma] {../../Results/reg_netflix_r0.500000.txt};

      \legend{$\lambda=0.0005$, $\lambda=0.005$, $\lambda=0.05$, $\lambda=0.5$}

    \end{axis}
  \end{tikzpicture}
  \begin{tikzpicture}[scale=0.68]
    \begin{axis}[minor tick num=1,
      title={Yahoo!, machines=8, cores=4, $k=100$},
      xlabel={seconds}, ylabel={test RMSE},
      xmin={0}, xmax={150}]
      \addplot[ultra thick, color=blue, mark=*] table [x index=3, y
      index=5, header=false, col sep=comma] {../../Results/reg_yahoo_r0.250000.txt};

      \addplot[ultra thick, color=green, mark=x] table [x index=3, y
      index=5, header=false, col sep=comma] {../../Results/reg_yahoo_r0.500000.txt};

      \addplot[ultra thick, color=red, mark=diamond*] table [x index=3, y
      index=5, header=false, col sep=comma] {../../Results/reg_yahoo_r1.000000.txt};

      \addplot[ultra thick, color=black, mark=square*] table [x index=3, y
      index=5, header=false, col sep=comma] {../../Results/reg_yahoo_r2.000000.txt};

      \legend{$\lambda=0.25$, $\lambda=0.5$, $\lambda=1$, $\lambda=2$}
    \end{axis}
  \end{tikzpicture}
  \begin{tikzpicture}[scale=0.68]
    \begin{axis}[minor tick num=1,
      title={Hugewiki, machines=8, cores=4, $k=100$},
      xlabel={seconds}, ylabel={test RMSE},
      xmin={0}, xmax={150}]
      \addplot[ultra thick, color=blue, mark=*] table [x index=3, y
      index=5, header=false, col sep=comma] {../../Results/reg_hugewiki_r0.002500.txt};

      \addplot[ultra thick, color=green, mark=x] table [x index=3, y
      index=5, header=false, col sep=comma] {../../Results/reg_hugewiki_r0.005000.txt};

      \addplot[ultra thick, color=red, mark=diamond*] table [x index=3, y
      index=5, header=false, col sep=comma] {../../Results/reg_hugewiki_r0.010000.txt};

      \addplot[ultra thick, color=black, mark=square*] table [x index=3, y
      index=5, header=false, col sep=comma] {../../Results/reg_hugewiki_r0.020000.txt};

      \legend{$\lambda=0.0025$, $\lambda=0.005$, $\lambda=0.01$, $\lambda=0.02$}
    \end{axis}
  \end{tikzpicture}

  \caption{Convergence behavior of NOMAD when the regularization
    parameter $\lambda$ is varied.}
  \label{fig:vary_regularization}
\end{figure*}
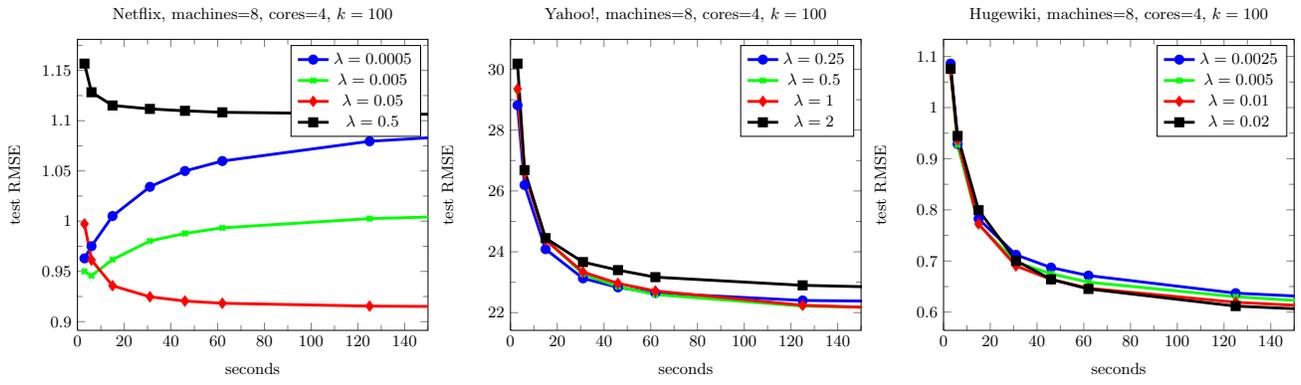

In this subsection, we study the convergence behavior of NOMAD as we
change the regularization parameter $\lambda$
(Figure~\ref{fig:vary_regularization}).  Note that in Netflix data
(left), for non-optimal choices of the regularization parameter the
test RMSE increases from the initial solution as the model overfits or
underfits to the training data.  While NOMAD reliably converges in all
cases, on Netflix the convergence is notably faster with
higher values of $\lambda$; this is expected because regularization
smooths the objective function and makes the optimization problem
easier to solve.  On other datasets, the speed of convergence was not
very sensitive to the selection of the regularization parameter.

\section{Effect of the Latent Dimension}
\label{sec:EffectLatentDimens}

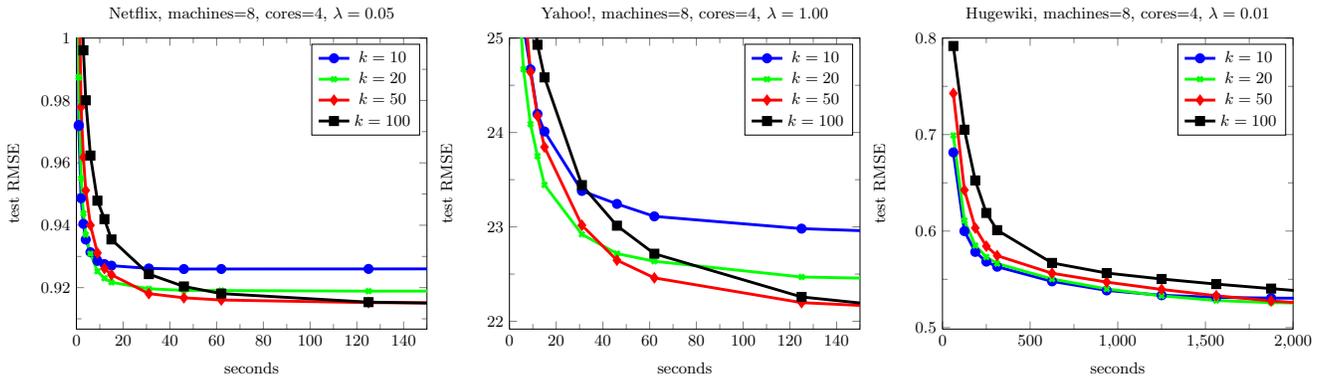
\begin{figure*}[htbp]
  \centering
  \begin{tikzpicture}[scale=0.68]
    \begin{axis}[minor tick num=1,
      title={Netflix, machines=8, cores=4, $\lambda=0.05$},
      xlabel={seconds}, ylabel={test RMSE},
      xmin={0}, xmax={150}, ymax={1.0}]
      \addplot[ultra thick, color=blue, mark=*] table [x index=3, y
      index=5, header=false, col sep=comma] {../../Results/dim_netflix_d10.txt};

      \addplot[ultra thick, color=green, mark=x] table [x index=3, y
      index=5, header=false, col sep=comma] {../../Results/dim_netflix_d20.txt};

      \addplot[ultra thick, color=red, mark=diamond*] table [x index=3, y
      index=5, header=false, col sep=comma] {../../Results/dim_netflix_d50.txt};

      \addplot[ultra thick, color=black, mark=square*] table [x index=3, y
      index=5, header=false, col sep=comma] {../../Results/dim_netflix_d100.txt};

      \legend{$k=10$, $k=20$, $k=50$, $k=100$}

    \end{axis}
  \end{tikzpicture}
  \begin{tikzpicture}[scale=0.68]
    \begin{axis}[minor tick num=1,
      title={Yahoo!, machines=8, cores=4, $\lambda=1.00$},
      xlabel={seconds}, ylabel={test RMSE},
      xmin={0}, xmax={150}, ymax={25}]
      \addplot[ultra thick, color=blue, mark=*] table [x index=3, y
      index=5, header=false, col sep=comma] {../../Results/dim_yahoo_d10.txt};

      \addplot[ultra thick, color=green, mark=x] table [x index=3, y
      index=5, header=false, col sep=comma] {../../Results/dim_yahoo_d20.txt};

      \addplot[ultra thick, color=red, mark=diamond*] table [x index=3, y
      index=5, header=false, col sep=comma] {../../Results/dim_yahoo_d50.txt};

      \addplot[ultra thick, color=black, mark=square*] table [x index=3, y
      index=5, header=false, col sep=comma] {../../Results/dim_yahoo_d100.txt};

      \legend{$k=10$, $k=20$, $k=50$, $k=100$}
    \end{axis}
  \end{tikzpicture}
  \begin{tikzpicture}[scale=0.68]
    \begin{axis}[minor tick num=1,
      title={Hugewiki, machines=8, cores=4, $\lambda=0.01$},
      xlabel={seconds}, ylabel={test RMSE},
      xmin={0}, xmax={2000}, ymax={0.8}]
      \addplot[ultra thick, color=blue, mark=*] table [x index=3, y
      index=5, header=false, col sep=comma] 
      {../../Plots/stampede_varydim_nomad_hugewiki_8_4_10_0.010000_0.000667_0.000000.txt};

      \addplot[ultra thick, color=green, mark=x] table [x index=3, y
      index=5, header=false, col sep=comma] 
      {../../Plots/stampede_varydim_nomad_hugewiki_8_4_20_0.010000_0.000667_0.000000.txt};

      \addplot[ultra thick, color=red, mark=diamond*] table [x index=3, y
      index=5, header=false, col sep=comma]
      {../../Plots/stampede_varydim_nomad_hugewiki_8_4_50_0.010000_0.000667_0.000000.txt};

      \addplot[ultra thick, color=black, mark=square*] table [x index=3, y
      index=5, header=false, col sep=comma]
      {../../Plots/stampede_varydim_nomad_hugewiki_8_4_100_0.010000_0.000667_0.000000.txt};

      \legend{$k=10$, $k=20$, $k=50$, $k=100$}
    \end{axis}
  \end{tikzpicture}

  \caption{Convergence behavior of NOMAD when the latent dimension $k$
    is varied.}
  \label{fig:vary_dimension}
\end{figure*}

In this subsection, we study the convergence behavior of NOMAD
as we change the dimensionality parameter $k$
(Figure~\ref{fig:vary_dimension}).  In general, the convergence is
faster for smaller values of $k$ as the computational cost of SGD
updates \eqref{eq:w_update} and \eqref{eq:h_update} is linear to $k$.
On the other hand, the model gets richer with higher values of $k$, as
its parameter space expands; it becomes capable of picking up weaker
signals in the data, with the risk of overfitting.  This is observed
in Figure~\ref{fig:vary_dimension} with Netflix (left) and
Yahoo!~Music (right).  In Hugewiki, however, small
values of $k$ were sufficient to fit the training data, and test RMSE
suffers from overfitting with higher values of $k$.  Nonetheless,
NOMAD reliably converged in all cases.

\section{Scaling on Commodity Hardware}
\label{app:ScalDiffHardw}

In this section, we augment Section~\ref{sec:ScalDiffHardw} by
providing actual plots of the experiment.  We increase the number of
machines from 1 to 32, and plot how the convergence of NOMAD is
affected by the number of machines.  As in
Section~\ref{sec:ScalDiffHardw}, we used \texttt{m1.xlarge} machines
from Amazon Web Servies (AWS) which have quad-core Intel Xeon E5430
CPU and 15GB of RAM per each.

\begin{figure*}[htbp]
  \centering
  \begin{tikzpicture}[scale=0.6]

    \begin{axis}[minor tick num=1,
      title={Netflix, cores=4, $\lambda=0.05$, $k=100$},
      xlabel={seconds $\times$ machines $\times$ cores}, ylabel={test RMSE},
      xmax={12000}, ymax={1}]
      
      \addplot
      table [x expr=\thisrowno{3} * 2, y
      index=5, header=false, col sep=comma]
      {../../Plots/aws_nomad_netflix_1_2_100_0.050000_0.008000_0.010000.txt};

      \addplot
      table [x expr=\thisrowno{3}*4, y
      index=5, header=false, col sep=comma]
      {../../Plots/aws_nomad_netflix_2_2_100_0.050000_0.008000_0.010000.txt};

      \addplot
      table [x expr=\thisrowno{3}*8, y
      index=5, header=false, col sep=comma]
      {../../Plots/aws_nomad_netflix_4_2_100_0.050000_0.008000_0.010000.txt};

      \addplot
      table [x expr=\thisrowno{3}*16, y
      index=5, header=false, col sep=comma]
      {../../Plots/aws_nomad_netflix_8_2_100_0.050000_0.008000_0.010000.txt};

      \addplot
      table [x expr=\thisrowno{3}*32, y
      index=5, header=false, col sep=comma]
      {../../Plots/aws_nomad_netflix_16_2_100_0.050000_0.008000_0.010000.txt};

      \addplot
      table [x expr=\thisrowno{3}*64, y
      index=5, header=false, col sep=comma]
      {../../Plots/aws_nomad_netflix_32_2_100_0.050000_0.008000_0.010000.txt};
      
      \legend{\# machines=1, \# machines=2, \# machines=4,\#
        machines=8,\# machines=16, \# machines=32}
    \end{axis}
  \end{tikzpicture}
  \begin{tikzpicture}[scale=0.6]
    \begin{axis}[minor tick num=1,
      title={Yahoo!, cores=4, $\lambda=1.00$, $k=100$},
      xlabel={seconds $\times$ machines $\times$ cores}, ylabel={test RMSE},
      xmax={20000}, ymax={27}]
      
      \addplot
      table [x expr=\thisrowno{3}*2, y
      index=5, header=false, col sep=comma]
      {../../Plots/aws_nomad_yahoo_1_2_100_1.000000_0.000500_0.050000.txt};

      \addplot 
      table [x expr=\thisrowno{3}*4, y
      index=5, header=false, col sep=comma]
      {../../Plots/aws_nomad_yahoo_2_2_100_1.000000_0.000500_0.050000.txt};

      \addplot 
      table [x expr=\thisrowno{3}*8, y
      index=5, header=false, col sep=comma]
      {../../Plots/aws_nomad_yahoo_4_2_100_1.000000_0.000500_0.050000.txt};

      \addplot 
      table [x expr=\thisrowno{3}*16, y
      index=5, header=false, col sep=comma]
      {../../Plots/aws_nomad_yahoo_8_2_100_1.000000_0.000500_0.050000.txt};

      \addplot 
      table [x expr=\thisrowno{3}*32, y
      index=5, header=false, col sep=comma]
      {../../Plots/aws_nomad_yahoo_16_2_100_1.000000_0.000500_0.050000.txt};

      \addplot 
      table [x expr=\thisrowno{3}*64, y
      index=5, header=false, col sep=comma]
      {../../Plots/aws_nomad_yahoo_32_2_100_1.000000_0.000500_0.050000.txt};
      
      \legend{\# machines=1, \# machines=2,\# machines=4,\#
        machines=8, \# machines=16, \# machines=32}
    \end{axis}
  \end{tikzpicture}
  \begin{tikzpicture}[scale=0.6]
    \begin{axis}[minor tick num=1,
      title={Hugewiki, cores=4, $\lambda=0.01$, $k=100$},
      xlabel={seconds $\times$ machines $\times$ cores}, ylabel={test RMSE}, xmax={100000}]
      
      \addplot
      table [x expr=\thisrowno{3}*16, y
      index=5, header=false, col sep=comma]
      {../../Plots/aws_nomad_hugewiki_8_2_100_0.010000_0.000667_0.000000.txt};

      \addplot
      table [x expr=\thisrowno{3}*32, y
      index=5, header=false, col sep=comma]
      {../../Plots/aws_nomad_hugewiki_16_2_100_0.010000_0.000667_0.000000.txt};

      \addplot
      table [x expr=\thisrowno{3}*64, y
      index=5, header=false, col sep=comma]
      {../../Plots/aws_nomad_hugewiki_32_2_100_0.010000_0.000667_0.000000.txt};

      \legend{\# machines=8,
      \# machinees=16, \#machines=32}
    \end{axis}
  \end{tikzpicture}
  \caption{Test RMSE of NOMAD as a function of computation time (time in
    seconds $\times$ the number of machines $\times$ the number of cores
    per each machine) on a commodity hardware cluster, when the number
    of machines is varied.}
  \label{fig:aws_converge_time}
\end{figure*}
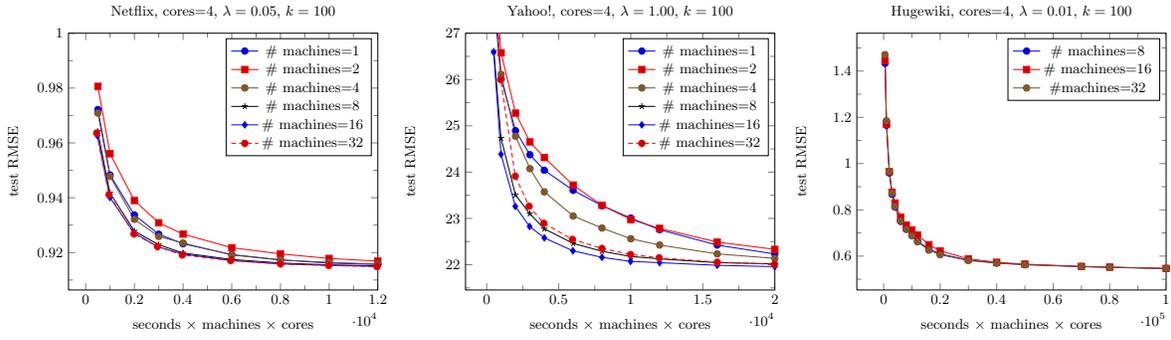

The overall pattern is identical to what was found in
Figure~\ref{fig:multi_converge_update}, \ref{fig:multi_collection} and
\ref{fig:multi_converge_time} of Section~\ref{sec:ScalingasFixed}.
Figure~\ref{fig:aws_converge_update} shows how the test RMSE decreases
as a function of the number of updates.  As in
Figure~\ref{fig:multi_converge_update}, the speed of convergence is
faster with larger number of machines as the updated information is more
frequently exchanged.  Figure~\ref{fig:aws_numupdates} shows the number
of updates performed per second in each computation core of each
machine; NOMAD exhibits linear scaling on Netflix and Hugewiki, but
slows down on Yahoo!~Music due to extreme sparsity of the data.
Figure~\ref{fig:aws_converge_time} compares the convergence speed of
different settings when the same amount of computational power is given
to each; on every dataset we observe linear to super-linear scaling up
to 32 machines.

\begin{figure*}[htbp]
  \centering
  \begin{tikzpicture}[scale=0.6]

    \begin{axis}[minor tick num=1,
      title={Netflix, cores=4, $\lambda=0.05$, $k=100$},
      xlabel={number of updates}, ylabel={test RMSE},
      xmax={15000000000}, ymax={1}]
      
      \addplot
      table [x index=4, y
      index=5, header=false, col sep=comma]
      {../../Plots/aws_nomad_netflix_1_2_100_0.050000_0.008000_0.010000.txt};

      \addplot
      table [x index=4, y
      index=5, header=false, col sep=comma]
      {../../Plots/aws_nomad_netflix_2_2_100_0.050000_0.008000_0.010000.txt};

      \addplot
      table [x index=4, y
      index=5, header=false, col sep=comma]
      {../../Plots/aws_nomad_netflix_4_2_100_0.050000_0.008000_0.010000.txt};

      \addplot
      table [x index=4, y
      index=5, header=false, col sep=comma]
      {../../Plots/aws_nomad_netflix_8_2_100_0.050000_0.008000_0.010000.txt};

      \addplot
      table [x index=4, y
      index=5, header=false, col sep=comma]
      {../../Plots/aws_nomad_netflix_16_2_100_0.050000_0.008000_0.010000.txt};

      \addplot
      table [x index=4, y
      index=5, header=false, col sep=comma]
      {../../Plots/aws_nomad_netflix_32_2_100_0.050000_0.008000_0.010000.txt};
      
      \legend{\# machines=1, \# machines=2, \# machines=4,\#
        machines=8,\# machines=16, \# machines=32}
    \end{axis}
  \end{tikzpicture}
  \begin{tikzpicture}[scale=0.6]
    \begin{axis}[minor tick num=1,
      title={Yahoo!, cores=4, $\lambda=1.00$, $k=100$},
      xlabel={number of updates}, ylabel={test RMSE},
      xmax={30000000000}, ymax={27}]
      
      \addplot
      table [x index=4, y
      index=5, header=false, col sep=comma]
      {../../Plots/aws_nomad_yahoo_1_2_100_1.000000_0.000500_0.050000.txt};

      \addplot 
      table [x index=4, y
      index=5, header=false, col sep=comma]
      {../../Plots/aws_nomad_yahoo_2_2_100_1.000000_0.000500_0.050000.txt};

      \addplot 
      table [x index=4, y
      index=5, header=false, col sep=comma]
      {../../Plots/aws_nomad_yahoo_4_2_100_1.000000_0.000500_0.050000.txt};

      \addplot 
      table [x index=4, y
      index=5, header=false, col sep=comma]
      {../../Plots/aws_nomad_yahoo_8_2_100_1.000000_0.000500_0.050000.txt};

      \addplot 
      table [x index=4, y
      index=5, header=false, col sep=comma]
      {../../Plots/aws_nomad_yahoo_16_2_100_1.000000_0.000500_0.050000.txt};

      \addplot 
      table [x index=4, y
      index=5, header=false, col sep=comma]
      {../../Plots/aws_nomad_yahoo_32_2_100_1.000000_0.000500_0.050000.txt};
      
      \legend{\# machines=1, \# machines=2,\# machines=4,\#
        machines=8, \# machines=16, \# machines=32}
    \end{axis}
  \end{tikzpicture}
  \begin{tikzpicture}[scale=0.6]
    \begin{axis}[minor tick num=1,
      title={Hugewiki, cores=4, $\lambda=0.01$, $k=100$},
      xlabel={number of updates}, ylabel={test RMSE}, xmax={200000000000}]
      
      \addplot
      table [x index=4, y
      index=5, header=false, col sep=comma]
      {../../Plots/aws_nomad_hugewiki_8_2_100_0.010000_0.000667_0.000000.txt};

      \addplot
      table [x index=4, y
      index=5, header=false, col sep=comma]
      {../../Plots/aws_nomad_hugewiki_16_2_100_0.010000_0.000667_0.000000.txt};

      \addplot
      table [x index=4, y
      index=5, header=false, col sep=comma]
      {../../Plots/aws_nomad_hugewiki_32_2_100_0.010000_0.000667_0.000000.txt};
      
      \legend{\# machines=8,
      \# machines=16, \#machines=32}
    \end{axis}
  \end{tikzpicture}
  \caption{Test RMSE of NOMAD as a function of the number of updates
    on a commodity hardware cluster, when the number of machines is
    varied.}
  \label{fig:aws_converge_update}
\end{figure*}
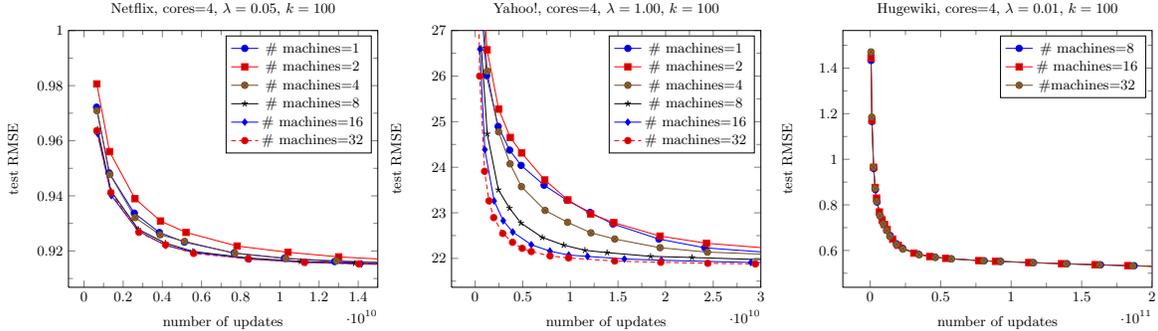

\begin{figure*}[htbp]
  \centering
  \begin{tikzpicture}[scale=0.6]
    \begin{axis}[minor tick num=1, only marks, 
      title={Netflix, cores=4, $\lambda=0.05$, $k=100$},
      xlabel={number of machines}, ylabel={updates per machine per core per sec},
      legend style={legend pos=north east}, ymin={0} ]
      
      \addplot table [x index=1, y
      expr={\thisrowno{4}/\thisrowno{3}/\thisrowno{2}/\thisrowno{1}}, header=false,
      col sep=comma]
      {../../Plots/aws_nomad_netflix_all_2_100_0.050000_0.008000_0.010000.txt};

    \end{axis}
  \end{tikzpicture}
  \begin{tikzpicture}[scale=0.6]
    \begin{axis}[minor tick num=1, only marks, 
      title={Yahoo!, cores=4, $\lambda=1.00$, $k=100$},
      xlabel={number of machines}, ylabel={updates per machine per core per sec},
      legend style={legend pos=north east}, ymin={0} ]
      
      \addplot table [x index=1, y
      expr={\thisrowno{4}/\thisrowno{3}/\thisrowno{2}/\thisrowno{1}}, header=false,
      col sep=comma]
      {../../Plots/aws_nomad_yahoo_all_2_100_1.000000_0.000500_0.050000.txt};

    \end{axis}
  \end{tikzpicture}
  \begin{tikzpicture}[scale=0.6]
    \begin{axis}[minor tick num=1, only marks, 
      title={Hugewiki, cores=4, $\lambda=1.00$, $k=100$},
      xlabel={number of machines}, ylabel={updates per machine per core per sec},
      legend style={legend pos=north east}, ymin={0} ]
      
      \addplot table [x index=1, y
      expr={\thisrowno{4}/\thisrowno{3}/\thisrowno{2}/\thisrowno{1}}, header=false,
      col sep=comma]
      {../../Plots/aws_nomad_hugewiki_all_2_100_0.010000_0.000667_0.000000.txt};

    \end{axis}
  \end{tikzpicture}

  \caption{Number of updates of NOMAD per machine per core per second
    as a function of the number of machines, on a commodity hardware
    cluster.}

  \label{fig:aws_numupdates}
\end{figure*}
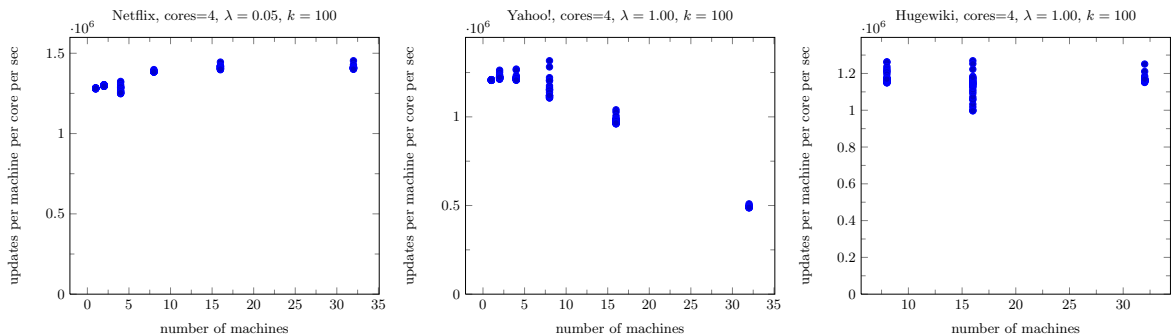

\section{Test RMSE as a function of the number of updates in HPC
  cluster}
\label{sec:test-rmse}

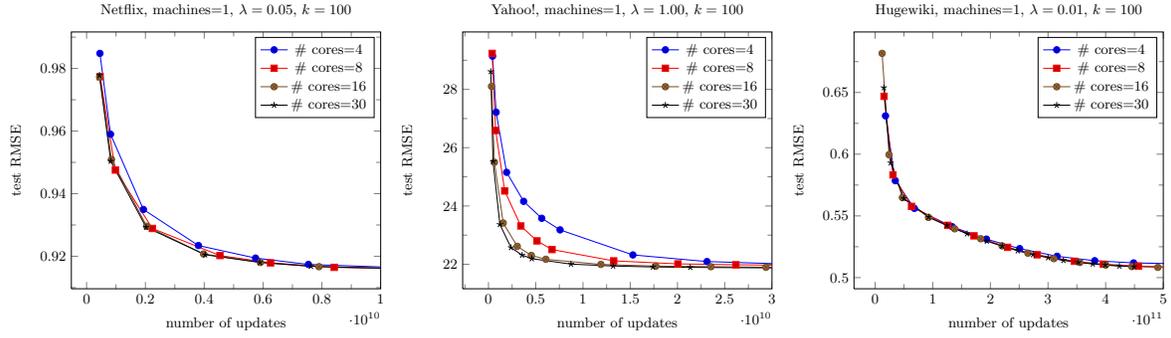
\begin{figure*}[htbp]
  \centering
  \begin{tikzpicture}[scale=0.6]

    \begin{axis}[minor tick num=1,
      title={Netflix, machines=1, $\lambda=0.05$, $k=100$},
      xlabel={number of updates}, ylabel={test RMSE}, xmax={10000000000}]
      
      \addplot
      table [x index=4, y
      index=5, header=false, col sep=comma]
      {../../Plots/stampede_single_nomad_netflix_1_4_100_0.050000_0.008000_0.010000.txt};

      \addplot
      table [x index=4, y
      index=5, header=false, col sep=comma]
      {../../Plots/stampede_single_nomad_netflix_1_8_100_0.050000_0.008000_0.010000.txt};

      \addplot
      table [x index=4, y
      index=5, header=false, col sep=comma]
      {../../Plots/stampede_single_nomad_netflix_1_16_100_0.050000_0.008000_0.010000.txt};

      \addplot
      table [x index=4, y
      index=5, header=false, col sep=comma]
      {../../Plots/stampede_single_nomad_netflix_1_30_100_0.050000_0.008000_0.010000.txt};
      
      \legend{\# cores=4, \# cores=8,\# cores=16,\# cores=30}
    \end{axis}
  \end{tikzpicture}
  \begin{tikzpicture}[scale=0.6]
    \begin{axis}[minor tick num=1,
      title={Yahoo!, machines=1, $\lambda=1.00$, $k=100$},
      xlabel={number of updates}, ylabel={test RMSE}, xmax={30000000000}]
      
      \addplot
      table [x index=4, y
      index=5, header=false, col sep=comma]
      {../../Plots/stampede_single_nomad_yahoo_1_4_100_1.000000_0.000500_0.050000.txt};

      \addplot 
      table [x index=4, y
      index=5, header=false, col sep=comma]
      {../../Plots/stampede_single_nomad_yahoo_1_8_100_1.000000_0.000500_0.050000.txt};

      \addplot 
      table [x index=4, y
      index=5, header=false, col sep=comma]
      {../../Plots/stampede_single_nomad_yahoo_1_16_100_1.000000_0.000500_0.050000.txt};

      \addplot 
      table [x index=4, y
      index=5, header=false, col sep=comma]
      {../../Plots/stampede_single_nomad_yahoo_1_30_100_1.000000_0.000500_0.050000.txt};
      
      \legend{\# cores=4, \# cores=8,\# cores=16,\# cores=30}
    \end{axis}
  \end{tikzpicture}
  \begin{tikzpicture}[scale=0.6]
    \begin{axis}[minor tick num=1,
      title={Hugewiki, machines=1, $\lambda=0.01$, $k=100$},
      xlabel={number of updates}, ylabel={test RMSE}, xmax={500000000000}]
      
      \addplot
      table [x index=4, y
      index=5, header=false, col sep=comma]
      {../../Plots/stampede_single_nomad_hugewiki_1_4_100_0.010000_0.000667_0.000000.txt};

      \addplot
      table [x index=4, y
      index=5, header=false, col sep=comma]
      {../../Plots/stampede_single_nomad_hugewiki_1_8_100_0.010000_0.000667_0.000000.txt};

      \addplot
      table [x index=4, y
      index=5, header=false, col sep=comma]
      {../../Plots/stampede_single_nomad_hugewiki_1_16_100_0.010000_0.000667_0.000000.txt};

      \addplot
      table [x index=4, y
      index=5, header=false, col sep=comma]
      {../../Plots/stampede_single_nomad_hugewiki_1_30_100_0.010000_0.000667_0.000000.txt};

      \legend{\# cores=4,\# cores=8,\# cores=16,\# cores=30}
    \end{axis}
  \end{tikzpicture}
  \caption{Test RMSE of NOMAD as a function of the number of updates,
    when the number of cores is varied.}
  \label{fig:single_converge_update}
\end{figure*}

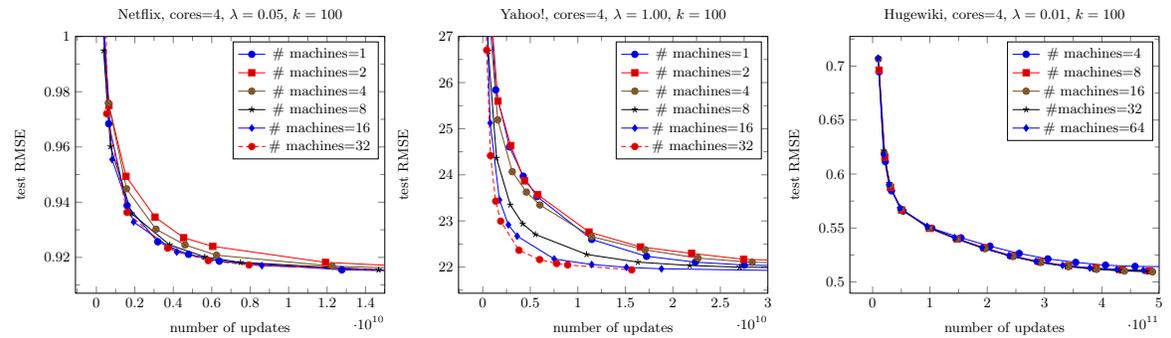
\begin{figure*}[htbp]
  \centering
  \begin{tikzpicture}[scale=0.6]

    \begin{axis}[minor tick num=1,
      title={Netflix, cores=4, $\lambda=0.05$, $k=100$},
      xlabel={number of updates}, ylabel={test RMSE},
      xmax={15000000000}, ymax={1}]
      
      \addplot
      table [x index=4, y
      index=5, header=false, col sep=comma]
      {../../Plots/stampede_multi_nomad_netflix_1_4_100_0.050000_0.008000_0.010000.txt};

      \addplot
      table [x index=4, y
      index=5, header=false, col sep=comma]
      {../../Plots/stampede_multi_nomad_netflix_2_4_100_0.050000_0.008000_0.010000.txt};

      \addplot
      table [x index=4, y
      index=5, header=false, col sep=comma]
      {../../Plots/stampede_multi_nomad_netflix_4_4_100_0.050000_0.008000_0.010000.txt};

      \addplot
      table [x index=4, y
      index=5, header=false, col sep=comma]
      {../../Plots/stampede_multi_nomad_netflix_8_4_100_0.050000_0.008000_0.010000.txt};

      \addplot
      table [x index=4, y
      index=5, header=false, col sep=comma]
      {../../Plots/stampede_multi_nomad_netflix_16_4_100_0.050000_0.008000_0.010000.txt};

      \addplot
      table [x index=4, y
      index=5, header=false, col sep=comma]
      {../../Plots/stampede_multi_nomad_netflix_32_4_100_0.050000_0.008000_0.010000.txt};
      
      \legend{\# machines=1, \# machines=2, \# machines=4,\#
        machines=8,\# machines=16, \# machines=32}
    \end{axis}
  \end{tikzpicture}
  \begin{tikzpicture}[scale=0.6]
    \begin{axis}[minor tick num=1,
      title={Yahoo!, cores=4, $\lambda=1.00$, $k=100$},
      xlabel={number of updates}, ylabel={test RMSE},
      xmax={30000000000}, ymax={27}]
      
      \addplot
      table [x index=4, y
      index=5, header=false, col sep=comma]
      {../../Plots/stampede_multi_nomad_yahoo_1_4_100_1.000000_0.000500_0.050000.txt};

      \addplot 
      table [x index=4, y
      index=5, header=false, col sep=comma]
      {../../Plots/stampede_multi_nomad_yahoo_2_4_100_1.000000_0.000500_0.050000.txt};

      \addplot 
      table [x index=4, y
      index=5, header=false, col sep=comma]
      {../../Plots/stampede_multi_nomad_yahoo_4_4_100_1.000000_0.000500_0.050000.txt};

      \addplot 
      table [x index=4, y
      index=5, header=false, col sep=comma]
      {../../Plots/stampede_multi_nomad_yahoo_8_4_100_1.000000_0.000500_0.050000.txt};

      \addplot 
      table [x index=4, y
      index=5, header=false, col sep=comma]
      {../../Plots/stampede_multi_nomad_yahoo_16_4_100_1.000000_0.000500_0.050000.txt};

      \addplot 
      table [x index=4, y
      index=5, header=false, col sep=comma]
      {../../Plots/stampede_multi_nomad_yahoo_32_4_100_1.000000_0.000500_0.050000.txt};
      
      \legend{\# machines=1, \# machines=2,\# machines=4,\#
        machines=8, \# machines=16, \# machines=32}
    \end{axis}
  \end{tikzpicture}
  \begin{tikzpicture}[scale=0.6]
    \begin{axis}[minor tick num=1,
      title={Hugewiki, cores=4, $\lambda=0.01$, $k=100$},
      xlabel={number of updates}, ylabel={test RMSE}, xmax={500000000000}]
      
      \addplot
      table [x index=4, y
      index=5, header=false, col sep=comma]
      {../../Plots/stampede_multi_nomad_hugewiki_4_4_100_0.010000_0.000667_0.000000.txt};

      \addplot
      table [x index=4, y
      index=5, header=false, col sep=comma]
      {../../Plots/stampede_multi_nomad_hugewiki_8_4_100_0.010000_0.000667_0.000000.txt};

      \addplot
      table [x index=4, y
      index=5, header=false, col sep=comma]
      {../../Plots/stampede_multi_nomad_hugewiki_16_4_100_0.010000_0.000667_0.000000.txt};

      \addplot
      table [x index=4, y
      index=5, header=false, col sep=comma]
      {../../Plots/stampede_multi_nomad_hugewiki_32_4_100_0.010000_0.000667_0.000000.txt};

      \addplot
      table [x index=4, y
      index=5, header=false, col sep=comma]
      {../../Plots/stampede_multi_nomad_hugewiki_64_4_100_0.010000_0.000667_0.000000.txt};
      
      \legend{\# machines=4,\# machines=8,
      \# machines=16, \#machines=32, \# machines=64}
    \end{axis}
  \end{tikzpicture}
  \caption{Test RMSE of NOMAD as a function of the number of updates
    on a HPC cluster, when the number of machines is varied.}
  \label{fig:multi_converge_update}
\end{figure*}

In this section we plot the test RMSE as a function of the number of
updates in HPC cluster, which were not included in the main text due
to page constraints.  Figure~\ref{fig:single_converge_update} shows
single-machine multi-threaded experiments, and
Figure~\ref{fig:multi_converge_update} shows multi-machine distributed
memory experiments.

\section{Comparison of Algorithms for Different Values of
    the Regularization Parameter}

\begin{figure*}[htbp]
  \centering
  \begin{tikzpicture}[scale=0.6]
    \begin{axis}[minor tick num=1,
      title={Netflix, machines=32, cores=4, $\lambda=0.0125$, $k=100$},
      xlabel={seconds}, ylabel={test RMSE},
      ymin={0.90}, ymax={1.1}, xmin={0}, xmax={80}
      ]
      
      \addplot[ultra thick, color=blue, mark=*] table [x index=3, y
      index=5, header=false, col sep=comma]
      {../../Plots/stampede_varyreg_nomad_netflix_32_4_100_0.012500_0.008000_0.010000.txt};
      
      \addplot[ultra thick, color=red, mark=none] table [x index=0, y index=1, header=false]
      {../../Plots/stampede_varyreg_dsgd_bold_wor_netflix_32_4_100_0.012500_0.008000_0.500000.txt};

      \addplot[ultra thick, color=brown, mark=none, dashed] table [x index=0, y index=1, header=false]
      {../../Plots/stampede_varyreg_ccdpp_netflix_32_4_100_0.012500.txt};

      \legend{NOMAD, DSGD, CCD++}
    \end{axis}
  \end{tikzpicture}
  \begin{tikzpicture}[scale=0.6]
    \begin{axis}[minor tick num=1,
      title={Yahoo!, machines=32, cores=4, $\lambda=0.25$, $k=100$},
      xlabel={seconds}, ylabel={test RMSE},
      ymin={21}, ymax={27}, xmin={0}, xmax={80}]
      
      \addplot[ultra thick, color=blue, mark=*] table [x index=3, y
      index=5, header=false, col sep=comma]
      {../../Plots/stampede_varyreg_nomad_yahoo_32_4_100_0.250000_0.000500_0.050000.txt};
      
      \addplot[ultra thick, color=red, mark=none] table [x index=0, y index=1, header=false]
      {../../Plots/stampede_varyreg_dsgd_bold_wor_yahoo_32_4_100_0.250000_0.000500_0.500000.txt};

      \addplot[ultra thick, color=brown, mark=none, dashed] table [x index=0, y index=1, header=false]
      {../../Plots/stampede_varyreg_ccdpp_yahoo_32_4_100_0.250000.txt};

      \legend{NOMAD, DSGD, CCD++}
    \end{axis}
  \end{tikzpicture}
  \begin{tikzpicture}[scale=0.6]
    \begin{axis}[minor tick num=1,
      title={Hugewiki, machines=64, cores=4, $\lambda=0.0025$, $k=100$},
      xlabel={seconds}, ylabel={test RMSE},
      ymin={0.5}, ymax={0.7}, xmin={0}, xmax={2000}]
      
      \addplot[ultra thick, color=blue, mark=*] table [x index=3, y
      index=5, header=false, col sep=comma]
      {../../Plots/stampede_varyreg_nomad_hugewiki_64_4_100_0.002500_0.000667_0.000000.txt};
      
      \addplot[ultra thick, color=red, mark=none] table [x index=0, y index=1, header=false]      
      {../../Plots/stampede_varyreg_dsgd_bold_wor_hugewiki_64_4_100_0.002500_0.008000_0.500000.txt};

      \addplot[ultra thick, color=brown, mark=none, dashed] table [x index=0, y index=1, header=false]
      {../../Plots/stampede_varyreg_ccdpp_hugewiki_64_4_100_0.002500.txt};

      \legend{NOMAD, DSGD, CCD++}
    \end{axis}
  \end{tikzpicture}

  \centering
  \begin{tikzpicture}[scale=0.6]
    \begin{axis}[minor tick num=1,
      title={Netflix, machines=32, cores=4, $\lambda=0.025$, $k=100$},
      xlabel={seconds}, ylabel={test RMSE},
      ymin={0.90}, ymax={1.00}, xmin={0}, xmax={80}
      ]
      
      \addplot[ultra thick, color=blue, mark=*] table [x index=3, y
      index=5, header=false, col sep=comma]
      {../../Plots/stampede_varyreg_nomad_netflix_32_4_100_0.025000_0.008000_0.010000.txt};
      
      \addplot[ultra thick, color=red, mark=none] table [x index=0, y index=1, header=false]
      {../../Plots/stampede_varyreg_dsgd_bold_wor_netflix_32_4_100_0.025000_0.008000_0.500000.txt};

      \addplot[ultra thick, color=brown, mark=none, dashed] table [x index=0, y index=1, header=false]
      {../../Plots/stampede_varyreg_ccdpp_netflix_32_4_100_0.025000.txt};

      \legend{NOMAD, DSGD, CCD++}
    \end{axis}
  \end{tikzpicture}
  \begin{tikzpicture}[scale=0.6]
    \begin{axis}[minor tick num=1,
      title={Yahoo!, machines=32, cores=4, $\lambda=0.5$, $k=100$},
      xlabel={seconds}, ylabel={test RMSE},
      ymin={21}, ymax={27}, xmin={0}, xmax={80}]
      
      \addplot[ultra thick, color=blue, mark=*] table [x index=3, y
      index=5, header=false, col sep=comma]
      {../../Plots/stampede_varyreg_nomad_yahoo_32_4_100_0.500000_0.000500_0.050000.txt};
      
      \addplot[ultra thick, color=red, mark=none] table [x index=0, y index=1, header=false]
      {../../Plots/stampede_varyreg_dsgd_bold_wor_yahoo_32_4_100_0.500000_0.000500_0.500000.txt};

      \addplot[ultra thick, color=brown, mark=none, dashed] table [x index=0, y index=1, header=false]
      {../../Plots/stampede_varyreg_ccdpp_yahoo_32_4_100_0.500000.txt};

      \legend{NOMAD, DSGD, CCD++}
    \end{axis}
  \end{tikzpicture}
  \begin{tikzpicture}[scale=0.6]
    \begin{axis}[minor tick num=1,
      title={Hugewiki, machines=64, cores=4, $\lambda=0.005$, $k=100$},
      xlabel={seconds}, ylabel={test RMSE},
      ymin={0.5}, ymax={0.7}, xmin={0}, xmax={2000}]
      
      \addplot[ultra thick, color=blue, mark=*] table [x index=3, y
      index=5, header=false, col sep=comma]
      {../../Plots/stampede_varyreg_nomad_hugewiki_64_4_100_0.005000_0.000667_0.000000.txt};
      
      \addplot[ultra thick, color=red, mark=none] table [x index=0, y index=1, header=false]      
      {../../Plots/stampede_varyreg_dsgd_bold_wor_hugewiki_64_4_100_0.005000_0.008000_0.500000.txt};

      \addplot[ultra thick, color=brown, mark=none, dashed] table [x index=0, y index=1, header=false]
      {../../Plots/stampede_varyreg_ccdpp_hugewiki_64_4_100_0.005000.txt};

      \legend{NOMAD, DSGD, CCD++}
    \end{axis}
  \end{tikzpicture}

  \centering
  \begin{tikzpicture}[scale=0.6]
    \begin{axis}[minor tick num=1,
      title={Netflix, machines=32, cores=4, $\lambda=0.05$, $k=100$},
      xlabel={seconds}, ylabel={test RMSE},
      ymin={0.91}, ymax={1.00}, xmin={0}, xmax={150}
      ]
      
      \addplot[ultra thick, color=blue, mark=*] table [x index=3, y
      index=5, header=false, col sep=comma]
      {../../Plots/stampede_multi_nomad_netflix_32_4_100_0.050000_0.008000_0.010000.txt};
      
      \addplot[ultra thick, color=red, mark=none] table [x index=0, y index=1, header=false]
      {../../Plots/stampede_multi_dsgd_bold_netflix_32_4_100_0.050000_0.008000_0.010000.txt};

      \addplot[ultra thick, color=brown, mark=none, dashed] table [x index=0, y index=1, header=false]
      {../../Plots/stampede_multi_ccdpp_netflix_32_4_100_0.050000.txt};

      \legend{NOMAD, DSGD, CCD++}
    \end{axis}
  \end{tikzpicture}
  \begin{tikzpicture}[scale=0.6]
    \begin{axis}[minor tick num=1,
      title={Yahoo!, machines=32, cores=4, $\lambda=1.00$, $k=100$},
      xlabel={seconds}, ylabel={test RMSE},
      ymin={21}, ymax={27}, xmin={0}, xmax={80}]
      
      \addplot[ultra thick, color=blue, mark=*] table [x index=3, y
      index=5, header=false, col sep=comma]
      {../../Plots/stampede_multi_nomad_yahoo_32_4_100_1.000000_0.000500_0.050000.txt};
      
      \addplot[ultra thick, color=red, mark=none] table [x index=0, y index=1, header=false]
      {../../Plots/stampede_multi_dsgd_bold_wor_yahoo_32_4_100_1.000000_0.000500_0.500000.txt};

      \addplot[ultra thick, color=brown, mark=none, dashed] table [x index=0, y index=1, header=false]
      {../../Plots/stampede_multi_ccdpp_yahoo_32_4_100_1.000000.txt};

      \legend{NOMAD, DSGD, CCD++}
    \end{axis}
  \end{tikzpicture}
  \begin{tikzpicture}[scale=0.6]
    \begin{axis}[minor tick num=1,
      title={Hugewiki, machines=64, cores=4, $\lambda=0.01$, $k=100$},
      xlabel={seconds}, ylabel={test RMSE},
      ymin={0.5}, ymax={0.7}, xmin={0}, xmax={2000}]
      
      \addplot[ultra thick, color=blue, mark=*] table [x index=3, y
      index=5, header=false, col sep=comma]
      {../../Plots/stampede_multi_nomad_hugewiki_64_4_100_0.010000_0.000667_0.000000.txt};
      
      \addplot[ultra thick, color=red, mark=none] table [x index=0, y index=1, header=false]      
      {../../Plots/stampede_multi_dsgd_bold_wor_hugewiki_64_4_100_0.010000_0.008000_0.500000.txt};

      \addplot[ultra thick, color=brown, mark=none, dashed] table [x index=0, y index=1, header=false]
      {../../Plots/stampede_multi_ccdpp_hugewiki_64_4_100_0.010000.txt};

      \legend{NOMAD, DSGD, CCD++}
    \end{axis}
  \end{tikzpicture}

  \centering
  \begin{tikzpicture}[scale=0.6]
    \begin{axis}[minor tick num=1,
      title={Netflix, machines=32, cores=4, $\lambda=0.1$, $k=100$},
      xlabel={seconds}, ylabel={test RMSE},
      ymin={0.93}, ymax={1.00}, xmin={0}, xmax={80}
      ]
      
      \addplot[ultra thick, color=blue, mark=*] table [x index=3, y
      index=5, header=false, col sep=comma]
      {../../Plots/stampede_varyreg_nomad_netflix_32_4_100_0.100000_0.008000_0.010000.txt};
      
      \addplot[ultra thick, color=red, mark=none] table [x index=0, y index=1, header=false]
      {../../Plots/stampede_varyreg_dsgd_bold_wor_netflix_32_4_100_0.100000_0.008000_0.500000.txt};

      \addplot[ultra thick, color=brown, mark=none, dashed] table [x index=0, y index=1, header=false]
      {../../Plots/stampede_varyreg_ccdpp_netflix_32_4_100_0.100000.txt};

      \legend{NOMAD, DSGD, CCD++}
    \end{axis}
  \end{tikzpicture}
  \begin{tikzpicture}[scale=0.6]
    \begin{axis}[minor tick num=1,
      title={Yahoo!, machines=32, cores=4, $\lambda=2$, $k=100$},
      xlabel={seconds}, ylabel={test RMSE},
      ymin={22}, ymax={27}, xmin={0}, xmax={80}]
      
      \addplot[ultra thick, color=blue, mark=*] table [x index=3, y
      index=5, header=false, col sep=comma]
      {../../Plots/stampede_varyreg_nomad_yahoo_32_4_100_2.000000_0.000500_0.050000.txt};
      
      \addplot[ultra thick, color=red, mark=none] table [x index=0, y index=1, header=false]
      {../../Plots/stampede_varyreg_dsgd_bold_wor_yahoo_32_4_100_2.000000_0.000500_0.500000.txt};

      \addplot[ultra thick, color=brown, mark=none, dashed] table [x index=0, y index=1, header=false]
      {../../Plots/stampede_varyreg_ccdpp_yahoo_32_4_100_2.000000.txt};

      \legend{NOMAD, DSGD, CCD++}
    \end{axis}
  \end{tikzpicture}
  \begin{tikzpicture}[scale=0.6]
    \begin{axis}[minor tick num=1,
      title={Hugewiki, machines=64, cores=4, $\lambda=0.02$, $k=100$},
      xlabel={seconds}, ylabel={test RMSE},
      ymin={0.5}, ymax={0.7}, xmin={0}, xmax={2000}]
      
      \addplot[ultra thick, color=blue, mark=*] table [x index=3, y
      index=5, header=false, col sep=comma]
      {../../Plots/stampede_varyreg_nomad_hugewiki_64_4_100_0.020000_0.000667_0.000000.txt};
      
      \addplot[ultra thick, color=red, mark=none] table [x index=0, y index=1, header=false]      
      {../../Plots/stampede_varyreg_dsgd_bold_wor_hugewiki_64_4_100_0.020000_0.008000_0.500000.txt};

      \addplot[ultra thick, color=brown, mark=none, dashed] table [x index=0, y index=1, header=false]
      {../../Plots/stampede_varyreg_ccdpp_hugewiki_64_4_100_0.020000.txt};

      \legend{NOMAD, DSGD, CCD++}
    \end{axis}
  \end{tikzpicture}

  \centering
  \begin{tikzpicture}[scale=0.6]
    \begin{axis}[minor tick num=1,
      title={Netflix, machines=32, cores=4, $\lambda=0.2$, $k=100$},
      xlabel={seconds}, ylabel={test RMSE},
      ymin={0.98}, ymax={1.1}, xmin={0}, xmax={80}
      ]
      
      \addplot[ultra thick, color=blue, mark=*] table [x index=3, y
      index=5, header=false, col sep=comma]
      {../../Plots/stampede_varyreg_nomad_netflix_32_4_100_0.200000_0.008000_0.010000.txt};
      
      \addplot[ultra thick, color=red, mark=none] table [x index=0, y index=1, header=false]
      {../../Plots/stampede_varyreg_dsgd_bold_wor_netflix_32_4_100_0.200000_0.008000_0.500000.txt};

      \addplot[ultra thick, color=brown, mark=none, dashed] table [x index=0, y index=1, header=false]
      {../../Plots/stampede_varyreg_ccdpp_netflix_32_4_100_0.200000.txt};

      \legend{NOMAD, DSGD, CCD++}
    \end{axis}
  \end{tikzpicture}
  \begin{tikzpicture}[scale=0.6]
    \begin{axis}[minor tick num=1,
      title={Yahoo!, machines=32, cores=4, $\lambda=4.00$, $k=100$},
      xlabel={seconds}, ylabel={test RMSE},
      ymin={23.5}, ymax={27}, xmin={0}, xmax={80}]
      
      \addplot[ultra thick, color=blue, mark=*] table [x index=3, y
      index=5, header=false, col sep=comma]
      {../../Plots/stampede_varyreg_nomad_yahoo_32_4_100_4.000000_0.000500_0.050000.txt};
      
      \addplot[ultra thick, color=red, mark=none] table [x index=0, y index=1, header=false]
      {../../Plots/stampede_varyreg_dsgd_bold_wor_yahoo_32_4_100_4.000000_0.000500_0.500000.txt};

      \addplot[ultra thick, color=brown, mark=none, dashed] table [x index=0, y index=1, header=false]
      {../../Plots/stampede_varyreg_ccdpp_yahoo_32_4_100_4.000000.txt};

      \legend{NOMAD, DSGD, CCD++}
    \end{axis}
  \end{tikzpicture}
  \begin{tikzpicture}[scale=0.6]
    \begin{axis}[minor tick num=1,
      title={Hugewiki, machines=64, cores=4, $\lambda=0.04$, $k=100$},
      xlabel={seconds}, ylabel={test RMSE},
      ymin={0.5}, ymax={0.7}, xmin={0}, xmax={2000}]
      
      \addplot[ultra thick, color=blue, mark=*] table [x index=3, y
      index=5, header=false, col sep=comma]
      {../../Plots/stampede_varyreg_nomad_hugewiki_64_4_100_0.040000_0.000667_0.000000.txt};
      
      \addplot[ultra thick, color=red, mark=none] table [x index=0, y index=1, header=false]      
      {../../Plots/stampede_varyreg_dsgd_bold_wor_hugewiki_64_4_100_0.040000_0.008000_0.500000.txt};

      \addplot[ultra thick, color=brown, mark=none, dashed] table [x index=0, y index=1, header=false]
      {../../Plots/stampede_varyreg_ccdpp_hugewiki_64_4_100_0.040000.txt};

      \legend{NOMAD, DSGD, CCD++}
    \end{axis}
  \end{tikzpicture}

  \caption{Comparison of NOMAD, DSGD and CCD++ on a HPC cluster when
    the regularization paramter $\lambda$ is varied.  The value of
    $\lambda$ increases from top to bottom.}
  \label{fig:varyreg_compare}
\end{figure*}
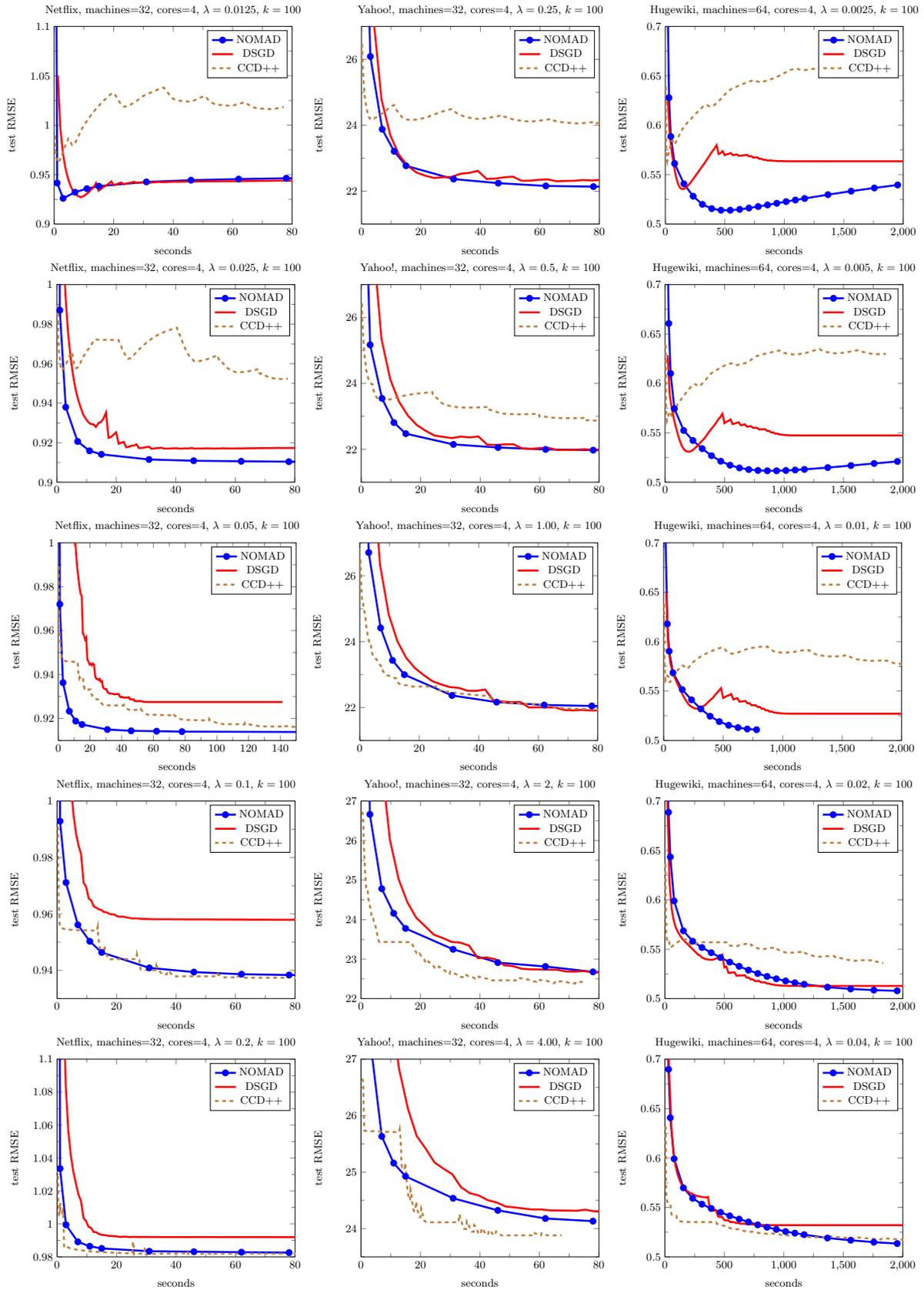

  In this section, we augment experiments in
  Section~\ref{sec:ScalingasFixed} by comparing the performance of
  NOMAD, CCD++, and DSGD on different values of the regularization
  parameter $\lambda$.  Figure~\ref{fig:varyreg_compare} shows the
  result of the experiment.  As NOMAD and DSGD are both stochastic
  gradient descent methods, they behave similarly to each other when
  the regularization parameter is changed.  On the other hand, CCD++,
  which decreases the objective function more greedily, behaves a
  differently.

  For small values of $\lambda$, CCD++ seems to overfit to the model
  due to its greedy strategy; it generally converges to a worse
  solution than others.  For high values of $\lambda$, however, the
  strategy of CCD++ is advantageous and it shows rapid initial
  convergence.  Note that in all cases, NOMAD is competitive with the
  better of the other two algorithms.

\section{Comparison with GraphLab }

\label{sec:comp-with-graphl}

Here we provide experimental comparison with GraphLab of
\citet{LowGonKyrBicetal12}.  GraphLab PowerGraph 2.2, which can be
downloaded from \url{https://github.com/graphlab-code/graphlab} was
used in our experiments.  Since GraphLab was not compatible with Intel
compiler, we had to compile it with \texttt{gcc}.  The rest of
experimental setting is identical to what was described in
Section~\ref{sec:ExperimentalSetup}.

Among a number of algorithms GraphLab provides for matrix completion
in its collaborative filtering toolkit, only Alternating Least Squares
(ALS) algorithm is suitable for solving the objective function
\eqref{eq:obj_fcn}; unfortunately, Stochastic Gradient Descent (SGD)
implementation of GraphLab does not converge.  According to private
conversations with GraphLab developers, this is because the
abstraction currently provided by GraphLab is not suitable for the SGD
algorithm.  Its \texttt{biassgd} algorithm, on the other hand, is
based on a model different from \eqref{eq:obj_fcn} and therefore not
directly comparable to NOMAD as an optimization algorithm.

Although each machine in HPC cluster is equipped with 32 GB of RAM and
we distribute the work into 32 machines in multi-machine experiments,
we had to tune \texttt{nfibers} parameter to avoid out of memory
problems, and still was not able to run GraphLab on Hugewiki data in
any setting.  We tried both synchronous and asynchronous engines of
GraphLab, and report the better of the two on each configuration.

\begin{figure*}
  \centering
  \centering
  \begin{tikzpicture}[scale=0.6]
    \begin{axis}[minor tick num=1,
      title={Netflix, machines=1, cores=30, $\lambda=0.05$, $k=100$},
      xlabel={seconds}, ylabel={test RMSE},
      ymin={0.91}, ymax={1.1}, xmin={0}, xmax={3000}]
      
      \addplot[ultra thick, color=blue, mark=*] table [x index=2, y index=1, header=false]
      {../../Results/stam_cpus_netflix_30.txt};
      
      \addplot[ultra thick, color=red, mark=none] table [x index=0, y index=1, header=false]
      {../../Plots/stampede_single_graphlab_try_netflix_1_30_100_0.050000_1.000000_5.000000_1_head.txt};
      
      \legend{NOMAD, GraphLab ALS}
    \end{axis}
  \end{tikzpicture}
  \begin{tikzpicture}[scale=0.6]
    \begin{axis}[minor tick num=1,
      title={Yahoo!, machines=1, cores=30, $\lambda=1.00$, $k=100$},
      xlabel={seconds}, ylabel={test RMSE},
      ymin={21}, ymax={30}, xmin={0}, xmax={6000}]
      
      \addplot[ultra thick, color=blue, mark=*] table [x index=2, y index=1, header=false]
      {../../Results/stam_cpus_yahoo_30.txt};
      
      \addplot[ultra thick, color=red, mark=none] table [x index=0, y index=1, header=false]
      {../../Plots/stampede_single_graphlab_try_yahoo_1_30_100_1.000000_0.000000_100.000000_2_head.txt};
      
      \legend{NOMAD, GraphLab ALS}
    \end{axis}
  \end{tikzpicture}

  \caption{Comparison of NOMAD and GraphLab on a single machine with
    30 computation cores. }
  \label{fig:graphlab_single_compare}
\end{figure*}
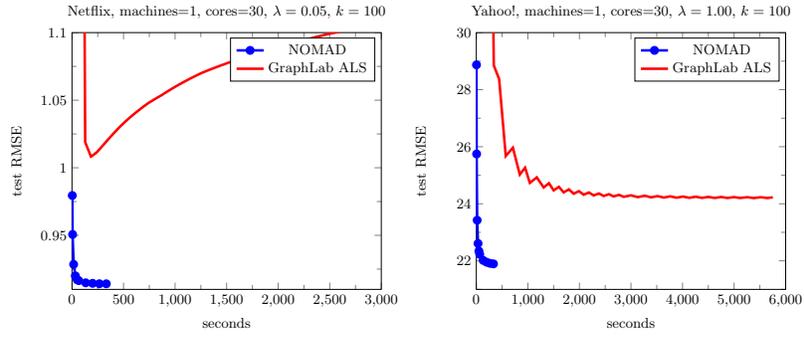

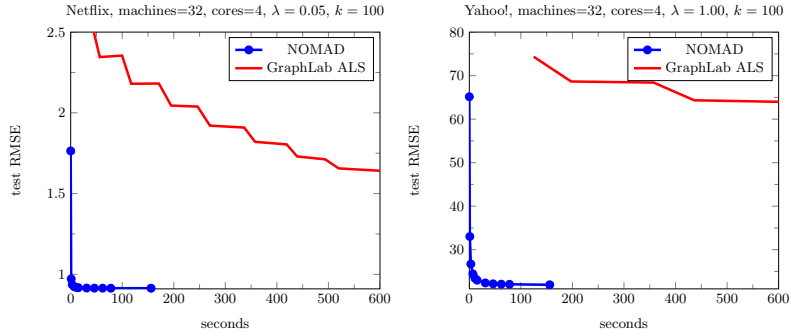
\begin{figure*}[htbp]
  \centering
  \begin{tikzpicture}[scale=0.6]
    \begin{axis}[minor tick num=1,
      title={Netflix, machines=32, cores=4, $\lambda=0.05$, $k=100$},
      xlabel={seconds}, ylabel={test RMSE},
      ymin={0.91}, ymax={2.50}, xmin={0}, xmax={600}
      ]
      
      \addplot[ultra thick, color=blue, mark=*] table [x index=3, y
      index=5, header=false, col sep=comma]
      {../../Plots/stampede_multi_nomad_netflix_32_4_100_0.050000_0.008000_0.010000.txt};
      
      \addplot[ultra thick, color=red, mark=none] table [x index=0, y index=1, header=false]
      {../../Plots/stampede_graphlab_netflix.txt};

      \legend{NOMAD, GraphLab ALS}
    \end{axis}
  \end{tikzpicture}
  \begin{tikzpicture}[scale=0.6]
    \begin{axis}[minor tick num=1,
      title={Yahoo!, machines=32, cores=4, $\lambda=1.00$, $k=100$},
      xlabel={seconds}, ylabel={test RMSE},
      ymin={21}, ymax={80}, xmin={0}, xmax={600}]
      
      \addplot[ultra thick, color=blue, mark=*] table [x index=3, y
      index=5, header=false, col sep=comma]
      {../../Plots/stampede_multi_nomad_yahoo_32_4_100_1.000000_0.000500_0.050000.txt};
      
      \addplot[ultra thick, color=red, mark=none] table [x index=0, y index=1, header=false]
      {../../Plots/stampede_graphlab_yahoo.txt};

      \legend{NOMAD, GraphLab ALS}
    \end{axis}
  \end{tikzpicture}

  \caption{Comparison of NOMAD and GraphLab on a HPC cluster.}
  \label{fig:graphlab_multi_compare}
\end{figure*}

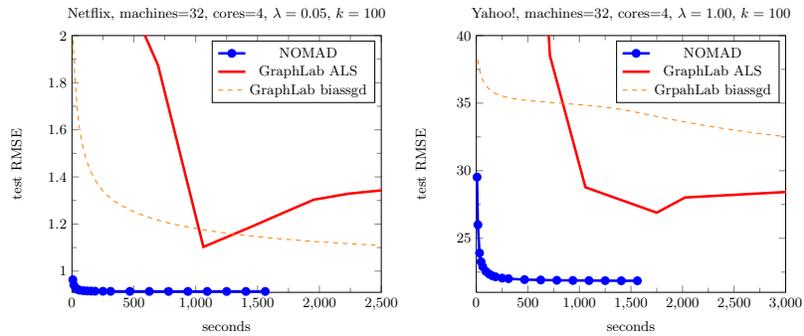
\begin{figure*}[htbp]
  \centering
  \begin{tikzpicture}[scale=0.6]
    \begin{axis}[minor tick num=1,
      title={Netflix, machines=32, cores=4, $\lambda=0.05$, $k=100$},
      xlabel={seconds}, ylabel={test RMSE},
      ymin={0.91}, ymax={2}, xmin={0}, xmax={2500}]
      
      \addplot[ultra thick, color=blue, mark=*] table [x index=3, y
      index=5, header=false, col sep=comma]
      {../../Plots/aws_nomad_netflix_32_2_100_0.050000_0.008000_0.010000.txt};

      \addplot[ultra thick, color=red, mark=none] table [x index=0, y index=1, header=false]      
      {../../Plots/aws_graphlab_als_netflix.txt};

      \addplot [dashed, color=orange, mark=none]
      table [x expr=\thisrowno{0}*0.5, y
      index=1, header=false]
      {../../Plots/aws_graphlab_biassgd_netflix.txt};
     
      \legend{NOMAD, GraphLab ALS, GraphLab biassgd}
    \end{axis}
  \end{tikzpicture}
  \begin{tikzpicture}[scale=0.6]
    \begin{axis}[minor tick num=1,
      title={Yahoo!, machines=32, cores=4, $\lambda=1.00$, $k=100$},
      xlabel={seconds}, ylabel={test RMSE},
      ymin={21}, ymax={40}, xmin={0}, xmax={3000}]
      
      \addplot[ultra thick, color=blue, mark=*] table [x index=3, y
      index=5, header=false, col sep=comma]
      {../../Plots/aws_nomad_yahoo_32_2_100_1.000000_0.000500_0.050000.txt};
      
      \addplot[ultra thick, color=red, mark=none] table [x index=0, y index=1, header=false]      
      {../../Plots/aws_graphlab_als_yahoo.txt};

      \addplot [dashed, color=orange, mark=none]
      table [x expr=\thisrowno{0}*0.5, y
      index=1, header=false]
      {../../Plots/aws_graphlab_biasssgd_yahoo.txt};

      \legend{NOMAD, GraphLab ALS, GrpahLab biassgd}
    \end{axis}
  \end{tikzpicture}

  \caption{Comparison of NOMAD and GraphLab on a commodity hardware
    cluster.  }

  \label{fig:graphlab_aws_compare}
\end{figure*}

Figure~\ref{fig:graphlab_single_compare} shows results of
single-machine multi-threaded experiments, while
Figure~\ref{fig:graphlab_multi_compare} and
Figure~\ref{fig:graphlab_aws_compare} shows multi-machine experiments
on HPC cluster and commodity cluster respectively.  Clearly, NOMAD
converges orders of magnitude faster than GraphLab in every setting,
and also converges to a better solution.  Note that GraphLab converges
faster in single-machine setting with large number of cores (30) than
in multi-machine setting with large number of machines (32) but small
number of cores (4) each.  We conjecture that this is because the
locking and unlocking of a variable has to be requested via network
communication in distributed memory setting; on the other hand, NOMAD
does not require a locking mechanism and thus scales better with the
number of machines.

Although GraphLab \texttt{biassgd} is based on a model different from
\eqref{eq:obj_fcn}, for the interest of readers we provide comparisons
with it on commodity hardware cluster.  Unfortunately, GraphLab
\texttt{biassgd} crashed when we ran it on more than 16 machines, so
we had to run it on only 16 machines and assumed GraphLab will
linearly scale up to 32 machines, in order to generate plots in
Figure~\ref{fig:graphlab_aws_compare}.  Again, NOMAD was orders of
magnitude faster than GraphLab and converges to a better solution.

\end{document}